\documentclass[aps,unsortedaddress,preprint]{revtex4-1}
\usepackage{graphicx}
\usepackage{dcolumn}
\usepackage{bm}
\usepackage{subfig}
\usepackage[utf8]{inputenc}
\usepackage{amssymb}
\usepackage{placeins}
\usepackage{amsmath}
\usepackage{amsthm}
\usepackage{commath}
\usepackage[margin=1in]{geometry}

\newcommand{\partder}[2]{\dfrac{\partial  #1}{\partial  #2}} 
\newcommand{\der}[2]{\dfrac{d #1}{d  #2}}

\begin{document}
\title{Rotation and Neoclassical Ripple Transport in ITER}
\author{E. J. Paul}
\affiliation{Department of Physics, University of Maryland, College Park, MD 20742, USA}
\email{ejpaul@umd.edu}

\author{M. Landreman}
\affiliation{Institute for Research in Electronics and Applied Physics, University of Maryland, College Park, MD 20742, USA}

\author{F. M. Poli}
\affiliation{Princeton Plasma Physics Laboratory, Princeton, NJ 08543, USA}

\author{D. A. Spong}
\affiliation{Oak Ridge National Laboratory, Oak Ridge, TN 37831, USA}

\author{H. M. Smith}
\affiliation{Max-Planck-Institut f\"{u}r Plasmaphysik, 17491 Greifswald, Germany}

\author{W. Dorland}
\affiliation{Department of Physics, University of Maryland, College Park, MD 20742, USA}

\begin{abstract}

Neoclassical transport in the presence of non-axisymmetric magnetic fields causes a toroidal torque known as neoclassical toroidal viscosity (NTV). The toroidal symmetry of ITER will be broken by the finite number of toroidal field coils and by test blanket modules (TBMs). The addition of ferritic inserts (FIs) will decrease the magnitude of the toroidal field ripple. 3D magnetic equilibria with toroidal field ripple and ferromagnetic structures are calculated for an ITER steady-state scenario using the Variational Moments Equilibrium Code (VMEC). Neoclassical transport quantities in the presence of these error fields are calculated using the Stellarator Fokker-Planck Iterative Neoclassical Conservative Solver (SFINCS). These calculations fully account for $E_r$, flux surface shaping, multiple species, magnitude of ripple, and collisionality rather than applying approximate analytic NTV formulae. As NTV is a complicated nonlinear function of $E_r$, we study its behavior over a plausible range of $E_r$. We estimate the toroidal flow, and hence $E_r$, using a semi-analytic turbulent intrinsic rotation model and NUBEAM calculations of neutral beam torque. The NTV from the $\abs{n} = 18$ ripple dominates that from lower $n$ perturbations of the TBMs. With the inclusion of FIs, the magnitude of NTV torque is reduced by about 75\% near the edge. We present comparisons of several models of tangential magnetic drifts, finding appreciable differences only for superbanana-plateau transport at small $E_r$. We find the scaling of calculated NTV torque with ripple magnitude to indicate that ripple-trapping may be a significant mechanism for NTV in ITER. The computed NTV torque without ferritic components is comparable in magnitude to the NBI and intrinsic turbulent torques and will likely damp rotation, but the NTV torque is significantly reduced by the planned ferritic inserts.
\end{abstract}

\maketitle

\section{Introduction}

Toroidal rotation is critical to the experimental control of tokamaks: the magnitude of rotation is known to affect resistive wall modes \cite{Bondeson1994, Garofalo2002}, while rotation shear can decrease microinstabilities and promote the formation of transport barriers \cite{Burrell1997, Terry2000}. As some ITER scenarios will be above the no-wall stability limit \cite{Liu2004}, it is important to understand the sources and sinks of angular momentum for stabilization of external kink modes. One such sink (or possible source) is the toroidal torque caused by 3D non-resonant error fields, known as neoclassical toroidal viscosity (NTV). Dedicated NTV experiments have been conducted in the Mega Amp Spherical Tokamak (MAST) \cite{Hua2010}, the Joint European Tokamak (JET) \cite{Lazzaro2002, DeVries2008b}, Alcator C-MOD \cite{Wolfe2005},  DIII-D \cite{Garofalo2008,Reimerdes2009}, JT-60U \cite{Honda2014}, and the National Spherical Tokamak Experiment (NSTX) \cite{Zhu2006}.

In addition to the ripple due to the finite number (18) of toroidal field (TF) coils, the magnetic field in ITER will be perturbed by ferromagnetic components including ferritic inserts (FIs) and test blanket modules (TBMs). TBMs will be installed in three equatorial ports to test tritium breeding and extraction of heat from the blanket. The structural material for these modules is ferritic steel and will produce additional error fields in response to the background field. The TBMs will be installed during the H/He phase in order to test their performance in addition to their possible effects on confinement and transport \cite{Chuyanov2010}. It is important to understand their effect on rotation during the early phases of ITER. Experiments at DIII-D using mock-ups of TBMs found a reduction in toroidal rotation by as much as 60\% due to an $n = 1$ locked mode \cite{Schaffer2011}. Here $n$ is the toroidal mode number. Further experiments showed compensation by $n=1$ control coils may enable access to low NBI torque (1.1 Nm) regimes relevant to ITER without rotation collapse \cite{Lanctot2017}. In addition to TBMs, ferritic steel plates (FIs) will be installed in each of the TF coil sections in order to mitigate energetic particle loss due to TF ripple \cite{Tobita2003}. Experiments including FIs on JT-60U \cite{Urano2007} and JFT-2M \cite{Kawashima2001} have found a reduction in counter-current rotation with the addition of FIs. As FIs will decrease TF ripple, they may decrease the NTV in ITER.

While  the bounce-averaged radial drift vanishes in a tokamak, trapped particles may wander off the flux surface in the presence of non-axisymmetric error fields. Particles trapped poloidally (bananas) can drift radially as the parallel adiabatic invariant, $J_{||} = \oint dl\, v_{||}$, becomes a function of toroidal angle in broken symmetry. Here $v_{||}$ is the velocity coordinate parallel to $\bm{b} = \bm{B}/B$ and integration is taken along the field between bounce points. If local ripple wells exist along a field line and the collisionality is small enough that helically trapped particles can complete their collisionless orbits, these trapped particles may grad-$B$ drift away from the flux surface \cite{Stringer1972}. The TF ripple in ITER causes local wells along the field line, corresponding to $\alpha = \epsilon/(qn\delta_B) < 1$ \cite{Stringer1972}. Here $\epsilon = r/R$ is the inverse aspect ratio, $r$ is the minor radius, $R$ is the major radius,  $q$ is the safety factor, and $\delta_B$ is a measure of the amplitude of the ripple. Because of ITER's low collisionality, $\nu_* \ll (\delta_B/\epsilon)^{3/2}$, ripple-trapped particles can complete their collisionless orbits \cite{Shaing2003}. Here the normalized collision frequency is $\nu_* = q R v_{ti}/(\nu_{ii} \epsilon^{3/2})$ where the ion-ion collision frequency is $\nu_{ii}$. The ion thermal velocity is $v_{ti} = \sqrt{2T_i/m_i}$ where $T_i$ is the ion temperature and $m_i$ is the ion mass. Therefore, both ripple trapping and banana diffusion should be considered for NTV in ITER. For a general electric field, the neoclassical electron and ion fluxes are not necessarily identical in broken symmetry. The resulting radial current induces a $\bm{J} \times \bm{B}$ torque which is often counter-current. 

Analytic expressions for neoclassical fluxes in several rippled tokamak regimes have been derived by various authors, making assumptions about the magnitude of the perturbing field, electric field, magnetic geometry, collisionality, and the collision operator. Multiple regimes are typically needed to describe all radial positions, classes of particles, and helicities of the magnetic field for a single discharge. When collisions set the radial step size of trapped particles, the transport scales as $1/\nu$ where $\nu$ is the collision frequency. The $1/\nu$ regime can be relevant for both ripple trapped and banana particles with small radial electric field. With a non-zero radial electric field, transport from the collisional trapped-passing boundary layer leads to fluxes that scale as $\sqrt{\nu}$. When the collisionality is sufficiently low, the collisionless detrapping/trapping layer becomes significant, where fluxes scale as $\nu$. Here bananas can become passing particles due to the variation of $B_{\max}$ along their drift trajectories \cite{Shaing2009}, and ripple trapped particles can experience collisionless detrapping from ripple wells to become bananas \cite{Shaing1982a, Shaing1982b}. If the collisionality is small compared with the typical toroidal precession frequency of trapped particles, the resonant velocity space layer where the bounce-averaged toroidal drift vanishes can dominate the neoclassical fluxes, leading to superbanana-plateau transport \cite{Shaing2009_sbp}. In the presence of a strong radial electric field, the resonance between the parallel bounce motion and drift motion of trapped particles can also result in enhanced transport, known as the bounce-harmonic resonance \cite{Linsker1982,Park2009}. The $1/\nu$ and $\sqrt{\nu}$ stellarator regimes for helically-trapped particles have been formulated by Galeev and Sagdeev \cite{Galeev1969}, Ho and Kulsrud \cite{Ho1987}, and Frieman \cite{Frieman1970}. These results were generalized to rippled tokamaks in the $1/\nu$ regime by Stringer \cite{Stringer1972}, Connor and Hastie \cite{Connor1973}, and Yushmanov \cite{Yushmanov1982}. Kadomtsev and Pogutse \cite{Kadomtsev1971} and Stringer \cite{Stringer1972} presented the scaling of ripple diffusion including trapping/detrapping by poloidal rotation, where fluxes scale as $\nu$. This regime is likely to be applicable for ITER's low collisionality and strong radial electric field. Banana diffusion in the $1/\nu$ regime has been evaluated by Davidson \cite{Davidson1976}, Linkser and Boozer \cite{Linsker1982}, and Tsang \cite{Tsang1977}. The corresponding $\nu$ transport was studied by Tsang \cite{Tsang1977} and Linsker and Boozer \cite{Linsker1982}. Shaing emphasized the relationship between nonaxisymmetric neoclassical transport and toroidal viscosity \cite{Shaing1983}. The theory for NTV torque due to banana diffusion has been formulated in the $1/\nu$ \cite{Shaing2003}, $\nu-\sqrt{\nu}$ \cite{Shaing2008}, $\nu$ \cite{Shaing2009}, and superbanana-plateau \cite{Shaing2009_sbp} regimes in addition to an approximate analytic formula which connects these regimes \cite{Shaing2010}.

The calculation of NTV torque requires two steps: (i) determine the equilibrium magnetic field in the presence of ripple and (ii) solve a drift kinetic equation (DKE) with the magnetohydrodynamic (MHD) equilibrium or apply reduced analytic formulae. The first step can be performed using various levels of approximation. The simplest method is to superimpose 3D ripple vacuum fields on an axisymmetric equilibrium, ignoring the plasma response.  A second level of approximation is to use a linearized 3D equilibrium code such as the Ideal Perturbed Equilibrium Code (IPEC) \cite{Park2009} or linear M3D-C1 \cite{Jardin2008}. A third level of approximation is to solve nonlinear MHD force balance using a code such as the Variational Moments Equilibrium Code (VMEC) \cite{Hirshman1986a} or M3D-C1 \cite{Ferraro2010} run in nonlinear mode. In this paper we use free-boundary VMEC to find the MHD equilibrium in the presence of TF ripple, FIs, and TBMs. 

Many previous NTV calculations \cite{Zhu2006,Hua2010,Cole2011,Park2009} have been performed using reduced analytic models with severe approximations. Solutions of the bounce-averaged kinetic equation have been found to agree with Shaing's analytic theory except in the transition between regimes \cite{Sun2010}. However, the standard bounce-averaged kinetic equation does not include contributions from bounce and transit resonances. Discrepancies have been found between numerical evaluation of NTV using the Monte Carlo neoclassical solver FORTEC-3D and analytic formulae for the $1/\nu$ and superbanana-plateau regimes \cite{Satake2011a,Satake2011b}. NTV calculations with quasilinear NEO-2 differ from Shaing's connected formulae \cite{Shaing2010}, especially in the edge where the large aspect ratio assumption breaks down \cite{Martitsch2016}. Rather than applying such reduced models, in this paper a DKE is solved using the Stellarator Fokker-Planck Iterative Neoclassical Conservative Solver (SFINCS) \cite{Landreman2014} to calculate neoclassical particle and heat fluxes for an ITER steady-state scenario. The SFINCS code does not exploit any expansions in collisionality, size of perturbing field, or magnitude of the radial electric field (beyond the assumption of small Mach number). It also allows for realistic experimental magnetic geometry rather than using simplified flux surface shapes. All trapped particle effects including ripple-trapping \cite{Stringer1972}, banana diffusion \cite{Linsker1982}, and bounce-resonance \cite{Linsker1982} are accounted for in these calculations. The DKE solved by SFINCS ensures intrinsic ambipolarity for axisymmetric or quasisymmetric flux surfaces in the presence of a radial electric field while this property is not satisfied by other codes such as DKES \cite{Hirshman1986b,Rij1989}. This prevents spurious NTV torque density, which is proportional to the radial current. As SFINCS makes no assumption about the size of ripple, it can account for non-quasilinear transport, such as ripple trapping, rather than assuming that the Fourier modes of the ripple can be decoupled. For TF ripple, the deviation from the quasilinear assumption has been found to be significant in benchmarks between SFINCS and NEO-2 \cite{Martitsch2016}.

In addition to NTV, neutral beams will provide an angular momentum source for ITER. As NBI torque scales as $P/E^{1/2}$ for input power $P$ and particle energy $E$, ITER's neutral beams, with $E = 1$ MeV and $P = 33$ MW, will provide less momentum than in other tokamaks such as JET, with $E = 125$ keV for $P = 34$ MW \cite{Ciric2011}. NBI-driven rotation will also be smaller in ITER because of its relatively large moment of inertia, with $R = 6$ m compared to 3 m for JET. However, spontaneous rotation may be significant in ITER. Turbulence can drive significant flows in the absence of external momentum injection, known as intrinsic or spontaneous rotation. This can be understood as a turbulent redistribution of toroidal angular momentum to produce large directed flows. For perturbed tokamaks this must be in the approximate symmetry direction. According to gyrokinetic orderings and inter-machine comparisons by Parra \textit{et al} \cite{Parra2012}, intrinsic toroidal rotation is expected to scale as $V_{\zeta} \sim  T_i/I_p$ where $I_p$ is the plasma current, and core rotations may be on the order of 100 km/s (ion sonic Mach number $M_i \approx 8\%$) in ITER. Scalings with $\beta_N = \beta_T a B_T/I_P$ by Rice \textit{et al} \cite{Rice2007} predict rotations of a slightly larger scale, $V_{\zeta} \approx 400$ km/s ($M_i \approx 30\%$). Here $\beta_T =  2\mu_0 p/B_T^2$, $B_T$ is the toroidal magnetic field in tesla, $a$ is the minor radius at the edge in meters, and $p$ is the plasma pressure. Co-current toroidal rotation appears to be a common feature of H-mode plasmas and has been observed in electron cyclotron (EC) \cite{DeGrassie2007}, ohmic \cite{DeGrassie2007}, and ion cyclotron range of frequencies (ICRF) \cite{Noterdaeme2003} heated plasmas. Gyrokinetic GS2 simulations with H-mode parameters find an inward intrinsic momentum flux, corresponding to a rotation profile peaked in the core toward the co-current direction \cite{Lee2014}. In an up-down symmetric tokamak, the radial intrinsic angular momentum flux can be shown to vanish to lowest order in $\rho_* = \rho_i/a$, but neoclassical departures from an equilibrium Maxwellian can break this symmetry and cause non-zero rotation in the absence of input momentum \cite{Barnes2013}. Here $\rho_i = v_{ti}m_i /{Z_ieB}$ is the gyroradius where $Z_i$ is the ion species charge.

In section \ref{vmec} we present the ITER steady state scenario and free boundary MHD equilibrium in the presence of field ripple. In section \ref{rotation} we estimate rotation driven by NBI and turbulence. This flow velocity is related to $E_r$ in section \ref{Erandv}. The NTV torque due to TF ripple, TBMs, and FIs is evaluated in section \ref{torque}. In section \ref{scaling} the scaling of transport calculated with SFINCS with ripple magnitude is compared with that predicted by NTV theory, and in section \ref{heatflux} neoclassical heat fluxes in the presence of ripple are presented. In section \ref{mds}, we assess several tangential magnetic drift models on the transport for this ITER scenario and a radial torque profile is presented. In section \ref{summary} we summarize the results and conclude.

\section{ITER Steady State Scenario and Free Boundary Equilibrium Calculations} \label{vmec}

\FloatBarrier

\begin{figure}[h!]
\centering
\includegraphics[width=0.7\textwidth]{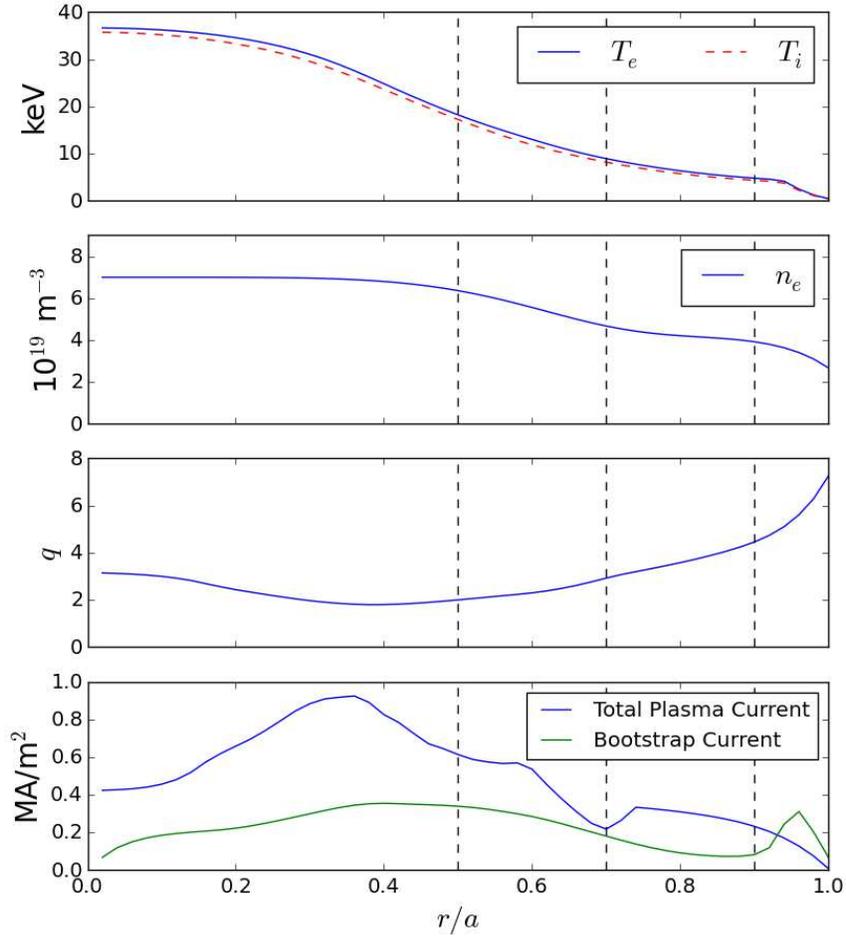}
\caption{\label{fig:profiles} Radial profiles of temperature, density, safety factor, total plasma current, and bootstrap current for the ITER steady state scenario \cite{Poli2014}. Black dashed lines indicate the radial locations that will be considered for neoclassical calculations.}
\end{figure}

We consider an advanced ITER steady state scenario with significant bootstrap current and reversed magnetic shear \cite{Poli2014}. The input power includes 33 MW NBI, 20 MW EC, and 20 MW lower hybrid (LH) heating for a fusion gain of $Q = 5$. This 9 MA non-inductive scenario is achieved with operation close to the Greenwald density limit. The discharge was simulated using the Tokamak Simulation Code (TSC) in the IPS \cite{Elwasif2010} framework for the calculation of the free-boundary equilibrium and the RF calculations, and TRANSP for calculations of the NBI heating and current and torque. The discharge was simulated using the Tokamak Simulation Code (TSC) \cite{Jardin1986} and TRANSP \cite{Hawryluk1980} using a current diffusive ballooning mode (CDBM) \cite{Fukuyama1995,Fukuyama1998} transport model and EPED1 \cite{Snyder2011} pedestal modeling. The NBI source is modeled using NUBEAM \cite{Goldston1981,Pankin2004} with 1 MeV particles. The beams are steered with one on-axis and one off-axis, which avoids heating on the midplane wall gap and excess heat deposition above or below the midplane. Further details of the steady state scenario modeling can be found in table 1 of \cite{Poli2014}.

The density ($n$), temperature ($T$), safety factor ($q$), total plasma current, and bootstrap current profiles are shown in figure \ref{fig:profiles}. Neoclassical transport will be analyzed in detail at the radial locations indicated by dashed horizontal lines ($r/a = 0.5, 0.7, 0.9$). Throughout we will use the radial coordinate $r/a \propto \sqrt{\Psi_{\mathrm{T}}}$ where $\Psi_{\mathrm{T}}$ is the toroidal flux.

\FloatBarrier

\begin{figure}[h!]
\centering
\includegraphics[width=0.7\textwidth]{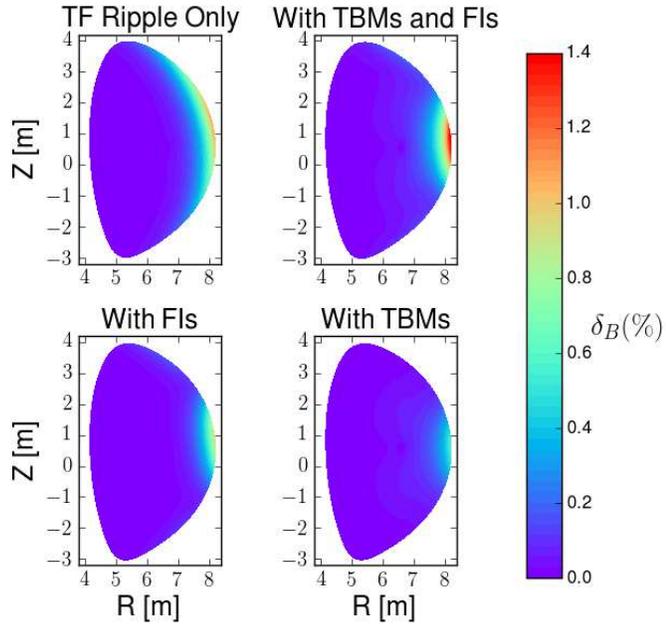}
\caption{\label{fig:ripplecontour} Magnetic field ripple, $\delta_B = (B_{\mathrm{max}}-B_{\mathrm{min}})/(B_{\mathrm{max}} + B_{\mathrm{min}})$, is plotted on the poloidal plane for VMEC free boundary equilibria including (i) only TF ripple (top left), (ii) TF ripple, TBMs, and FIs (top right), (iii) TF ripple and FIs (bottom left), and (iv) with TBMs only (bottom right). FIs decrease the poloidal extent of the ripple, while TBMs add an additional ripple near the outboard midplane.}
\end{figure}

The VMEC free boundary \cite{Hirshman1986a} magnetic equilibrium was computed using the TRANSP profiles along with filamentary models of the toroidal field (TF), poloidal field (PF), and central solenoid (CS) coils and their corresponding currents. The vacuum fields produced by the three TBMs and the FIs have been modeled using FEMAG \cite{Shinohara2009}. The equilibrium is computed for four geometries: (i) including only the TF ripple, (ii) including TF ripple, TBMs, and FIs, (iii) TF ripple and FIs, and (iv) axisymmetric geometry. We define the magnitude of the magnetic field ripple to be
\begin{gather}
\delta_B = (B_{\mathrm{max}}-B_{\mathrm{min}})/(B_{\mathrm{max}} + B_{\mathrm{min}}), 
\end{gather}
where the maximum and minimum are evaluated at fixed radius and VMEC poloidal angle $\theta$. In figure \ref{fig:ripplecontour}, $\delta_B$ is plotted on the poloidal plane for the three rippled VMEC equilibria. A fourth case is also shown in which the component of $\bm{B}$ with $\abs{ n } = 18$ was removed from the geometry with TBMs and FIs in order to consider the $\,\abs{ n } < 18$ ripple from the TBMs (bottom right). When only TF ripple is present, significant ripple persists over the entire outboard side, while in the configurations with FIs the ripple is much more localized in $\theta$. When TBMs are present, the ripple is higher in magnitude near the outboard midplane ($\delta_B \approx 1.4\%$), while in the other magnetic configurations $\delta_B \approx$ 1\% near the outboard midplane. For comparison, the TF ripple during standard operations is $0.08\%$ in JET \cite{DeVries2008b} and $0.6\%$ in ASDEX Upgrade \cite{Martitsch2016}. In JT-60U the amplitude of TF ripple is reduced from $\delta_B \approx 1.7\%$ to $\delta_B \approx 1\%$ by FIs \cite{Urano2007}.

\begin{figure}[h!]
\centering
\includegraphics[width=0.7\textwidth]{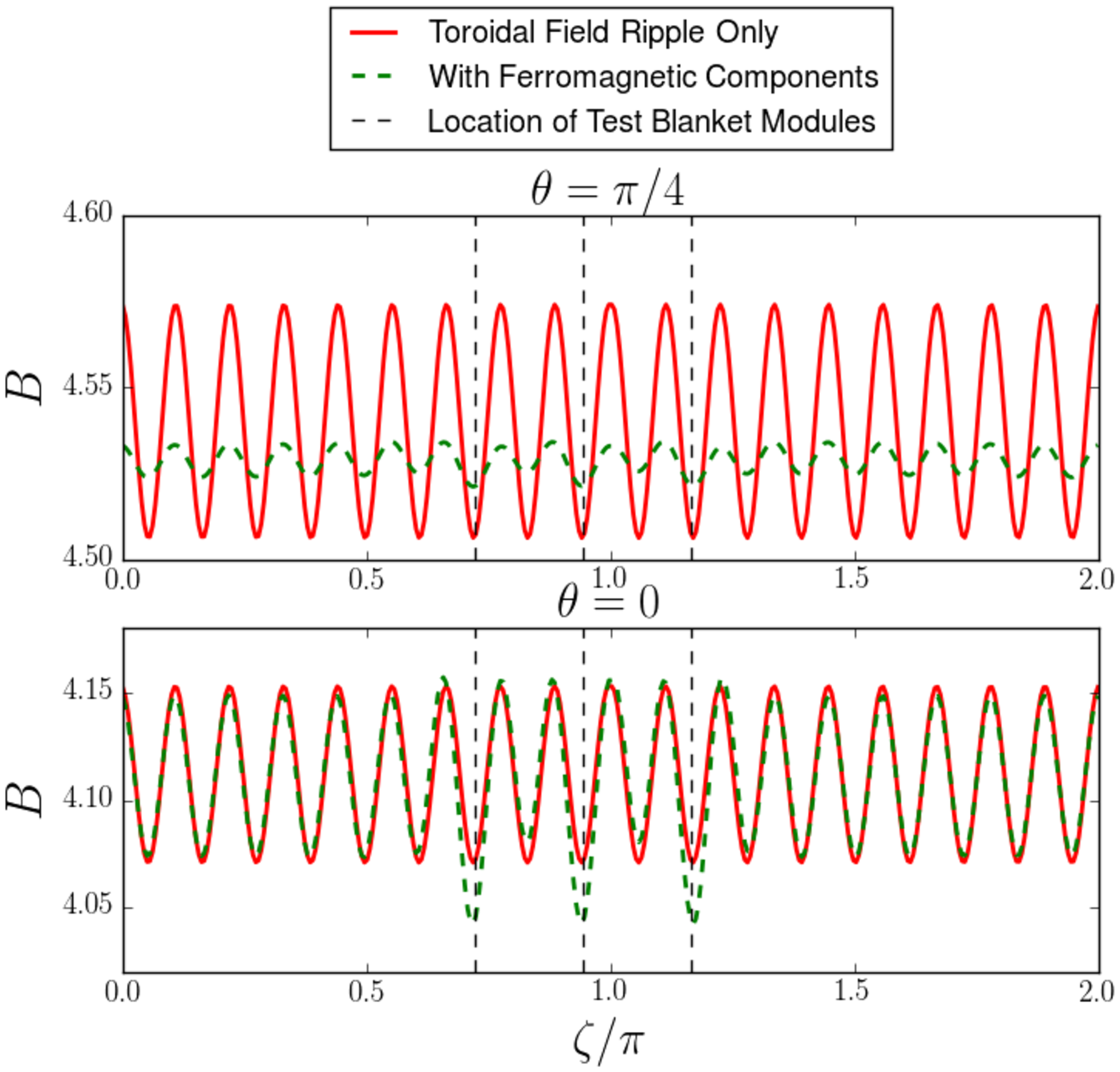}
\caption{\label{fig:toroidalripple} The magnitude of $\bm{B}$ as a function of toroidal angle ($\zeta$) at $r/a = 1$, $\theta = 0$ and $\pi/4$. Vertical dashed lines indicate the toroidal locations of the TBM ports. The mitigating effect of the FIs is stronger away from the midplane, where an increased number of steel plates are inserted. The TBMs add an additional ripple near their locations at $\theta = 0$. }
\end{figure}

In figure \ref{fig:toroidalripple}, the magnitude of $\bm{B}$ is plotted as a function of toroidal angle $\zeta$ at $\theta = 0$ and $\theta = \pi/4$. Away from the midplane ($\theta = \pi/4$) the FIs greatly decrease the magnitude of the TF ripple. Near the midplane the FIs do not decrease the magnitude of the toroidal ripple as strongly, as the number of steel plates is reduced near the midplane \cite{Shinohara2009}. The ferromagnetic steel of the TBMs concentrates magnetic flux and locally decreases $B$ in the plasma near their location. This causes enhancement of $\delta_B$  near $\theta = 0$. 
\FloatBarrier

\FloatBarrier

\section{Estimating Toroidal Rotation}\label{rotation}

In order to predict the ripple transport in ITER, the radial electric field, $E_r = - \Phi'(r) $, must be estimated, as particle and heat fluxes are nonlinear functions of $E_r$. This is equivalent to predicting the parallel flow velocity, $V_{||}$, which scales monotonically with $E_r$.  As we simply wish to determine a plausible value of $E_r$, the difference between $V_{||}$ and $V_{\zeta}$, the toroidal flow, will be unimportant for our estimates. We define $V_{\zeta}$ in terms of the toroidal rotation frequency, $V_{\zeta} = \Omega_{\zeta} R$, where $\Omega_{\zeta} \approx \Omega_{\zeta}(r)$. As $I_P$ and the toroidal magnetic field are both directed clockwise when viewed from above, $V_{\zeta}$ and $V_{||}$ will point in the same direction. Here we use the convention that positive $V_{\zeta}$ corresponds to co-current rotation. For this rotation calculation, angular momentum transport due to neutral beams and turbulence will be considered. There is an additional torque caused by the radial current of orbit-lost alphas \cite{Rosenbluth1996}, but it will be negligible ($\approx 0.006$ Nm/m$^3$). The following time-independent momentum balance equation is considered in determining $\Omega_{\zeta}(r)$,
\begin{gather}
\nabla \cdot \Pi_{\zeta}^{\mathrm{turb}}(\Omega_{\zeta}) + \nabla \cdot \Pi_{\zeta}^{\mathrm{NC}}(\Omega_{\zeta}) = \tau^{\mathrm{NBI}},
\end{gather}
where $\Pi^{\mathrm{turb}}_{\zeta}$ and $\Pi^{\mathrm{NC}}_{\zeta}$ are the toroidal angular momentum flux densities due to turbulent and neoclassical transport and $\tau^{\mathrm{NBI}}$ is the NBI torque density. For this paper the feedback of $\Pi_{\zeta}^{\mathrm{NC}}$ on $\Omega_{\zeta}$ will not be calculated. Determining the change in rotation due to NTV would require iteratively solving this equation for $\Omega_{\zeta}$, as $\Pi_{\zeta}^{\mathrm{NC}}$ is a nonlinear function of $\Omega_{\zeta}$. 

The quantity $\Pi_{\zeta}^{\mathrm{turb}}$ consists of a diffusive term as well as a term independent of $\Omega_{\zeta}$ which accounts for turbulent intrinsic rotation, 
\begin{gather}
\Pi_{\zeta}^{\mathrm{turb}} = -m_i n_i \chi_{\zeta} \langle R^2 \rangle \partder{\Omega_{\zeta}}{r} + \Pi_{\mathrm{int}}.
\end{gather}
For simplicity, an angular momentum pinch, $P_{\zeta}$, will not be considered for this analysis. As $R P_{\zeta}/\chi_{\zeta} \approx 2$, there would be a factor of 2 difference in rotation peaking at the core due to the turbulent momentum source at the edge \cite{LeeThesis}. Here $\chi_{\zeta}$ is the toroidal ion angular momentum diffusivity. The flux surface average is denoted by $\langle ... \rangle$,
\begin{gather}
\langle ... \rangle = \frac{1}{V'} \int_0^{2 \pi} d \theta \int_0^{2 \pi} d \zeta \sqrt{g} (...)
\\ V' = \int_0^{2\pi} d \theta \int_0^{2 \pi} d \zeta \sqrt{g},
\end{gather}
where $\sqrt{g}$ is the Jacobian.
Ignoring NTV torque, we will solve the following angular momentum balance equation,
\begin{gather}
-m_i \frac{1}{V'} \partder{}{r} \left( V' n_i \chi_{\zeta} \langle R^2 \rangle \partder{\Omega_{\zeta}}{r} \right) =  -\frac{1}{V'} \partder{}{r} \left( V' \Pi_{\mathrm{int}} \right) + \tau^{\mathrm{NBI}}.
\label{eq:angularmomentum}
\end{gather}
Equation \ref{eq:angularmomentum} is a linear inhomogeneous equation for $\Omega_{\zeta}$, as the right hand side is independent of $\Omega_{\zeta}$. We can therefore solve for the rotation due to each of the source terms individually and add the results to obtain the rotation due to both NBI torque and turbulent intrinsic torque. 

The NBI-driven rotation profile is evolved by TRANSP assuming $\chi_{\zeta} = \chi_{i}$, the ion heat diffusivity. The total beam torque density, $\tau^{\mathrm{NBI}}$, is calculated by NUBEAM including collisional, $\bm{J} \times \bm{B}$, thermalization, and recombination torques. The following momentum balance equation is solved to compute $\Omega_{\zeta}$ driven by NBI,
\begin{gather}
\tau^{\mathrm{NBI}} = -\frac{1}{V'} \partder{}{r} \left( V' m_i n_i \chi_{i} \langle R^2 \rangle \partder{\Omega_{\zeta}}{r} \right).
\end{gather} 

We consider a semi-analytic intrinsic rotation model to determine the turbulent-driven rotation \cite{Hillesheim2015},
\begin{gather}
\Omega_{\zeta}(r) = - \int_{r}^a \frac{v_{ti} \rho_{*,\theta}} {2 P_r L_T^2} \widetilde{\Pi} (\nu_*) \, d r',
\label{eq:Hillesheim}
\end{gather} 
where $\rho_{*,\theta} = v_{ti} m_i/(Z_i e B_{\theta} \langle R \rangle)$ is the poloidal normalized gyroradius, $B_{\theta} = \bm{B} \cdot \partial \bm{r}/\partial \theta$, and $L_T = - \left( \partial \ln T_i/ \partial r \right)^{-1}$ is the temperature gradient scale length. The Prandtl number $P_r = \chi_{\zeta}/\chi_i$ is again taken to be 1. Equation \ref{eq:Hillesheim} is obtained assuming that $\Pi_{\mathrm{int}}$ balances turbulent momentum diffusion in steady state, $\Pi_{\mathrm{int}} = m_i n_i \chi_{\zeta} \langle R^2 \rangle \partial \Omega_{\zeta}/\partial r$. This model considers the intrinsic torque driven by the neoclassical diamagnetic flows, such that $\Omega_{\zeta} \sim \rho_{*,\theta} v_{ti}/\langle R \rangle$ and $\Omega_{\zeta} \Pi_{\mathrm{int}}/Q_i \sim \rho_{*, \theta}$ where $Q_i = n_i T_i \chi_i L_T$ is the turbulent energy flux. We also take $\Omega_{\zeta}(a) = 0$. The quantity $\widetilde{\Pi} (\nu_*)$ is an order unity function which characterizes the collisionality dependence of rotation reversals, determined from gyrokinetic turbulence simulations \cite{Barnes2013},
\begin{gather}
\widetilde{\Pi} (\nu_*) = \frac{(\nu_*/\nu_c -1)}{1 + (\nu_*/\nu_c)},
\end{gather}
where $\nu_c = 1.7$. Because of ITER's low collisionality, we do not expect a rotation reversal, which is correlated with transitioning between the banana and plateau regimes. Equation \ref{eq:Hillesheim} was integrated using profiles for the ITER steady state scenario. 

\begin{figure}[h!]
\centering
\includegraphics[width=0.7\textwidth]{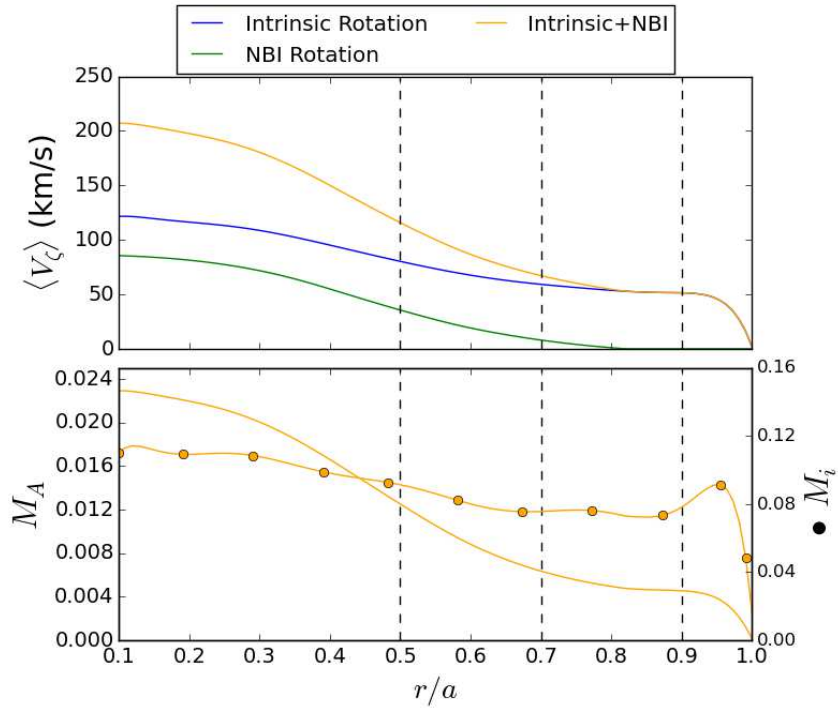}
\caption{\label{fig:rotation_estimate} Flux-surface averaged toroidal rotation, $\langle V_{\zeta} \rangle$, due to turbulence and NBI (top) is shown along with  corresponding Alfv\`{e}n Mach number (bottom, solid), and ion sonic Mach number (bottom, bulleted). The intrinsic rotation calculation uses a semi-analytic model of turbulent momentum redistribution \cite{Hillesheim2015}. The NBI rotation is calculated from turbulent diffusion of NBI torque using NUBEAM and TRANSP \cite{Poli2014}. Dashed vertical lines indicate the radial positions where SFINCS calculations are performed. }
\end{figure}

The flux-surface averaged toroidal rotation, $\langle V_{\zeta} \rangle = \Omega_{\zeta}(r) \langle R \rangle$, predicted by these models is shown in figure \ref{fig:rotation_estimate}. NBI torque contributes to significant rotation at $r/a \lesssim 0.4$ where the torque density also peaks (see figure \ref{fig:alltorque}), while turbulent torque produces rotation in the pedestal due to the $L_T^{-2}$ scaling of our model.  The intrinsic rotation calculated is comparable to that predicted from theoretical scaling arguments by Parra \textit{et al} \cite{Parra2012}, $V_{\zeta} \approx 100$ km/s. At the radii that will be considered for neoclassical calculations (indicated by dashed vertical lines), intrinsic turbulent rotation may dominate over that due to NBI. However, we emphasize that it is an estimate based on scaling arguments, as much uncertainty is inherent in predicting turbulent rotation. The volume-averaged toroidal rotation due to both NBI and turbulent torques, 113 km/s, is slightly larger than that  predicted from dimensionless parameter scans on DIII-D, 87 km/s \cite{Chrystal2017}.

\FloatBarrier

For stabilization of the resistive wall mode (RWM) in ITER, it has been estimated \cite{Liu2004} that a critical central Mach number, $M_A = \Omega_{\zeta}(0)/\omega_A \gtrsim 5\%$, must be achieved given a peaked rotation profile. Here $\omega_A = B/(\langle R\rangle\sqrt{\mu_0 m_i n_i})$ is the Alfv\`{e}n frequency. With a central rotation frequency $\Omega_{\zeta}(0) \approx 2\% \, \omega_A$ as shown in figure \ref{fig:rotation_estimate}, it may be difficult to suppress the RWM in ITER with rotation alone. As this calculation does not take into account NTV torque, $M_A$ is likely to be smaller than what is shown. Additionally, the TBM are known to increase the critical rotation frequency as they have a much shorter resistive time scale than the wall \cite{Liu2004}. More recent analysis has shown that even above such a critical rotation value, the plasma can become unstable due to resonances between the drift frequency and bounce frequency \cite{Berkery2010, Liu2009}.

\FloatBarrier

\section{Relationship Between $E_r$ and $V_{||}$}\label{Erandv}
Neoclassical theory predicts a specific linear-plus-offset relationship between $V_{||}$ and $E_r$, but it does not predict a particular value for either $V_{||}$ or $E_r$ in a tokamak. Neoclassical calculations of $V_{||}$ are made in order to determine an $E_r$ profile consistent with our estimate of $V_{\zeta} \approx \langle V_{||} B \rangle/\langle B^2 \rangle^{1/2}$ made in section \ref{rotation}. The parallel flow velocity for species $a$ is computed from the neoclassical distribution function,
\begin{gather}
V^a_{||} = \left(\frac{1}{n_a}\right) \int d^3 v \, v_{||} f_a,
\label{eq:parallelflow}
\end{gather}
which we calculate with the SFINCS \cite{Landreman2014} code. SFINCS is used to solve a radially-local DKE for the gyro-averaged distribution function, $f_{a1}$, on a single flux surface including coupling between species. 
\begin{gather}
( v_{||} \bm{b} + \bm{v}_E + \bm{v}_{\mathrm{m}a}) \cdot (\nabla f_{a1})  - C(f_{a1}) = - \bm{v}_{\mathrm{m}a} \cdot \nabla \psi \left( \partder{f_{a0}}{\psi} \right) + \frac{Z_a e v_{||} B \langle E_{||} B \rangle}{T_a \langle B^2 \rangle } f_{a0}
\label{kineticequation}
\end{gather} 
\hspace{-1mm}
Here $a$ indicates species, $f_{a0}$ is an equilibrium Maxwellian, $\psi = \Psi_{\mathrm{T}}/2\pi$, $Z$ indicates charge, and $C$ is the linearized Fokker-Planck collision operator. Gradients are performed at constant $W = m_a v^2/2 + Z_a e \Phi$ and $\mu = v_{\perp}^2/(2B)$. The $\bm{E} \times \bm{B}$ drift is 
\begin{gather}
\bm{v}_E = \frac{1}{B^2} \bm{B} \times \nabla \Phi
\end{gather} 
and the radial magnetic drift is
\begin{gather}
\bm{v}_{\mathrm{m}a} \cdot \nabla \psi = \frac{1}{\Omega_a B} \left(v_{||}^2 + \frac{v_{\perp}^2}{2} \right) \bm{b} \times \nabla B \cdot \nabla \psi,
\label{magneticdrift}
\end{gather} 
where $v_{\perp}$ is the velocity coordinate perpendicular to $\bm{b}$. The quantity $\Omega_a = Z_aeB/m_a$ is the gyrofrequency. Transport quantities have been calculated using the steady state scenario ion and electron profiles and VMEC geometry. We consider a three species plasma (D, T, and electrons), and we assume that $n_D = n_T = n_e/2$. The second term on the right hand side of equation \ref{kineticequation} proportional to $E_{||}$ is negligible for this non-inductive scenario with loop voltage $ \approx 10^{-4}$ V. For the calculations presented in sections \ref{Erandv}, \ref{torque}, \ref{scaling}, and \ref{heatflux}, $\bm{v}_{\mathrm{m}a} \cdot \nabla f_{a1}$ is not included. The effect of keeping this term is shown to be small in section \ref{mds}.

\begin{figure}[h!]
\centering
\includegraphics[width=.7\textwidth]{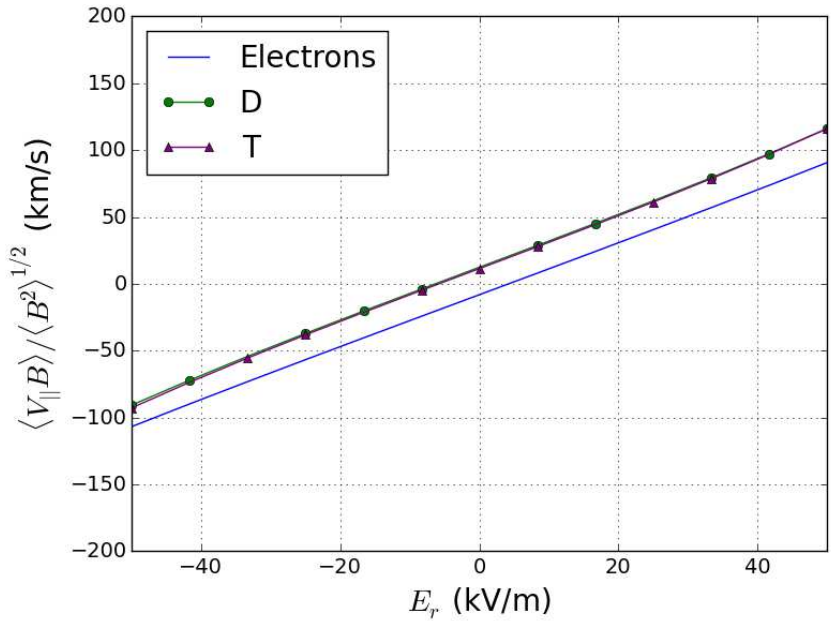}
\caption{\label{fig:Er_flow} SFINCS calculation of the flux surface averaged parallel flow, $\langle B V_{||} \rangle/\langle B^2 \rangle^{1/2}$, at $r/a = 0.9$ for ions and electrons. The addition of ripple does not change the tokamak neoclassical relationship between $E_r$ and $V_{||}$ by a discernible amount on this scale although the radial particle fluxes, $\Gamma_{\psi}$, are sensitive to the perturbing field.}
\end{figure}

The relationship between $E_r$ and $\langle V_{||} B \rangle/\langle B^2 \rangle^{1/2}$ for electrons and ions at $r/a = 0.9$ is shown in figure \ref{fig:Er_flow}. Only one curve is shown for each species as the addition of ripple fields does not change the dependence of $V_{||}$ on $E_r$ significantly ($\lesssim 5 \%$). While radial transport of heat and particles changes substantially in the presence of small ripple fields (see sections \ref{torque}, \ref{scaling}, and \ref{heatflux}), the parallel flow is much less sensitive to the perturbing field. Note that the parallel flow is non-zero in axisymmetry while the radial current vanishes without symmetry-breaking.

In a tokamak we can write $\langle V_{||}^a B \rangle$ in terms of a dimensionless parallel flow coefficient, $k_{||}$, and thermodynamic drives,
\begin{gather}
\langle  V_{||}^a  B\rangle = -\frac{G}{Z_a e n_a} \left[ \frac{1}{n} \der{(nT)}{\psi_P} + Z_a e \der{\Phi}{\psi_P} - k_{||} \der{T}{\psi_P} \right],
\end{gather}
where $2 \pi \psi_P$ is the poloidal flux, $G(\psi) = R B_{\zeta}$, and $B_{\zeta} = \bm{B} \cdot \partial \bm{r}/\partial \zeta$. The low collisionality, large aspect ratio limit \cite{Hinton1976, Hirshman1981} $k_{||} \approx 1.17$ is often assumed in NTV theory \cite{Callen2011, Sun2011} in relating analytic expressions of torque density to toroidal rotation frequency. The value of the ion $k_{||}$ calculated by SFINCS for ITER parameters varies between 0.5 near the edge and 0.9 near the core. The bootstrap current computed with SFINCS,
\begin{gather}
J_{BS} = \sum_a n_a Z_a e \langle V_{||}^a B \rangle,
\end{gather}
is consistent with that computed by TRANSP within 10\% for $r/a \geq 0.5$. Though there is some discrepancy in the core, they have the same qualitative behavior and similar maxima. The bootstrap current in TRANSP is computed using a Sauter model \cite{Sauter1999}, an analytic fit to numerical solutions of the Fokker-Planck equation.

\FloatBarrier

\section{Torque Calculation}\label{torque}

The NTV torque density, $\tau^{\mathrm{NTV}}$, is calculated from radial particle fluxes, $\Gamma_{\psi}$, 
\begin{gather}
\Gamma_{\psi,a} = \left \langle \int d^3v (\bm{v}_{\mathrm{m}a} \cdot \nabla \psi) f_a \right \rangle,
\label{eq:particleflux}
\end{gather}
using the flux-force relation,
\begin{gather}
\tau^{\mathrm{NTV}} = - B^{\theta} \sqrt{g} \sum_a Z_a e \Gamma_{\psi, a},
\end{gather}
where $B^{\theta} = \bm{B} \cdot \nabla \theta$ and the summation is performed over species. This expression relates radial particle transport to a toroidal angular momentum source caused by the non-axisymmetric field. This relationship can be derived from action-angle coordinates \cite{Albert2016}, neoclassical moment equations \cite{Shaing1986}, or from the definition of the drift-driven flux \cite{Shaing2006}. 

\begin{figure}[h!]
\centering
\includegraphics[width=0.7\textwidth]{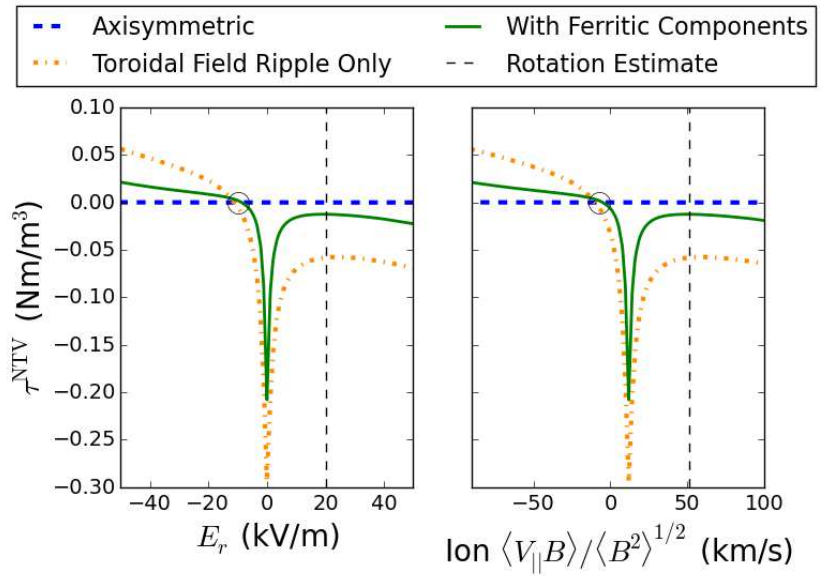}
\caption{\label{fig:Torque_ErandV} SFINCS calculation of NTV torque density as a function of $E_r$ and ion $\langle V_{||} B \rangle/\langle B^2 \rangle^{1/2}$ at $r/a = 0.9$ is shown for 3 VMEC geometries: (i) axisymmetric (blue dashed), (ii) with TF ripple only (orange dash-dot), and (iii) TF ripple with FIs and TBMs (green solid). The vertical dashed line indicates the estimate of $E_r$ and $V_{\zeta} \approx \langle V_{||} B \rangle/\langle B^2 \rangle^{1/2}$ based on the intrinsic and NBI rotation model. The circle denotes the offset rotation at $V_{||} \approx -10$ km/s. The magnitude of $\tau^{\mathrm{NTV}}$ at this radius is of similar magnitude to the NBI and turbulent torques but is opposite in direction (see figure \ref{fig:alltorque}).}
\end{figure}

\begin{figure}[h!]
\centering
\includegraphics[width=0.7\textwidth]{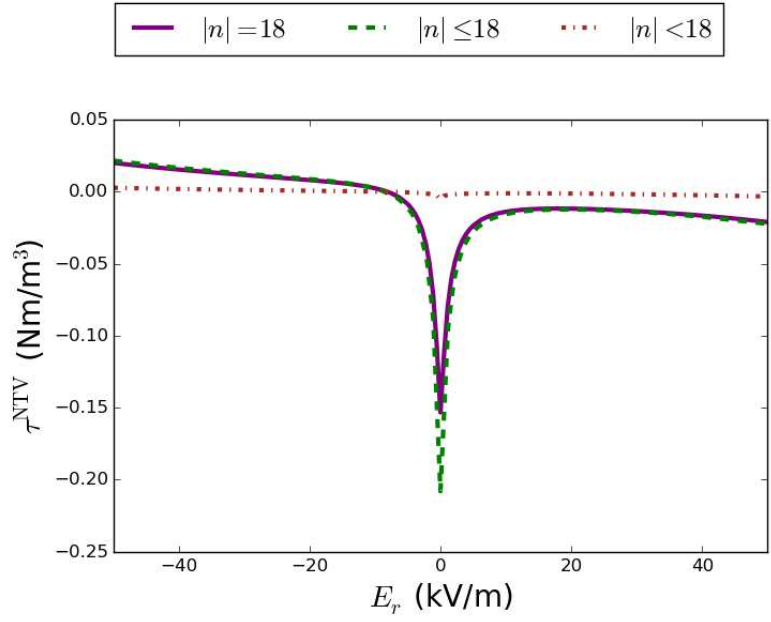}
\caption{\label{fig:Torque_comparingTBMandFI} The NTV torque density at $r/a = 0.9$ for toroidal mode numbers (i) $\abs{n} = 18$ (purple solid), (ii) $\abs{n} < 18$ (brown dash dot), and (iii) $\abs{n} \leq 18$ (green dashed). The TBM ripple contributes most strongly to low $\abs{n}$, while the FIs and TF ripple only contribute to $\abs{n} = 18$. The low $n$ TBM ripple does not contribute as strongly to the NTV torque density as the $\abs{n} = 18$ ripple does.}
\end{figure}

The calculation of $\tau^{\mathrm{NTV}}$ for three geometries at $r/a = 0.9$ is shown in figure \ref{fig:Torque_ErandV}. Here positive corresponds to the co-current direction. The numerically computed NTV torque is found to vanish in axisymmetric geometry, as expected. Overall, the magnitude of $\tau^{\mathrm{NTV}}$ with only TF ripple is larger than that with the addition of both the FIs and the TBMs.  In figure \ref{fig:Torque_comparingTBMandFI} we show that the $\abs{n} < 18$ TBM  ripple produces much less torque than the $\abs{n} = 18$ ripple, so the decrease in $\tau^{\mathrm{NTV}}$ magnitude with both FIs and TBMs can be attributed to the decrease in ripple in the presence of FIs. As will be discussed in section \ref{scaling}, neoclassical ripple transport in most regimes scales positively with $\delta_B$. The addition of FIs significantly decreases the magnitude of $\delta_B$ across most of the outboard side, and as a result the magnitude of $\tau^{\mathrm{NTV}}$ is reduced. The dashed vertical line indicates the value of $\langle V_{||} B\rangle/\langle B^2 \rangle^{1/2}$ and $E_r$ predicted from the intrinsic and NBI rotation model. At this value of $E_r$ the presence of ferritic components decreases the magnitude of the torque density by about $75\%$. 

The circle indicates the offset rotation at the ambipolar $E_r$. If no other angular momentum source were present in the system, $\tau^{\mathrm{NTV}}$ would drive the plasma to rotate at this velocity. Although $\tau^{\mathrm{NTV}}$ differs significantly between the two geometries they have similar offset rotation velocities, $V_{\zeta}$ = -10 km/s with TF ripple only and -6 km/s with TBMs and FIs. Note that for $E_r$ greater than this ambipolar value, $\tau^{\mathrm{NTV}}$ is counter-current while neutral beams and turbulence drive rotation in the co-current direction, so $\tau^{\mathrm{NTV}}$ is a damping torque. The NTV torque due to TF ripple only is larger in magnitude than $\tau^{\mathrm{NBI}}$ and $\tau^{\mathrm{turb}}$ while that with TBMs and FIs is of similar magnitude (see figure \ref{fig:alltorque}). Therefore, NTV torque may be key in determining the edge rotation in ITER. 

The magnitude of $\tau^{\mathrm{NTV}}$ peaks at $E_r = 0$ where $1/\nu$ transport becomes dominant. Although $\nu_*$ is sufficiently small such that the superbanana-plateau regime becomes relevant, the physics of superbanana formation is not accounted for in these SFINCS calculations which do not include $\bm{v}_{\mathrm{m}} \cdot \nabla f_1$. Superbanana-plateau transport will be considered in section \ref{mds}. At $r/a = 0.9$, the $1/\nu$ regime applies for $\abs{E_r} \lesssim 0.2$ kV/m where the effective collision frequency of trapped particles is larger than the $E \times B$ precession frequency. The peak at small $\abs{E_r}$ also corresponds to the region of $1/\nu$ transport of particles trapped in local ripple wells. Much NTV literature is based on banana diffusion and ripple trapping in the $1/\nu$ regime \cite{Stringer1972, Connor1974}, which is not applicable for the range of $E_r$ predicted for ITER. For the range of applicable $E_r$, bounce-harmonic resonance may occur. The $l =1$, $n = 18$ resonance condition, $\omega_b - n(\omega_E + \omega_B) = 0$, will be satisfied for $v_{||} \approx v_{ti}$ at $E_r \approx 7$ kV/m. Here $\omega_b$ is the bounce frequency, $\omega_E$ is the $E \times B$ precession frequency, and $\omega_B$ is the toroidal magnetic drift precession \cite{Park2009}. Note that here  $\bm{v}_{\text{m}} \cdot \nabla f_1$ is not included in the kinetic equation ($\omega_B =0$), but the physics of the bounce harmonic resonance between $\omega_E$ and $\omega_b$ is still accounted for in our calculation. However, we see no evidence of enhanced $\tau^{\mathrm{NTV}}$ near this $E_r$ that would be indicative of a bounce-harmonic resonance.

NTV torque is often expressed in terms of a toroidal damping frequency, $\nu_{\zeta}$,
\begin{gather}
\tau^{\mathrm{NTV}} = - \nu_{\zeta} \langle R^2 \rangle m n ( \Omega_{\zeta} - \Omega_{\zeta, \mathrm{offset}}),
\label{eq:dampingfreq}
\end{gather}
where $\Omega_{\zeta, \mathrm{offset}}$ is the offset rotation frequency. We note that $\tau^{\mathrm{NTV}}$ does appear to scale linearly with $E_r$ (and thus $\Omega_{\zeta}$) for $\abs{E_r} \gtrsim 30$ kV/m. However, $\tau^{\mathrm{NTV}}$ is a complicated nonlinear function of $\Omega_{\zeta}$ for $\abs{E_r} \lesssim 30$ kV/m at the transition between collision-limited $1/\nu$ transport and $\nu-\sqrt{\nu}$ transport, so equation \ref{eq:dampingfreq} is not a very useful representation in this context. 

\FloatBarrier

\begin{figure}[h!]
\centering
\includegraphics[width=0.7\textwidth]{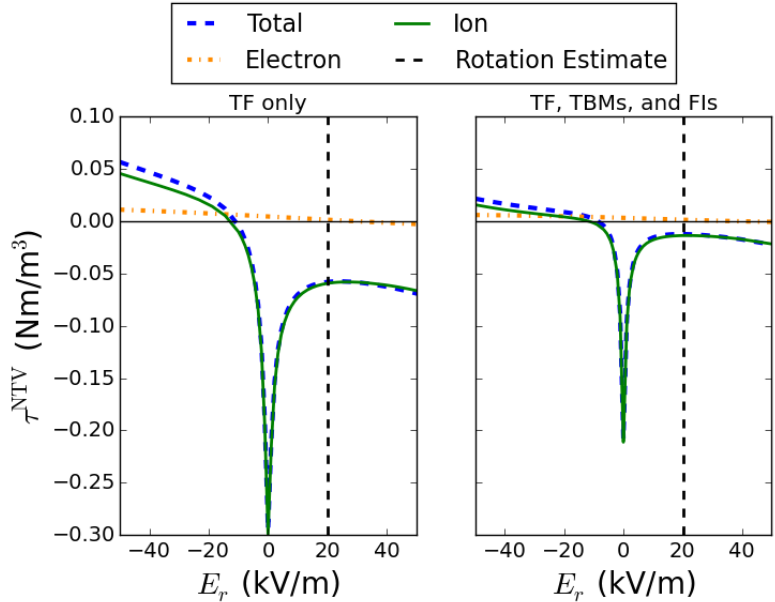}
\caption{\label{fig:Torque_eandi} Total (blue dashed), electron (yellow dash-dot), and ion (green solid) contributions to NTV torque density at $r/a = 0.9$ for TF ripple only geometry (left) and TF ripple with ferromagnetic components (right). The dashed vertical line indicates the $E_r$ predicted by the intrinsic and NBI rotation model. The electrons have a co-current neoclassical offset rotation and contribute a small co-current NTV torque density at the $E_r$ predicted by the rotation model.}
\end{figure}

In figure \ref{fig:Torque_eandi} we present $\tau^{\mathrm{NTV}}$ at $r/a = 0.9$ due to the electron and ion radial current in the presence of TF ripple only (left) and TF ripple with ferromagnetic components (right). The $E_r$ corresponding to the offset rotation frequency for the electrons is positive while that of the ions is negative. At the predicted $E_r$, $\tau^{\mathrm{NTV}}$ due to the electron particle flux is positive while that due to ion particle flux is negative. At all radial locations the electron contribution to $\tau^{\mathrm{NTV}}$ is less than 10\% of the total torque density. 

In order to decouple the influence of the FI ripple and the TBM ripple, $\tau^{\mathrm{NTV}}$ at $r/a = 0.9$ is calculated for toroidal modes (i) $\abs{n} \leq 18$, (ii) $\abs{n} = 18$, and (iii) $\abs{n} < 18$, shown in figure \ref{fig:Torque_comparingTBMandFI}. For $\abs{n} \leq 18$ and $\abs{n} = 18$, VMEC free boundary equilibria were computed including these toroidal modes. For $\abs{n} < 18$, the SFINCS calculation was performed including the desired $n$ from the VMEC fields. Here $B$ is decomposed as,
\begin{gather}
B = \sum_{m,n} b_{mn}^c \cos(m\theta-n\zeta) + b_{mn}^s \sin(m\theta-n\zeta),
\end{gather}
where $\theta$ and $\zeta$ are VMEC angles. The covariant and contravariant components of $B$ along with their partial derivatives and $\sqrt{g}$ are similarly decomposed such that the DKE can be solved for the desired toroidal modes.

The TBM produces a wide spectrum of toroidal perturbations, including $\abs{n} = 1$ and $\abs{n} = 18$. While the FIs decrease the magnitude of the $\abs{n} =18$ ripple, the TBM contributes most strongly to low mode numbers. As SFINCS is not linearized in the perturbing field, the torque due to $\abs{n} \leq 18$ is the not the sum of the torques due to $\abs{n} = 18$ and $\abs{n} < 18$.  We find that the $\abs{n} = 18$ ripple drives about 100 times more torque than the lower $n$ ripple. This result is in agreement with most relevant rippled tokamak transport regimes, which feature positive scaling with $n$ \cite{Shaing2003, Shaing2008}. For tokamak banana diffusion, in the $\sqrt{\nu}$ boundary layer \cite{Shaing2008} ion transport scales as $\Gamma_{\psi} \sim \sqrt{n}$ and in the $1/\nu$ regime \cite{Shaing2003} $\Gamma_{\psi} \sim n^2$. Moreover, it is more difficult to form ripple wells along a field line from low-$n$ ripple, so ripple trapping cannot contribute as strongly to transport. This matches our findings that the higher harmonic $\abs{n} = 18$ ripple contributes more strongly to $\tau^{\mathrm{NTV}}$ than the $\abs{n} < 18$ ripple. 

\begin{figure}[h!]
\centering
\includegraphics[width=0.7\textwidth]{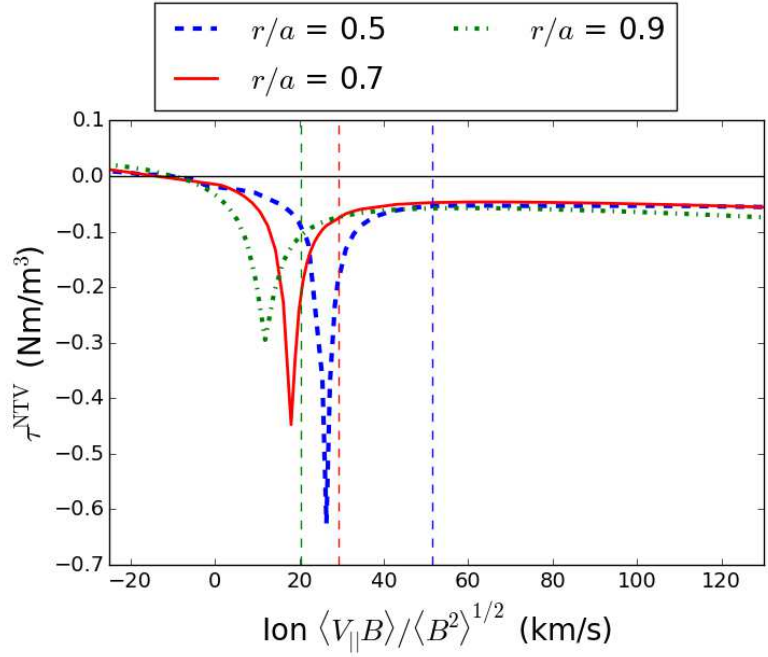}
\caption{\label{fig:Torque_radiusscaling} SFINCS calculation of NTV torque density ($\tau^{\mathrm{NTV}}$) as a function of ion $\langle V_{||} B \rangle/\langle B^2 \rangle^{1/2}$ for VMEC geometry with TF ripple only at $r/a$ = 0.5 (blue dashed), 0.7 (red solid), and 0.9 (green dash-dot). Although the field ripple decreases with radius (maximum $\delta_B = 0.82\%$ at $r/a = 0.9$, $\delta_B = 0.51\%$ at $r/a = 0.7$, $\delta_B = 0.26\%$ at $r/a = 0.5$), transport near $E_r = 0$ increases with decreasing radius because of strong scaling of $\tau^{\mathrm{NTV}}$ with $T_i$ \cite{Stringer1972,Shaing2003}.}
\end{figure}

In figure \ref{fig:Torque_radiusscaling}, the SFINCS calculation of $\tau^{\mathrm{NTV}}$ with TF ripple only is shown at $r/a$ = 0.5, 0.7, and 0.9. For these three radii the maximum $\delta_B = 0.26\%$,  0.51\%, and 0.82\% respectively. As $\tau^{\mathrm{NTV}}$ scales with a positive power of $\delta_B$ in most rippled tokamak regimes, it is reasonable to expect that the magnitude of $\tau^{\mathrm{NTV}}$ would decrease with decreasing radius. On the other hand, transport scales strongly with $T_i$. In the $\sqrt{\nu}$ banana diffusion regime \cite{Shaing2008} $\Gamma_{\psi} \sim v_{ti}^4 \sqrt{\nu_{ii}} \sim T_i^{5/4}$. The combined effect of decreased ripple and increased temperature with decreasing radius leads to comparable torques with decreasing radius in the presence of significant $E_r$.  The scaling with $T_i$ is even stronger in the $1/\nu$ regime \cite{Stringer1972, Shaing2003}, where $\Gamma_{\psi} \sim v_{ti}^4/\nu_{ii} \sim T_i^{7/2}$. Indeed, we find that the magnitude of $\tau^{\mathrm{NTV}}$ at $E_r = 0$ increases with decreasing radius.
 
\FloatBarrier

\section{Scaling with Ripple Magnitude}\label{scaling}
\begin{figure}[h!]
\centering
\includegraphics[width=0.7\textwidth]{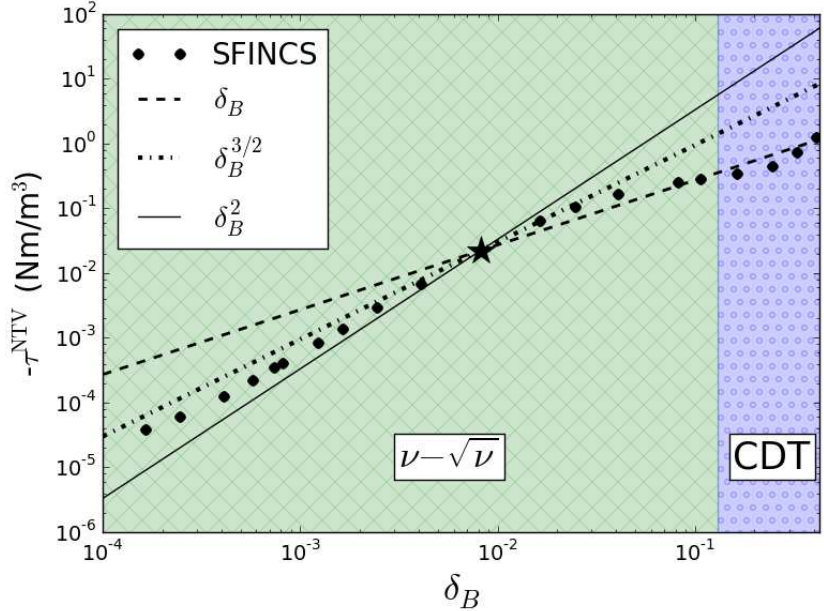}
\caption{\label{fig:scalescan} SFINCS calculations of NTV torque density as a function of $\delta_B$ at $r/a = 0.9$. A single value of $E_r = 30$ kV/m is used corresponding to the intrinsic rotation estimate. The color-shading indicates the approximate regions of applicability for rippled tokamak banana diffusion $\nu-\sqrt{\nu}$ regime \cite{Shaing2008} where $\tau^{\mathrm{NTV}} \sim \delta_B^2$ and the collisionless detrapping/trapping $\nu$ regime \cite{Shaing2009} where $\tau^{\mathrm{NTV}} \sim \delta_B$.}
\end{figure}

In figure \ref{fig:scalescan}, the NTV torque density calculated by SFINCS is shown as a function of the magnitude of the ripple, $\delta_B$, for TF only geometry. The additional ferromagnetic ripple is not included, while the $\abs{n}=18$ components of $\bm{B}$, its derivatives, and $\sqrt{g}$ are rescaled as described above. The quantity $\tau^{\mathrm{NTV}}$ is calculated at $r/a = 0.9$ with $E_r = 30$ kV/m, corresponding to the intrinsic rotation estimate. The color-shaded background indicates the approximate regions of applicability of the collisional boundary layer ($\nu-\sqrt{\nu}$) and the collisionless detrapping/trapping ($\nu$) regimes. The boundary between these regimes corresponds to the $\delta_B$ for which the width in pitch angle of the detrapping/retrapping layer is similar to the width of the collisional boundary layer, $(\delta_B/\epsilon) \sim (\nu/(\epsilon \omega_E))^{1/2}$. The $1/\nu$ regime \cite{Shaing2003} does not apply at this $E_r$, as $\omega_E \gg \nu/\epsilon$ where $\omega_E = E_r/B^{\theta}$ is the $E\times B$ precession frequency. The radial electric field is also large enough that the resonance between $\bm{v}_{E}$ and $\bm{v}_{\mathrm{m}}$ cannot occur, so the superbanana-plateau \cite{Shaing2009_sbp} and superbanana \cite{Shaing2009_sb} regimes are avoided. This significant $E_r$ may also allow the bounce-harmonic resonance to occur \cite{Park2009}. Transport from ripple-trapped particles in the $\nu- \sqrt{\nu}$ regime may also be significant for these parameters. 

The observed scaling appears somewhat consistent with ripple trapping in the stellarator $\sqrt{\nu}$ regime \cite{Ho1987} which predicts $\Gamma_{\psi} \sim \delta_B^{3/2}$. However, this result is inconsistent with predictions for tokamak ripple transport in the $\nu$ regime, $\Gamma_{\psi} \sim \delta_B^0$ \cite{Tsang1977,Linsker1982}. Contributions from other transport regimes may also influence the observed scaling. In the banana diffusion $\sqrt{\nu}$ regime $\tau^{\mathrm{NTV}} \sim \delta_B^2$ and in the $\nu$ regime $\tau^{\mathrm{NTV}} \sim \delta_B$. Bounce-harmonic resonant fluxes scale as $\delta_B^2$ \cite{Park2009}. A scaling between $\delta_B^0$ and $\delta_B^{2}$ has been predicted for plasmas close to symmetry with large gradient ripple in the absence of $E_r$ \cite{Calvo2014}. For $\delta_B$ smaller than $0.82\%$, the actual value of ripple at $r/a=0.9$ for ITER geometry, the scaling of $\tau^{\mathrm{NTV}}$ with $\delta_B$ appears similar to $\delta_B^{3/2}$. The disagreement between the SFINCS calculations and the quasilinear prediction, $\Gamma_{\psi} \sim \delta_B^2$, indicates the presence of nonlinear effects such as local ripple trapping and collisionless detrapping. The departure from quasilinear scaling increases with $\delta_B$, which is consistent with comparisons of SFINCS with quasilinear NEO-2 \cite{Martitsch2016}. We see that $\tau^{\text{NTV}}$ shows very shallow scaling between $\delta_B = 0.05$ and $\delta_B = 0.2$. This could be in agreement with  a scaling of $\delta_B^0$ predicted for $\nu$ regime ripple-trapping in tokamaks \cite{Kadomtsev1971,Stringer1972}. In this region the collisionless detrapping boundary layer and collisional boundary layer are of comparable widths, so it is possible that the transport here is not described well by any of the displayed scalings. Furthermore, near the collisionless detrapping-trapping regime, $\delta_B$ becomes comparable to the inverse aspect ratio and the assumptions made for rippled tokamak theory are not satisfied.

\FloatBarrier

\section{Heat Flux Calculation}\label{heatflux}
\begin{figure}[h!]
\centering
\includegraphics[width=0.7\textwidth]{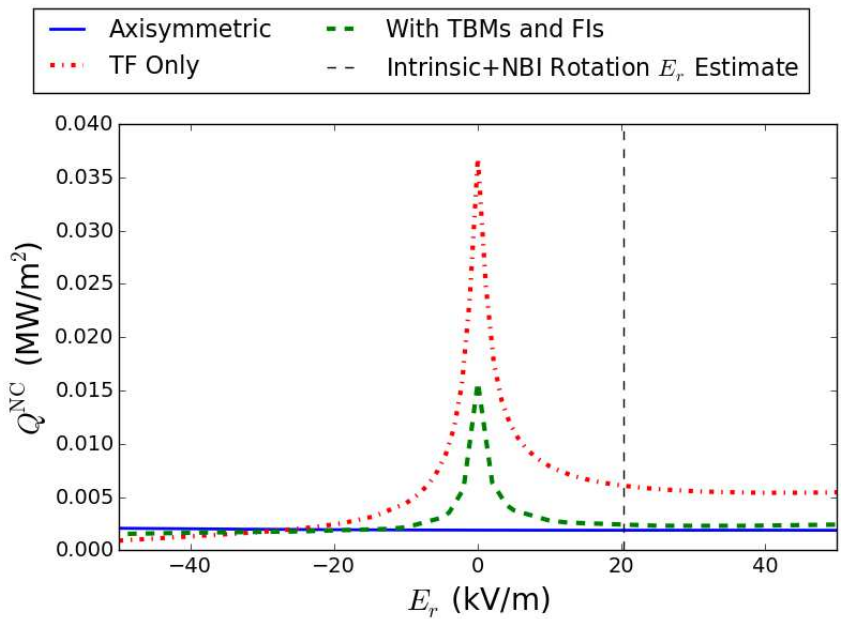}
\caption{\label{fig:HeatFlux} SFINCS calculation of neoclassical heat flux, $Q^{\mathrm{NC}}$ at $r/a = 0.9$ for three magnetic geometries: (i) axisymmetric (blue solid), (ii) with TF ripple only (red dash-dot), and (iii) TF ripple with TBMs and FIs (green dashed). The vertical dashed line corresponds to the intrinsic and NBI rotation estimate for $E_r$. Note that $Q^{\mathrm{NC}}$ is much smaller than the estimated anomalous heat transport, $Q\approx 0.2$ MW/m$^2$,}
\end{figure}

As well as driving non-ambipolar particle fluxes, the breaking of toroidal symmetry drives an additional neoclassical heat flux. In figure \ref{fig:HeatFlux}, the SFINCS calculation of the heat flux, $Q^{\mathrm{NC}}$, is shown for three magnetic geometries: (i) axisymmetric (blue solid), (ii) with TF ripple only (red dash-dot), and (iii) TF ripple with TBMs and FIs (green dashed). In the presence of TF ripple, the ripple drives an additional heat flux that is comparable to the axisymmetric heat flux. However, with the addition of the FIs the heat flux is reduced to the magnitude of the axisymmetric value, except near $E_r = 0$ where $1/\nu$ transport dominates. 

While the radial ripple-driven particle fluxes will significantly alter the ITER angular momentum transport, the neoclassical heat fluxes are insignificant in comparison to the turbulent heat flux. Note that the neoclassical heat flux is $\lesssim 5\%$  of the heat flux calculated from heating and fusion rate profiles (see appendix \ref{turbQ}), $Q\approx 0.2$ MW/m$^2$. Thus we can attribute $\gtrsim 95\%$ of the heat transport to turbulence. If ITER ripple were scaled up to $\delta_B \gtrsim 30\%$, the neoclassical ripple heat transport would be comparable to the anomalous transport at this radius.

\FloatBarrier

\section{Tangential Magnetic Drifts}\label{mds}
Although $(\bm{v}_E + \bm{v}_{\mathrm{m}}) \cdot \nabla f_1$ is formally of higher order than the other terms in equation \ref{kineticequation}, it has been found to be important when $\nu_* \lesssim \rho_*$ \cite{Calvo2016, Matsuoka2015} and has been included in other calculations of 3D neoclassical transport. In the SFINCS calculations shown in sections \ref{Erandv}, \ref{torque}, \ref{scaling}, and \ref{heatflux}, $\bm{v}_{\mathrm{m}} \cdot \nabla f_1$ has not been included, but now we examine the effect of including parts of this term. As SFINCS does not maintain radial coupling of $f_1$, only the poloidal and toroidal components of this magnetic drift term can be retained while the radial component cannot. Note that the radial magnetic drift is retained in $\bm{v}_{\mathrm{m}} \cdot \nabla f_0$. We first implement $\bm{v}_{\mathrm{m}} \cdot \nabla \theta$ and $\bm{v}_{\mathrm{m}}  \cdot \nabla \zeta$ using the following form of the magnetic drifts,
\begin{gather}
\bm{v}_{\mathrm{m} a} = \frac{v^2}{2 \Omega_a B^2} (1 + \xi^2) \bm{B} \times \nabla B + \frac{v^2}{\Omega_a B} \xi^2 \nabla \times \bm{B},
\label{eq:mds1}
\end{gather}
where $\xi = v_{||}/v$. However, a coordinate-dependence can be introduced as we simply drop one component of $\bm{v}_{\mathrm{m}}$. For a coordinate-independent form, one must project $\bm{v}_{\mathrm{m}}$ onto the flux surface. Additionally, when poloidal and toroidal drifts are retained, the effective particle trajectories do not necessarily conserve $\mu$ when $\mu = 0$. The drifts can be regularized in order to satisfy $\dot{\xi} (\xi = \pm 1) = 0$. Regularization also eliminates the need for additional particle and heat sources due to the radially local assumption and preserves ambipolarity of axisymmetric systems \cite{Sugama2016}. To this end, we also implement a coordinate-independent magnetic drift perpendicular to $\nabla \psi$,
\begin{gather}
\bm{v}_{\mathrm{m}a}^{\perp} = \frac{\nabla \psi \times (\bm{v}_{\mathrm{m}a} \times \nabla \psi)}{\rvert \nabla \psi \rvert^2}= \frac{v^2}{2\Omega_a B^2 } \frac{(\bm{B} \times \nabla \psi)}{\rvert \nabla \psi \rvert^2} \nabla \psi \cdot\left[(1-\xi^2)\nabla B + 2B\xi^2 (\bm{b} \cdot \nabla \bm{b}) \right].
\end{gather}
Note that the $\nabla B$ drift term is regularized while the curvature drift term is not. As tangential drifts are important for the trapped portion of velocity space, we can consider $\xi^2 \ll 1$. For this reason we drop the curvature drift for regularization,
\begin{gather}
\bm{v}_{\mathrm{m}a}^{\perp} = \frac{v^2}{2\Omega_aB^2} (\bm{B} \times \nabla \psi) (1 - \xi^2) \frac{(\nabla \psi \cdot \nabla B)}{\rvert \nabla \psi \rvert^2}.
\label{eq:mds5}
\end{gather}
This is similar to the form presented by Sugama \cite{Sugama2016}, but we have chosen a different form of regularization. This choice for $\bm{v}_{\mathrm{m}a}^{\perp}$ does not alter the conservation properties shown by Sugama, as it remains in the $\bm{B} \times \nabla \psi$ direction and vanishes at $\xi = \pm 1$. We note that the phase space conservation properties rely on the choice of a modified Jacobian in the presence of tangential magnetic drifts. In SFINCS we have not implemented such a modification. However, as Sugama shows, the correction to the Jacobian is an order $\rho_*$ correction. As particle and heat sources have been implemented in SFINCS, we have confirmed that the addition of tangential magnetic drifts does not necessitate the use of appreciable source terms. We note that this form of the tangential magnetic drifts we have chosen does not include a magnetic shear term which is present in the bounce-averaged radial drift. This non-local modification has been found to significantly alter superbanana transport \cite{Albert2016,Shaing2015} and the drift-orbit resonance \cite{Martitsch2016}.

\begin{figure}[h!]
\centering
\includegraphics[width=0.7\textwidth]{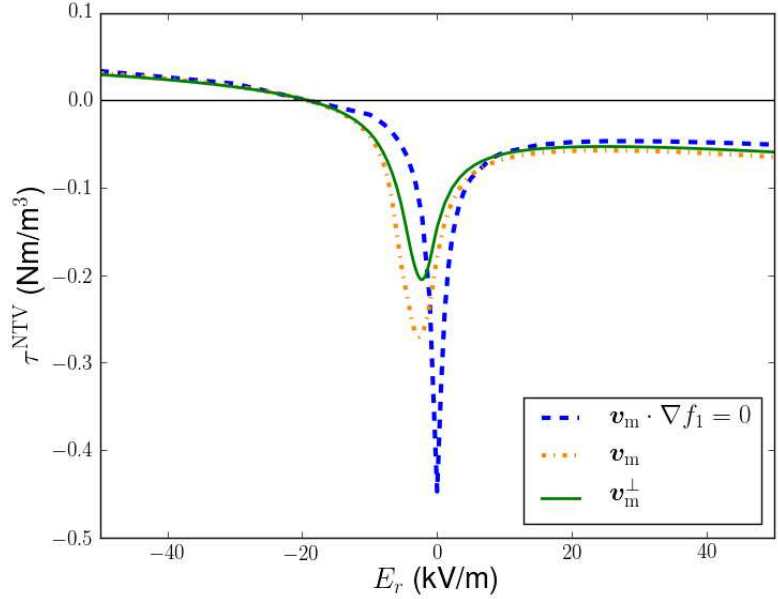}
\caption{\label{fig:driftschemes} Calculation of NTV torque density, $\tau^{\mathrm{NTV}}$, as a function of $E_r$ at $r/a = 0.7$. The blue dashed curve corresponds to a SFINCS calculation without $\bm{v}_m \cdot \nabla f_1$ in the DKE. The orange dash-dot curve corresponds to the addition of $\bm{v}_m \cdot \nabla f_1$ as given in equation \ref{eq:mds1}. The green solid curve corresponds to the addition of the projected and regularized drift, $\bm{v}^{\perp}_{\text{m}} \cdot \nabla f_1$, as given in equation \ref{eq:mds5}. }
\end{figure}

An $E_r$ scan at $r/a = 0.7$, where $\rho_*$ becomes comparable to $\nu_*$, is shown in figure \ref{fig:driftschemes}. When $\bm{v}_{\mathrm{m}} \cdot \nabla f_1$ is added to the kinetic equation, the $1/\nu$ peak at $E_r = 0$ is shifted toward a slightly negative $E_r$, corresponding to the region where $(\bm{v}_E + \bm{v}_\mathrm{M})\cdot \nabla \zeta \approx 0$, where superbanana-plateau transport takes place. For ITER parameters at this radius, the collisionality is large enough that superbananas cannot complete their collisionless trajectories but small enough that non-resonant trapped particles precess, $\nu_*^{\mathrm{SB}} \ll \nu_* \ll \nu_*^{\mathrm{SBP}}$, where $\nu_*^{\mathrm{SBP}} = \rho_*q^2/\epsilon^{1/2}$ and $\nu_*^{\mathrm{SB}} = \rho_* \delta_B^{3/2} q^2/\epsilon^2$ \cite{Shaing2009_sb, Shaing2009_sbp}, thus superbanana-plateau transport is relevant.

When the in-surface magnetic drifts are present, the depth of the resonant peak is diminished. In the absence of tangential drifts, the bounce-averaged toroidal drift vanishes at $E_r = 0$ for all particles regardless of pitch angle and energy. When tangential drifts are added to the DKE, the resonant peak will occur at the $E_r$ for which thermal trapped particles satisfy the resonance condition. However, only particles above a certain energy and at the resonant pitch angle will participate in the superbanana-plateau transport, thus the depth of the peak is diminished. Note that local ripple trapping might also contribute to the $1/\nu$ transport at small $\abs{E_r}$. For $\abs{E_r} > 20$ kV/m, the range relevant for ITER, the addition of $\bm{v}_{\mathrm{m}}\cdot \nabla f_1$ has a negligible effect on $\tau^{\mathrm{NTV}}$. The addition of tangential magnetic drifts would not dramatically change the results in previous sections.  

\begin{figure}[h!]
\subfloat[]{\includegraphics[width=0.6\textwidth]{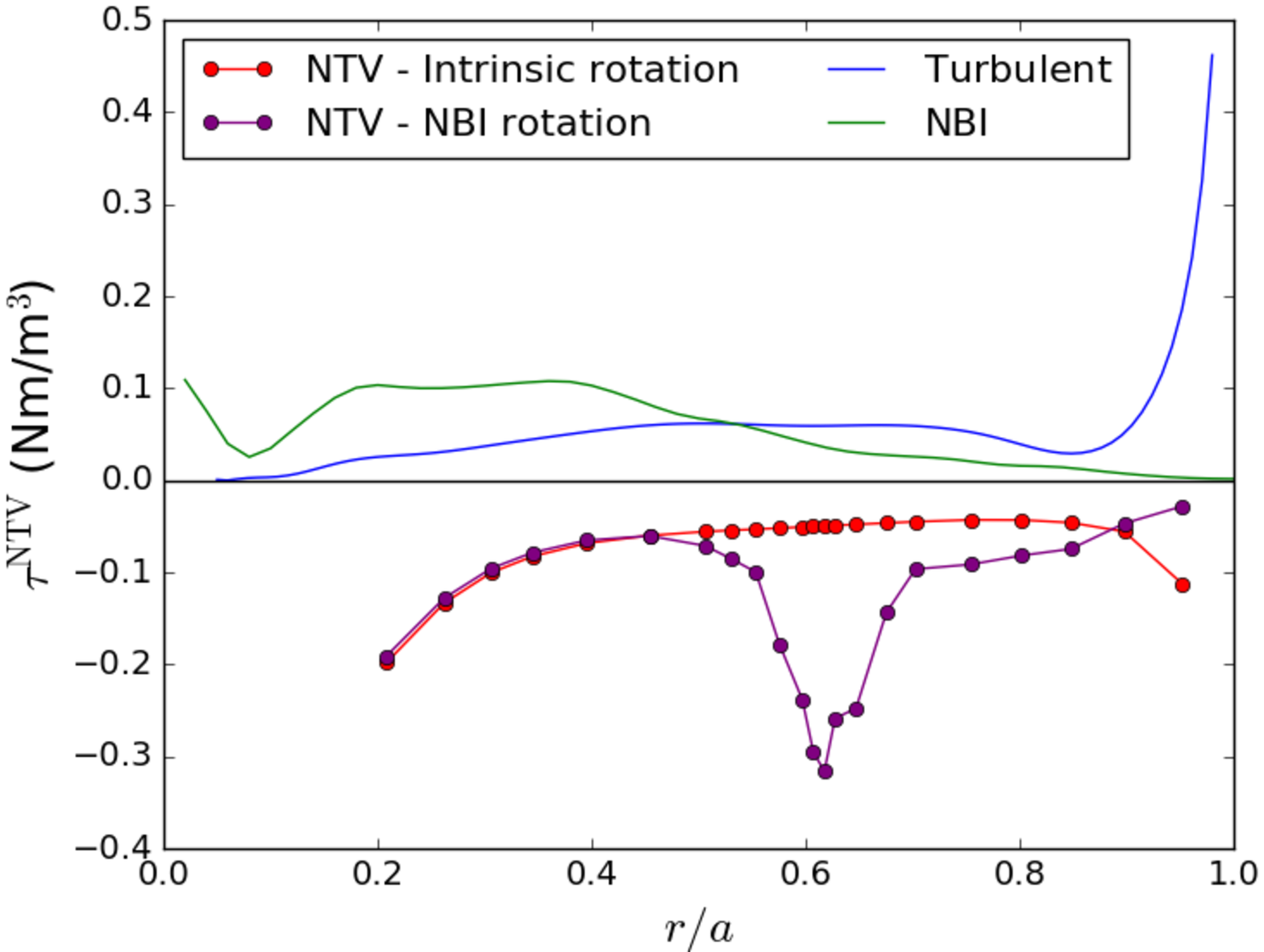}}\\
\vspace{-5mm}
\subfloat[]{\includegraphics[width=0.6\textwidth]{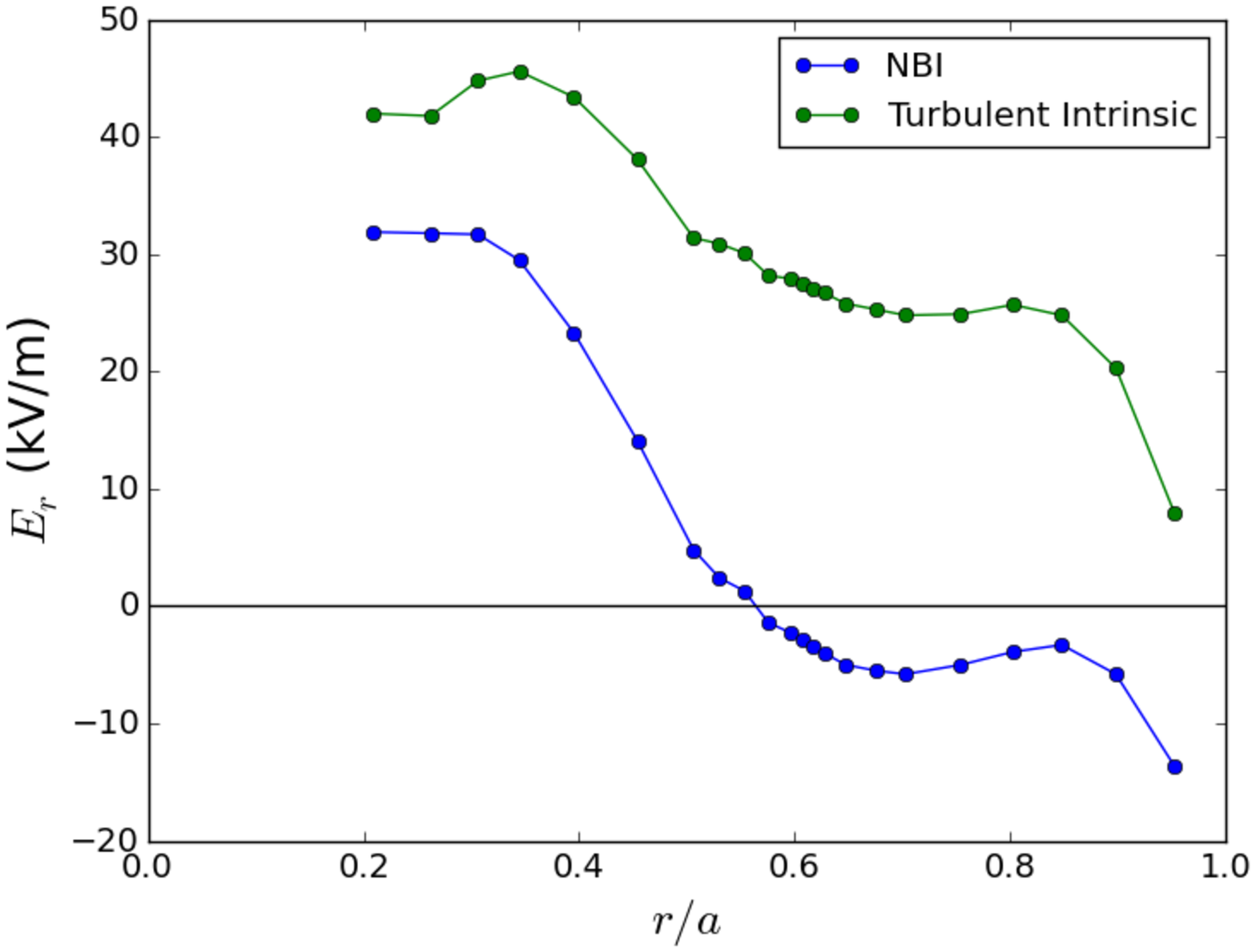}}
\vspace{-5mm}
\caption{\label{fig:alltorque} (a) Profiles of NTV torque density ($\tau^{\mathrm{NTV}}$) due to TF ripple without ferritic components calculated with SFINCS, NBI torque density calculated from NUBEAM ($\tau^{\mathrm{NBI}}$), and estimate of turbulent intrinsic rotation momentum source ($\tau^{\mathrm{turb}}$). The quantity $\tau^{\mathrm{NTV}}$ is calculated using the $E_r$ determined by the intrinsic rotation model and NBI rotation model described in section \ref{rotation}. Turbulent torque is estimated using $\tau^{\mathrm{turb}} \sim -\Pi_{\mathrm{int}}/a$ where $\Pi_{\mathrm{int}} \sim \rho_{*, \theta} \widetilde{\Pi}(\nu_*) Q \langle R \rangle/v_{ti}$ (see appendix \ref{turbQ} for details). (b) Profiles of radial electric field ($E_r$) due to NBI torque and turbulent intrinsic rotation. Here toroidal rotation is computed with the model described in section \ref{rotation} and $E_r$ is computed with SFINCS as described in section \ref{Erandv}.}
\end{figure}

\FloatBarrier
We compute a radial profile of $\tau^{\mathrm{NTV}}$ due to TF ripple including $\bm{v}_{\text{m}}^{\perp} \cdot \nabla f_1$. The intrinsic rotation model and NBI rotation model are used to estimate $E_r$ at each radius, as shown in the $E_r$ profile in figure \ref{fig:alltorque}. As $E_r$ crosses through 0 at $r/a = 0.56$ for the NBI rotation model, tangential drifts will affect the transport. In figure \ref{fig:alltorque}, we compare the magnitude of $\tau^{\mathrm{NTV}}$ due to TF ripple with $\tau^{\mathrm{NBI}}$ and $\tau^{\mathrm{turb}} = -\nabla \cdot \Pi_{\mathrm{int}}$, the turbulent momentum source causing intrinsic rotation. The $\tau^{\mathrm{NBI}}$ profile was computed by NUBEAM as used in section \ref{rotation}, and $\tau^{\mathrm{turb}}$ is estimated using $\Pi_{\mathrm{int}} \sim (\rho_{\theta}/L_T) \widetilde{\Pi}(\nu_*) Q (\langle R \rangle/v_{ti})$ (see appendix \ref{turbQ}). At $r/a = 0.62$ superbanana-plateau transport dominates when NBI rotation is considered, and $\tau^{NTV}$ is about 6 times larger than when the higher-rotation turbulent torque $E_r$ is considered. For both $E_r$ estimates $\abs{\tau^{\text{NTV}}}$ increases with decreasing radius due to the scaling with $T_i$ as discussed in section \ref{torque}. Note that the turbulent torque produces much rotation in the pedestal according to this model as $\tau^{\mathrm{turb}} \propto 1/L_T$. The integrated NTV torque, -45.6 Nm with the turbulent rotation model and -71 Nm with the NBI rotation model, is larger in magnitude than the NBI torque, 35 Nm, but smaller than the turbulent torque, 93 Nm. Here the integrated $\tau^{\text{turb}}$ is significantly larger than that obtained from dimensionless parameter scans on DIII-D, 33 Nm \cite{Chrystal2017}. This is possibly due to the assumed scaling in our turbulent rotation model, which may not be physical near the edge.

In the region $0.5 \lesssim r/a \lesssim 0.9$, the magnitude of $\tau^{\mathrm{NTV}}$ is comparable to $\tau^{\mathrm{turb}}$ and greater than $\tau^{\mathrm{NBI}}$. The NTV torque will likely significantly damp rotation in the absence of inserts, decreasing MHD stability. However, the resulting rotation profile may be sheared because of the significant counter-current NTV source at the edge and co-current NBI source in the core. We estimate the rotation shear, $\gamma = \Delta V_{\zeta}/ \Delta r \approx 0.4 (v_{ti}/R)$, using the neoclassical offset at $r/a = 0.7$ and the NBI-driven rotation at $r/a = 0.4$. Assuming the maximum linear growth rate for drift wave instabilities, $\gamma_{\mathrm{Lin}} \approx v_{ti}/R$ \cite{Connor2004}, this rotation shear may be large enough to suppress microturbulence. In concert with reversed magnetic shear sustained by heating and current drive sources \cite{Poli2014}, rotation shear may support the formation of an ITB \cite{Waltz1994} for this steady state scenario. 

\FloatBarrier
\section{Summary}\label{summary}

We calculate neoclassical transport in the presence of 3D magnetic fields, including toroidal field ripple and ferromagnetic components, for an ITER steady state scenario. We use an NBI and intrinsic turbulent rotation model to estimate $E_r$ for neoclassical calculations. We find that without considering $\tau^{\mathrm{NTV}}$, toroidal rotation with $M_A \lesssim 2\% \,$ is to be expected, which is likely not large enough to suppress resistive wall modes \cite{Liu2004}. We use VMEC free boundary equilibria in the presence of ripple fields to calculate neoclassical particle and heat fluxes using the drift-kinetic solver, SFINCS. At large radii $r/a \gtrsim 0.5$, $\tau^{\mathrm{NTV}}$ due to TF ripple without ferritic components is comparable to $\tau^{\mathrm{NBI}}$ and $\tau^{\mathrm{turb}}$ in magnitude but opposite in sign, which may result in flow damping at the edge and a decrease in MHD stability. As the integral NTV torque is similar in magnitude to the NBI torque, non-resonant magnetic braking cannot be ignored in analysis of ITER rotation. The torque profile may also result in a significant rotation shear which could suppress turbulent transport. While the addition of FIs significantly reduces the transport ($\approx 75\%$ reduction at $r/a = 0.9$), the low $n$ perturbation of the TBM produces very little NTV torque. The neoclassical heat flux caused by ripple is insignificant in comparison to the turbulent heat flux. Though NTV torque has been shown to be important for ITER angular momentum balance, iteratively solving for the rotation profile with $\tau^{\mathrm{NTV}}$ will be left for future consideration.  

Several transport regimes must be considered for ITER NTV: the $\nu-\sqrt{\nu}$ banana diffusion, bounce-resonance, and $\nu-\sqrt{\nu}$ ripple trapping regimes. The calculated scaling of $\tau^{\mathrm{NTV}}$ with $\delta_B$ is between the $\delta_B^{3/2}$ scaling of the ripple trapping $\sqrt{\nu}$ regime and the $\delta_B^2$ scaling predicted in the $\nu-\sqrt{\nu}$ regime at small $\delta_B$. There is room for further comparison between SFINCS calculations of $\tau^{\mathrm{NTV}}$ and analytic fomulae. However, we note that the analytic theory for transport of ripple-trapped particles in a tokamak close to axisymmetry in the presence of $E_r$ is not fully developed. 

\appendix

\section{Approximate Turbulent Heat Flux and Torque}\label{turbQ}

\begin{figure}[h!]
\centering
\includegraphics[width=0.7\textwidth]{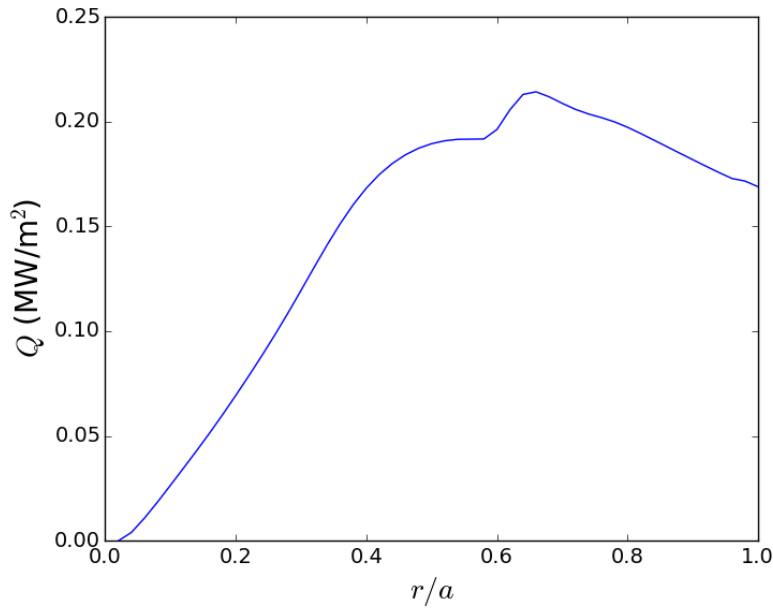}
\caption{\label{fig:turbHeatFlux} Heat flux $Q$ calculated with input heating and fusion rate profiles from TRANSP and TSC.}
\end{figure}

As $\Pi_{\mathrm{int}}$ is proportional to $Q$ in our model, we must estimate $Q$ using the input heating power and D-T fusion rates calculated with TRANSP. The LH, NBI, and ECH power densities ($P_{\mathrm{LH}}$, $P_{\mathrm{NBI}}$, and $P_{\mathrm{ECH}}$) are integrated along with the fusion reaction rate density ($R_{\mathrm{DT}}$) to calculate the total integrated heating source, $H(r)$,
\begin{gather}
\int_0^r dV(r') \, H(r') = \int_0^r dV(r') \left(\, P_{\mathrm{LH}} + P_{\mathrm{NBI}} + P_{\mathrm{ECH}} + R_{\mathrm{DT}} (3.5 \mathrm{MeV}) \right).
\end{gather}
As $\int Q dS = \int H dV$, 
\begin{gather}
Q(r) = \frac{\int_0^{r} dV(r') \left(\, P_{\mathrm{LH}} + P_{\mathrm{NBI}} + P_{\mathrm{ECH}} + R_{\mathrm{DT}} (3.5 \mathrm{MeV}) \right)}{A(r)},
\end{gather}
where $A(r)$ is the flux surface area and $V(r)$ is the volume enclosed by flux surface $r$. We have shown in section \ref{heatflux} that the neoclassical heat flux is insignificant in comparison to $Q(r)$, so we can attribute $Q(r)$ to turbulent heat transport. The calculated $Q$ is shown in figure \ref{fig:turbHeatFlux}.

We estimate $\tau^{\mathrm{turb}} = - \nabla \cdot \Pi_{\mathrm{int}} \sim -\Pi_{\mathrm{int}}/a$ using 
\begin{gather}
\Pi_{\mathrm{int}} \sim \frac{\rho_{\theta} \widetilde{\Pi}(\nu_*) Q \langle R \rangle}{v_{ti} L_T}.
\end{gather}
The quantity $\tau^{\mathrm{turb}}$ is shown in figure \ref{fig:alltorque}.

\FloatBarrier

\section*{Acknowledgements}
The authors would like to thank I. Calvo, F. Parra, J. Hillesheim, J. Lee, G. Papp, S. Satake, and J. Harris for helpful input and discussions. This work was supported by the US Department of Energy through grants DE-FG02-93ER-54197 and DE-FC02-08ER-54964. The computations presented in this paper have used resources at the National Energy Research Scientific Computing Center (NERSC). 

\raggedright
\bibliographystyle{apsrev4-1}
\bibliography{NTV_paper}

%merlin.mbs apsrev4-1.bst 2010-07-25 4.21a (PWD, AO, DPC) hacked
%Control: key (0)
%Control: author (72) initials jnrlst
%Control: editor formatted (1) identically to author
%Control: production of article title (-1) disabled
%Control: page (0) single
%Control: year (1) truncated
%Control: production of eprint (0) enabled
\begin{thebibliography}{89}%
\makeatletter
\providecommand \@ifxundefined [1]{%
 \@ifx{#1\undefined}
}%
\providecommand \@ifnum [1]{%
 \ifnum #1\expandafter \@firstoftwo
 \else \expandafter \@secondoftwo
 \fi
}%
\providecommand \@ifx [1]{%
 \ifx #1\expandafter \@firstoftwo
 \else \expandafter \@secondoftwo
 \fi
}%
\providecommand \natexlab [1]{#1}%
\providecommand \enquote  [1]{``#1''}%
\providecommand \bibnamefont  [1]{#1}%
\providecommand \bibfnamefont [1]{#1}%
\providecommand \citenamefont [1]{#1}%
\providecommand \href@noop [0]{\@secondoftwo}%
\providecommand \href [0]{\begingroup \@sanitize@url \@href}%
\providecommand \@href[1]{\@@startlink{#1}\@@href}%
\providecommand \@@href[1]{\endgroup#1\@@endlink}%
\providecommand \@sanitize@url [0]{\catcode `\\12\catcode `\$12\catcode
  `\&12\catcode `\#12\catcode `\^12\catcode `\_12\catcode `\%12\relax}%
\providecommand \@@startlink[1]{}%
\providecommand \@@endlink[0]{}%
\providecommand \url  [0]{\begingroup\@sanitize@url \@url }%
\providecommand \@url [1]{\endgroup\@href {#1}{\urlprefix }}%
\providecommand \urlprefix  [0]{URL }%
\providecommand \Eprint [0]{\href }%
\providecommand \doibase [0]{http://dx.doi.org/}%
\providecommand \selectlanguage [0]{\@gobble}%
\providecommand \bibinfo  [0]{\@secondoftwo}%
\providecommand \bibfield  [0]{\@secondoftwo}%
\providecommand \translation [1]{[#1]}%
\providecommand \BibitemOpen [0]{}%
\providecommand \bibitemStop [0]{}%
\providecommand \bibitemNoStop [0]{.\EOS\space}%
\providecommand \EOS [0]{\spacefactor3000\relax}%
\providecommand \BibitemShut  [1]{\csname bibitem#1\endcsname}%
\let\auto@bib@innerbib\@empty
%</preamble>
\bibitem [{\citenamefont {Bondeson}\ and\ \citenamefont
  {Ward}(1994)}]{Bondeson1994}%
  \BibitemOpen
  \bibfield  {author} {\bibinfo {author} {\bibfnamefont {A.}~\bibnamefont
  {Bondeson}}\ and\ \bibinfo {author} {\bibfnamefont {D.~J.}\ \bibnamefont
  {Ward}},\ }\href {\doibase 10.1103/PhysRevLett.72.2709} {\bibfield  {journal}
  {\bibinfo  {journal} {Physical Review Letters}\ }\textbf {\bibinfo {volume}
  {72}},\ \bibinfo {pages} {2709} (\bibinfo {year} {1994})}\BibitemShut
  {NoStop}%
\bibitem [{\citenamefont {Garofalo}\ \emph {et~al.}(2002)\citenamefont
  {Garofalo}, \citenamefont {Strait}, \citenamefont {Johnson}, \citenamefont
  {{La Haye}}, \citenamefont {Lazarus}, \citenamefont {Navratil}, \citenamefont
  {Okabayashi}, \citenamefont {Scoville}, \citenamefont {Taylor},\ and\
  \citenamefont {Turnbull}}]{Garofalo2002}%
  \BibitemOpen
  \bibfield  {author} {\bibinfo {author} {\bibfnamefont {A.~M.}\ \bibnamefont
  {Garofalo}}, \bibinfo {author} {\bibfnamefont {E.~J.}\ \bibnamefont
  {Strait}}, \bibinfo {author} {\bibfnamefont {L.~C.}\ \bibnamefont {Johnson}},
  \bibinfo {author} {\bibfnamefont {R.~J.}\ \bibnamefont {{La Haye}}}, \bibinfo
  {author} {\bibfnamefont {E.~A.}\ \bibnamefont {Lazarus}}, \bibinfo {author}
  {\bibfnamefont {G.~A.}\ \bibnamefont {Navratil}}, \bibinfo {author}
  {\bibfnamefont {M.}~\bibnamefont {Okabayashi}}, \bibinfo {author}
  {\bibfnamefont {J.~T.}\ \bibnamefont {Scoville}}, \bibinfo {author}
  {\bibfnamefont {T.~S.}\ \bibnamefont {Taylor}}, \ and\ \bibinfo {author}
  {\bibfnamefont {A.~D.}\ \bibnamefont {Turnbull}},\ }\href {\doibase
  10.1103/Physrevlett.89.235001} {\bibfield  {journal} {\bibinfo  {journal}
  {Physical Review Letters}\ }\textbf {\bibinfo {volume} {89}},\ \bibinfo
  {pages} {235001} (\bibinfo {year} {2002})}\BibitemShut {NoStop}%
\bibitem [{\citenamefont {Burrell}(1997)}]{Burrell1997}%
  \BibitemOpen
  \bibfield  {author} {\bibinfo {author} {\bibfnamefont {K.~H.}\ \bibnamefont
  {Burrell}},\ }\href {\doibase doi:10.1063/1.872367} {\bibfield  {journal}
  {\bibinfo  {journal} {Physics of Plasmas}\ }\textbf {\bibinfo {volume} {4}},\
  \bibinfo {pages} {1499} (\bibinfo {year} {1997})}\BibitemShut {NoStop}%
\bibitem [{\citenamefont {Terry}(2000)}]{Terry2000}%
  \BibitemOpen
  \bibfield  {author} {\bibinfo {author} {\bibfnamefont {P.~W.}\ \bibnamefont
  {Terry}},\ }\href {\doibase 10.1103/RevModPhys.72.109} {\bibfield  {journal}
  {\bibinfo  {journal} {Reviews of Modern Physics}\ }\textbf {\bibinfo {volume}
  {72}},\ \bibinfo {pages} {109} (\bibinfo {year} {2000})}\BibitemShut
  {NoStop}%
\bibitem [{\citenamefont {Liu}\ \emph {et~al.}(2004)\citenamefont {Liu},
  \citenamefont {Bondeson}, \citenamefont {Gribov},\ and\ \citenamefont
  {Polevoi}}]{Liu2004}%
  \BibitemOpen
  \bibfield  {author} {\bibinfo {author} {\bibfnamefont {Y.}~\bibnamefont
  {Liu}}, \bibinfo {author} {\bibfnamefont {A.}~\bibnamefont {Bondeson}},
  \bibinfo {author} {\bibfnamefont {Y.}~\bibnamefont {Gribov}}, \ and\ \bibinfo
  {author} {\bibfnamefont {A.}~\bibnamefont {Polevoi}},\ }\href
  {http://stacks.iop.org/0029-5515/44/i=2/a=003} {\bibfield  {journal}
  {\bibinfo  {journal} {Nuclear Fusion}\ }\textbf {\bibinfo {volume} {44}},\
  \bibinfo {pages} {232} (\bibinfo {year} {2004})}\BibitemShut {NoStop}%
\bibitem [{\citenamefont {Hua}\ \emph {et~al.}(2010)\citenamefont {Hua},
  \citenamefont {Chapman}, \citenamefont {Field}, \citenamefont {Hastie},
  \citenamefont {Pinches},\ and\ \citenamefont {the MAST~Team}}]{Hua2010}%
  \BibitemOpen
  \bibfield  {author} {\bibinfo {author} {\bibfnamefont {M.}~\bibnamefont
  {Hua}}, \bibinfo {author} {\bibfnamefont {I.~T.}\ \bibnamefont {Chapman}},
  \bibinfo {author} {\bibfnamefont {A.~R.}\ \bibnamefont {Field}}, \bibinfo
  {author} {\bibfnamefont {R.~J.}\ \bibnamefont {Hastie}}, \bibinfo {author}
  {\bibfnamefont {S.~D.}\ \bibnamefont {Pinches}}, \ and\ \bibinfo {author}
  {\bibnamefont {the MAST~Team}},\ }\href {\doibase
  10.1088/0741-3335/52/3/035009} {\bibfield  {journal} {\bibinfo  {journal}
  {Plasma Physics and Controlled Fusion}\ }\textbf {\bibinfo {volume} {52}},\
  \bibinfo {pages} {035009} (\bibinfo {year} {2010})}\BibitemShut {NoStop}%
\bibitem [{\citenamefont {Lazzaro}\ \emph {et~al.}(2002)\citenamefont
  {Lazzaro}, \citenamefont {Buttery}, \citenamefont {Hender}, \citenamefont
  {Zanca}, \citenamefont {Fitzpatrick}, \citenamefont {Bigi}, \citenamefont
  {Bolzonella}, \citenamefont {Coelho}, \citenamefont {DeBenedetti},
  \citenamefont {Nowak}, \citenamefont {Sauter},\ and\ \citenamefont
  {Stamp}}]{Lazzaro2002}%
  \BibitemOpen
  \bibfield  {author} {\bibinfo {author} {\bibfnamefont {E.}~\bibnamefont
  {Lazzaro}}, \bibinfo {author} {\bibfnamefont {R.~J.}\ \bibnamefont
  {Buttery}}, \bibinfo {author} {\bibfnamefont {T.~C.}\ \bibnamefont {Hender}},
  \bibinfo {author} {\bibfnamefont {P.}~\bibnamefont {Zanca}}, \bibinfo
  {author} {\bibfnamefont {R.}~\bibnamefont {Fitzpatrick}}, \bibinfo {author}
  {\bibfnamefont {M.}~\bibnamefont {Bigi}}, \bibinfo {author} {\bibfnamefont
  {T.}~\bibnamefont {Bolzonella}}, \bibinfo {author} {\bibfnamefont
  {R.}~\bibnamefont {Coelho}}, \bibinfo {author} {\bibfnamefont
  {M.}~\bibnamefont {DeBenedetti}}, \bibinfo {author} {\bibfnamefont
  {S.}~\bibnamefont {Nowak}}, \bibinfo {author} {\bibfnamefont
  {O.}~\bibnamefont {Sauter}}, \ and\ \bibinfo {author} {\bibfnamefont
  {M.}~\bibnamefont {Stamp}},\ }\href {\doibase 10.1063/1.1499495} {\bibfield
  {journal} {\bibinfo  {journal} {Physics of Plasmas}\ }\textbf {\bibinfo
  {volume} {9}},\ \bibinfo {pages} {3906} (\bibinfo {year} {2002})}\BibitemShut
  {NoStop}%
\bibitem [{\citenamefont {de~Vries}\ \emph {et~al.}(2008)\citenamefont
  {de~Vries}, \citenamefont {Salmi}, \citenamefont {Parail}, \citenamefont
  {Giroud}, \citenamefont {Andrew}, \citenamefont {Biewer}, \citenamefont
  {Cromb{\'{e}}}, \citenamefont {Jenkins}, \citenamefont {Johnson},
  \citenamefont {Kiptily}, \citenamefont {Loarte}, \citenamefont
  {L{\"{o}}nnroth}, \citenamefont {Meigs}, \citenamefont {Oyama}, \citenamefont
  {Sartori}, \citenamefont {Saibene}, \citenamefont {Urano}, \citenamefont
  {Zastrow},\ and\ \citenamefont {{JET EFDA contributors}}}]{DeVries2008b}%
  \BibitemOpen
  \bibfield  {author} {\bibinfo {author} {\bibfnamefont {P.}~\bibnamefont
  {de~Vries}}, \bibinfo {author} {\bibfnamefont {A.}~\bibnamefont {Salmi}},
  \bibinfo {author} {\bibfnamefont {V.}~\bibnamefont {Parail}}, \bibinfo
  {author} {\bibfnamefont {C.}~\bibnamefont {Giroud}}, \bibinfo {author}
  {\bibfnamefont {Y.}~\bibnamefont {Andrew}}, \bibinfo {author} {\bibfnamefont
  {T.}~\bibnamefont {Biewer}}, \bibinfo {author} {\bibfnamefont
  {K.}~\bibnamefont {Cromb{\'{e}}}}, \bibinfo {author} {\bibfnamefont
  {I.}~\bibnamefont {Jenkins}}, \bibinfo {author} {\bibfnamefont
  {T.}~\bibnamefont {Johnson}}, \bibinfo {author} {\bibfnamefont
  {V.}~\bibnamefont {Kiptily}}, \bibinfo {author} {\bibfnamefont
  {A.}~\bibnamefont {Loarte}}, \bibinfo {author} {\bibfnamefont
  {J.}~\bibnamefont {L{\"{o}}nnroth}}, \bibinfo {author} {\bibfnamefont
  {A.}~\bibnamefont {Meigs}}, \bibinfo {author} {\bibfnamefont
  {N.}~\bibnamefont {Oyama}}, \bibinfo {author} {\bibfnamefont
  {R.}~\bibnamefont {Sartori}}, \bibinfo {author} {\bibfnamefont
  {G.}~\bibnamefont {Saibene}}, \bibinfo {author} {\bibfnamefont
  {H.}~\bibnamefont {Urano}}, \bibinfo {author} {\bibfnamefont
  {K.}~\bibnamefont {Zastrow}}, \ and\ \bibinfo {author} {\bibnamefont {{JET
  EFDA contributors}}},\ }\href {\doibase 10.1088/0029-5515/48/3/035007}
  {\bibfield  {journal} {\bibinfo  {journal} {Nuclear Fusion}\ }\textbf
  {\bibinfo {volume} {48}},\ \bibinfo {pages} {035007} (\bibinfo {year}
  {2008})}\BibitemShut {NoStop}%
\bibitem [{\citenamefont {Wolfe}\ \emph {et~al.}(2005)\citenamefont {Wolfe},
  \citenamefont {Hutchinson}, \citenamefont {Granetz}, \citenamefont {Rice},
  \citenamefont {Hubbard}, \citenamefont {Lynn}, \citenamefont {Phillips},
  \citenamefont {Hender}, \citenamefont {Howell}, \citenamefont {{La Haye}},\
  and\ \citenamefont {Scoville}}]{Wolfe2005}%
  \BibitemOpen
  \bibfield  {author} {\bibinfo {author} {\bibfnamefont {S.~M.}\ \bibnamefont
  {Wolfe}}, \bibinfo {author} {\bibfnamefont {I.~H.}\ \bibnamefont
  {Hutchinson}}, \bibinfo {author} {\bibfnamefont {R.~S.}\ \bibnamefont
  {Granetz}}, \bibinfo {author} {\bibfnamefont {J.}~\bibnamefont {Rice}},
  \bibinfo {author} {\bibfnamefont {A.}~\bibnamefont {Hubbard}}, \bibinfo
  {author} {\bibfnamefont {A.}~\bibnamefont {Lynn}}, \bibinfo {author}
  {\bibfnamefont {P.}~\bibnamefont {Phillips}}, \bibinfo {author}
  {\bibfnamefont {T.~C.}\ \bibnamefont {Hender}}, \bibinfo {author}
  {\bibfnamefont {D.~F.}\ \bibnamefont {Howell}}, \bibinfo {author}
  {\bibfnamefont {R.~J.}\ \bibnamefont {{La Haye}}}, \ and\ \bibinfo {author}
  {\bibfnamefont {J.~T.}\ \bibnamefont {Scoville}},\ }\href {\doibase
  10.1063/1.1883665} {\bibfield  {journal} {\bibinfo  {journal} {Physics of
  Plasmas}\ }\textbf {\bibinfo {volume} {12}},\ \bibinfo {pages} {1} (\bibinfo
  {year} {2005})}\BibitemShut {NoStop}%
\bibitem [{\citenamefont {Garofalo}\ \emph {et~al.}(2008)\citenamefont
  {Garofalo}, \citenamefont {Burrell}, \citenamefont {Deboo}, \citenamefont
  {Degrassie}, \citenamefont {Jackson}, \citenamefont {Lanctot}, \citenamefont
  {Reimerdes}, \citenamefont {Schaffer}, \citenamefont {Solomon},\ and\
  \citenamefont {Strait}}]{Garofalo2008}%
  \BibitemOpen
  \bibfield  {author} {\bibinfo {author} {\bibfnamefont {A.~M.}\ \bibnamefont
  {Garofalo}}, \bibinfo {author} {\bibfnamefont {K.~H.}\ \bibnamefont
  {Burrell}}, \bibinfo {author} {\bibfnamefont {J.~C.}\ \bibnamefont {Deboo}},
  \bibinfo {author} {\bibfnamefont {J.~S.}\ \bibnamefont {Degrassie}}, \bibinfo
  {author} {\bibfnamefont {G.~L.}\ \bibnamefont {Jackson}}, \bibinfo {author}
  {\bibfnamefont {M.}~\bibnamefont {Lanctot}}, \bibinfo {author} {\bibfnamefont
  {H.}~\bibnamefont {Reimerdes}}, \bibinfo {author} {\bibfnamefont {M.~J.}\
  \bibnamefont {Schaffer}}, \bibinfo {author} {\bibfnamefont {W.~M.}\
  \bibnamefont {Solomon}}, \ and\ \bibinfo {author} {\bibfnamefont {E.~J.}\
  \bibnamefont {Strait}},\ }\href
  {http://link.aps.org/doi/10.1103/PhysRevLett.101.195005} {\bibfield
  {journal} {\bibinfo  {journal} {Physical Review Letters}\ }\textbf {\bibinfo
  {volume} {101}} (\bibinfo {year} {2008})}\BibitemShut {NoStop}%
\bibitem [{\citenamefont {Reimerdes}\ \emph {et~al.}(2009)\citenamefont
  {Reimerdes}, \citenamefont {Garofalo}, \citenamefont {Strait}, \citenamefont
  {Buttery}, \citenamefont {Chu}, \citenamefont {In}, \citenamefont {Jackson},
  \citenamefont {{La Haye}}, \citenamefont {Lanctot}, \citenamefont {Liu},
  \citenamefont {Okabayashi}, \citenamefont {Park}, \citenamefont {Schaffer},\
  and\ \citenamefont {Solomon}}]{Reimerdes2009}%
  \BibitemOpen
  \bibfield  {author} {\bibinfo {author} {\bibfnamefont {H.}~\bibnamefont
  {Reimerdes}}, \bibinfo {author} {\bibfnamefont {A.~M.}\ \bibnamefont
  {Garofalo}}, \bibinfo {author} {\bibfnamefont {E.}~\bibnamefont {Strait}},
  \bibinfo {author} {\bibfnamefont {R.}~\bibnamefont {Buttery}}, \bibinfo
  {author} {\bibfnamefont {M.}~\bibnamefont {Chu}}, \bibinfo {author}
  {\bibfnamefont {Y.}~\bibnamefont {In}}, \bibinfo {author} {\bibfnamefont
  {G.}~\bibnamefont {Jackson}}, \bibinfo {author} {\bibfnamefont
  {R.}~\bibnamefont {{La Haye}}}, \bibinfo {author} {\bibfnamefont
  {M.}~\bibnamefont {Lanctot}}, \bibinfo {author} {\bibfnamefont
  {Y.}~\bibnamefont {Liu}}, \bibinfo {author} {\bibfnamefont {M.}~\bibnamefont
  {Okabayashi}}, \bibinfo {author} {\bibfnamefont {J.~K.}\ \bibnamefont
  {Park}}, \bibinfo {author} {\bibfnamefont {M.}~\bibnamefont {Schaffer}}, \
  and\ \bibinfo {author} {\bibfnamefont {W.}~\bibnamefont {Solomon}},\ }\href
  {\doibase 10.1088/0029-5515/49/11/115001} {\bibfield  {journal} {\bibinfo
  {journal} {Nuclear Fusion}\ }\textbf {\bibinfo {volume} {49}},\ \bibinfo
  {pages} {115001} (\bibinfo {year} {2009})}\BibitemShut {NoStop}%
\bibitem [{\citenamefont {Honda}\ \emph {et~al.}(2014)\citenamefont {Honda},
  \citenamefont {Satake}, \citenamefont {Suzuki}, \citenamefont {Matsunaga},
  \citenamefont {Shinohara}, \citenamefont {Yoshida}, \citenamefont
  {Matsuyama}, \citenamefont {Ide},\ and\ \citenamefont {Urano}}]{Honda2014}%
  \BibitemOpen
  \bibfield  {author} {\bibinfo {author} {\bibfnamefont {M.}~\bibnamefont
  {Honda}}, \bibinfo {author} {\bibfnamefont {S.}~\bibnamefont {Satake}},
  \bibinfo {author} {\bibfnamefont {Y.}~\bibnamefont {Suzuki}}, \bibinfo
  {author} {\bibfnamefont {G.}~\bibnamefont {Matsunaga}}, \bibinfo {author}
  {\bibfnamefont {K.}~\bibnamefont {Shinohara}}, \bibinfo {author}
  {\bibfnamefont {M.}~\bibnamefont {Yoshida}}, \bibinfo {author} {\bibfnamefont
  {A.}~\bibnamefont {Matsuyama}}, \bibinfo {author} {\bibfnamefont
  {S.}~\bibnamefont {Ide}}, \ and\ \bibinfo {author} {\bibfnamefont
  {H.}~\bibnamefont {Urano}},\ }\href {\doibase 10.1088/0029-5515/54/11/114005}
  {\bibfield  {journal} {\bibinfo  {journal} {Nuclear Fusion}\ }\textbf
  {\bibinfo {volume} {54}},\ \bibinfo {pages} {114005} (\bibinfo {year}
  {2014})}\BibitemShut {NoStop}%
\bibitem [{\citenamefont {Zhu}\ \emph {et~al.}(2006)\citenamefont {Zhu},
  \citenamefont {Sabbagh}, \citenamefont {Bell}, \citenamefont {Bialek},
  \citenamefont {Bell}, \citenamefont {Leblanc}, \citenamefont {Kaye},
  \citenamefont {Levinton}, \citenamefont {Menard}, \citenamefont {Shaing},
  \citenamefont {Sontag},\ and\ \citenamefont {Yuh}}]{Zhu2006}%
  \BibitemOpen
  \bibfield  {author} {\bibinfo {author} {\bibfnamefont {W.}~\bibnamefont
  {Zhu}}, \bibinfo {author} {\bibfnamefont {S.~A.}\ \bibnamefont {Sabbagh}},
  \bibinfo {author} {\bibfnamefont {R.~E.}\ \bibnamefont {Bell}}, \bibinfo
  {author} {\bibfnamefont {J.~M.}\ \bibnamefont {Bialek}}, \bibinfo {author}
  {\bibfnamefont {M.~G.}\ \bibnamefont {Bell}}, \bibinfo {author}
  {\bibfnamefont {B.~P.}\ \bibnamefont {Leblanc}}, \bibinfo {author}
  {\bibfnamefont {S.~M.}\ \bibnamefont {Kaye}}, \bibinfo {author}
  {\bibfnamefont {F.~M.}\ \bibnamefont {Levinton}}, \bibinfo {author}
  {\bibfnamefont {J.~E.}\ \bibnamefont {Menard}}, \bibinfo {author}
  {\bibfnamefont {K.~C.}\ \bibnamefont {Shaing}}, \bibinfo {author}
  {\bibfnamefont {A.~C.}\ \bibnamefont {Sontag}}, \ and\ \bibinfo {author}
  {\bibfnamefont {H.}~\bibnamefont {Yuh}},\ }\href {\doibase
  10.1103/PhysRevLett.96.225002} {\bibfield  {journal} {\bibinfo  {journal}
  {Physical Review Letters}\ }\textbf {\bibinfo {volume} {96}},\ \bibinfo
  {pages} {225002} (\bibinfo {year} {2006})}\BibitemShut {NoStop}%
\bibitem [{\citenamefont {Chuyanov}\ \emph {et~al.}(2010)\citenamefont
  {Chuyanov}, \citenamefont {Campbell},\ and\ \citenamefont
  {Giancarli}}]{Chuyanov2010}%
  \BibitemOpen
  \bibfield  {author} {\bibinfo {author} {\bibfnamefont {V.~A.}\ \bibnamefont
  {Chuyanov}}, \bibinfo {author} {\bibfnamefont {D.~J.}\ \bibnamefont
  {Campbell}}, \ and\ \bibinfo {author} {\bibfnamefont {L.~M.}\ \bibnamefont
  {Giancarli}},\ }\href {\doibase 10.1016/j.fusengdes.2010.07.005} {\bibfield
  {journal} {\bibinfo  {journal} {Fusion Engineering and Design}\ }\textbf
  {\bibinfo {volume} {85}},\ \bibinfo {pages} {2005} (\bibinfo {year}
  {2010})}\BibitemShut {NoStop}%
\bibitem [{\citenamefont {Schaffer}\ \emph {et~al.}(2011)\citenamefont
  {Schaffer}, \citenamefont {Snipes}, \citenamefont {Gohil}, \citenamefont
  {de~Vries}, \citenamefont {Evans}, \citenamefont {Fenstermacher},
  \citenamefont {Gao}, \citenamefont {Garofalo}, \citenamefont {Gates},
  \citenamefont {Greenfield}, \citenamefont {Heidbrink}, \citenamefont
  {Kramer}, \citenamefont {{La Haye}}, \citenamefont {Liu}, \citenamefont
  {Loarte}, \citenamefont {Nave}, \citenamefont {Osborne}, \citenamefont
  {Oyama}, \citenamefont {Park}, \citenamefont {Ramasubramanian}, \citenamefont
  {Reimerdes}, \citenamefont {Saibene}, \citenamefont {Salmi}, \citenamefont
  {Shinohara}, \citenamefont {Spong}, \citenamefont {Solomon}, \citenamefont
  {Tala}, \citenamefont {Zhu}, \citenamefont {Boedo}, \citenamefont {Chuyanov},
  \citenamefont {Doyle}, \citenamefont {Jakubowski}, \citenamefont {Jhang},
  \citenamefont {Nazikian}, \citenamefont {Pustovitov}, \citenamefont
  {Schmitz}, \citenamefont {Srinivasan}, \citenamefont {Taylor}, \citenamefont
  {Wade}, \citenamefont {You},\ and\ \citenamefont {Zeng}}]{Schaffer2011}%
  \BibitemOpen
  \bibfield  {author} {\bibinfo {author} {\bibfnamefont {M.}~\bibnamefont
  {Schaffer}}, \bibinfo {author} {\bibfnamefont {J.}~\bibnamefont {Snipes}},
  \bibinfo {author} {\bibfnamefont {P.}~\bibnamefont {Gohil}}, \bibinfo
  {author} {\bibfnamefont {P.}~\bibnamefont {de~Vries}}, \bibinfo {author}
  {\bibfnamefont {T.}~\bibnamefont {Evans}}, \bibinfo {author} {\bibfnamefont
  {M.}~\bibnamefont {Fenstermacher}}, \bibinfo {author} {\bibfnamefont
  {X.}~\bibnamefont {Gao}}, \bibinfo {author} {\bibfnamefont {A.}~\bibnamefont
  {Garofalo}}, \bibinfo {author} {\bibfnamefont {D.}~\bibnamefont {Gates}},
  \bibinfo {author} {\bibfnamefont {C.}~\bibnamefont {Greenfield}}, \bibinfo
  {author} {\bibfnamefont {W.}~\bibnamefont {Heidbrink}}, \bibinfo {author}
  {\bibfnamefont {G.}~\bibnamefont {Kramer}}, \bibinfo {author} {\bibfnamefont
  {R.}~\bibnamefont {{La Haye}}}, \bibinfo {author} {\bibfnamefont
  {S.}~\bibnamefont {Liu}}, \bibinfo {author} {\bibfnamefont {A.}~\bibnamefont
  {Loarte}}, \bibinfo {author} {\bibfnamefont {M.}~\bibnamefont {Nave}},
  \bibinfo {author} {\bibfnamefont {T.}~\bibnamefont {Osborne}}, \bibinfo
  {author} {\bibfnamefont {N.}~\bibnamefont {Oyama}}, \bibinfo {author}
  {\bibfnamefont {J.-K.}\ \bibnamefont {Park}}, \bibinfo {author}
  {\bibfnamefont {N.}~\bibnamefont {Ramasubramanian}}, \bibinfo {author}
  {\bibfnamefont {H.}~\bibnamefont {Reimerdes}}, \bibinfo {author}
  {\bibfnamefont {G.}~\bibnamefont {Saibene}}, \bibinfo {author} {\bibfnamefont
  {A.}~\bibnamefont {Salmi}}, \bibinfo {author} {\bibfnamefont
  {K.}~\bibnamefont {Shinohara}}, \bibinfo {author} {\bibfnamefont
  {D.}~\bibnamefont {Spong}}, \bibinfo {author} {\bibfnamefont
  {W.}~\bibnamefont {Solomon}}, \bibinfo {author} {\bibfnamefont
  {T.}~\bibnamefont {Tala}}, \bibinfo {author} {\bibfnamefont {Y.}~\bibnamefont
  {Zhu}}, \bibinfo {author} {\bibfnamefont {J.}~\bibnamefont {Boedo}}, \bibinfo
  {author} {\bibfnamefont {V.}~\bibnamefont {Chuyanov}}, \bibinfo {author}
  {\bibfnamefont {E.}~\bibnamefont {Doyle}}, \bibinfo {author} {\bibfnamefont
  {M.}~\bibnamefont {Jakubowski}}, \bibinfo {author} {\bibfnamefont
  {H.}~\bibnamefont {Jhang}}, \bibinfo {author} {\bibfnamefont
  {R.}~\bibnamefont {Nazikian}}, \bibinfo {author} {\bibfnamefont
  {V.}~\bibnamefont {Pustovitov}}, \bibinfo {author} {\bibfnamefont
  {O.}~\bibnamefont {Schmitz}}, \bibinfo {author} {\bibfnamefont
  {R.}~\bibnamefont {Srinivasan}}, \bibinfo {author} {\bibfnamefont
  {T.}~\bibnamefont {Taylor}}, \bibinfo {author} {\bibfnamefont
  {M.}~\bibnamefont {Wade}}, \bibinfo {author} {\bibfnamefont {K.}~\bibnamefont
  {You}}, \ and\ \bibinfo {author} {\bibfnamefont {L.}~\bibnamefont {Zeng}},\
  }\href {\doibase 10.1088/0029-5515/51/10/103028} {\bibfield  {journal}
  {\bibinfo  {journal} {Nuclear Fusion}\ }\textbf {\bibinfo {volume} {51}},\
  \bibinfo {pages} {103028} (\bibinfo {year} {2011})}\BibitemShut {NoStop}%
\bibitem [{\citenamefont {Lanctot}\ \emph {et~al.}(2017)\citenamefont
  {Lanctot}, \citenamefont {Snipes}, \citenamefont {Reimerdes}, \citenamefont
  {Paz-Soldan}, \citenamefont {Logan}, \citenamefont {Hanson}, \citenamefont
  {Buttery}, \citenamefont {DeGrassie}, \citenamefont {Garofalo}, \citenamefont
  {Gray}, \citenamefont {Grierson}, \citenamefont {King}, \citenamefont
  {Kramer}, \citenamefont {{La Haye}}, \citenamefont {Pace}, \citenamefont
  {Park}, \citenamefont {Salmi}, \citenamefont {Shiraki}, \citenamefont
  {Strait}, \citenamefont {Solomon}, \citenamefont {Tala},\ and\ \citenamefont
  {{Van Zeeland}}}]{Lanctot2017}%
  \BibitemOpen
  \bibfield  {author} {\bibinfo {author} {\bibfnamefont {M.}~\bibnamefont
  {Lanctot}}, \bibinfo {author} {\bibfnamefont {J.}~\bibnamefont {Snipes}},
  \bibinfo {author} {\bibfnamefont {H.}~\bibnamefont {Reimerdes}}, \bibinfo
  {author} {\bibfnamefont {C.}~\bibnamefont {Paz-Soldan}}, \bibinfo {author}
  {\bibfnamefont {N.}~\bibnamefont {Logan}}, \bibinfo {author} {\bibfnamefont
  {J.}~\bibnamefont {Hanson}}, \bibinfo {author} {\bibfnamefont
  {R.}~\bibnamefont {Buttery}}, \bibinfo {author} {\bibfnamefont
  {J.}~\bibnamefont {DeGrassie}}, \bibinfo {author} {\bibfnamefont
  {A.}~\bibnamefont {Garofalo}}, \bibinfo {author} {\bibfnamefont
  {T.}~\bibnamefont {Gray}}, \bibinfo {author} {\bibfnamefont {B.}~\bibnamefont
  {Grierson}}, \bibinfo {author} {\bibfnamefont {J.}~\bibnamefont {King}},
  \bibinfo {author} {\bibfnamefont {G.}~\bibnamefont {Kramer}}, \bibinfo
  {author} {\bibfnamefont {R.}~\bibnamefont {{La Haye}}}, \bibinfo {author}
  {\bibfnamefont {D.}~\bibnamefont {Pace}}, \bibinfo {author} {\bibfnamefont
  {J.-K.}\ \bibnamefont {Park}}, \bibinfo {author} {\bibfnamefont
  {A.}~\bibnamefont {Salmi}}, \bibinfo {author} {\bibfnamefont
  {D.}~\bibnamefont {Shiraki}}, \bibinfo {author} {\bibfnamefont
  {E.}~\bibnamefont {Strait}}, \bibinfo {author} {\bibfnamefont
  {W.}~\bibnamefont {Solomon}}, \bibinfo {author} {\bibfnamefont
  {T.}~\bibnamefont {Tala}}, \ and\ \bibinfo {author} {\bibfnamefont
  {M.}~\bibnamefont {{Van Zeeland}}},\ }\href {\doibase
  10.1088/1741-4326/57/3/036004} {\bibfield  {journal} {\bibinfo  {journal}
  {Nuclear Fusion}\ }\textbf {\bibinfo {volume} {57}},\ \bibinfo {pages}
  {036004} (\bibinfo {year} {2017})}\BibitemShut {NoStop}%
\bibitem [{\citenamefont {Tobita}\ \emph {et~al.}(2003)\citenamefont {Tobita},
  \citenamefont {Nakayama}, \citenamefont {Konovalov},\ and\ \citenamefont
  {Sato}}]{Tobita2003}%
  \BibitemOpen
  \bibfield  {author} {\bibinfo {author} {\bibfnamefont {K.}~\bibnamefont
  {Tobita}}, \bibinfo {author} {\bibfnamefont {T.}~\bibnamefont {Nakayama}},
  \bibinfo {author} {\bibfnamefont {S.~V.}\ \bibnamefont {Konovalov}}, \ and\
  \bibinfo {author} {\bibfnamefont {M.}~\bibnamefont {Sato}},\ }\href {\doibase
  10.1088/0741-3335/45/2/305} {\bibfield  {journal} {\bibinfo  {journal}
  {Plasma Physics and Controlled Fusion}\ }\textbf {\bibinfo {volume} {45}},\
  \bibinfo {pages} {133} (\bibinfo {year} {2003})}\BibitemShut {NoStop}%
\bibitem [{\citenamefont {Urano}\ \emph {et~al.}(2007)\citenamefont {Urano},
  \citenamefont {Oyama}, \citenamefont {Kamiya}, \citenamefont {Koide},
  \citenamefont {Takenaga}, \citenamefont {Takizuka}, \citenamefont {Yoshida},
  \citenamefont {Kamada},\ and\ \citenamefont {the JT-60~Team}}]{Urano2007}%
  \BibitemOpen
  \bibfield  {author} {\bibinfo {author} {\bibfnamefont {H.}~\bibnamefont
  {Urano}}, \bibinfo {author} {\bibfnamefont {N.}~\bibnamefont {Oyama}},
  \bibinfo {author} {\bibfnamefont {K.}~\bibnamefont {Kamiya}}, \bibinfo
  {author} {\bibfnamefont {Y.}~\bibnamefont {Koide}}, \bibinfo {author}
  {\bibfnamefont {H.}~\bibnamefont {Takenaga}}, \bibinfo {author}
  {\bibfnamefont {T.}~\bibnamefont {Takizuka}}, \bibinfo {author}
  {\bibfnamefont {M.}~\bibnamefont {Yoshida}}, \bibinfo {author} {\bibfnamefont
  {Y.}~\bibnamefont {Kamada}}, \ and\ \bibinfo {author} {\bibnamefont {the
  JT-60~Team}},\ }\href {\doibase 10.1088/0029-5515/47/7/022} {\bibfield
  {journal} {\bibinfo  {journal} {Nuclear Fusion}\ }\textbf {\bibinfo {volume}
  {47}},\ \bibinfo {pages} {706} (\bibinfo {year} {2007})}\BibitemShut
  {NoStop}%
\bibitem [{\citenamefont {Kawashima}\ \emph {et~al.}(2001)\citenamefont
  {Kawashima}, \citenamefont {Sato}, \citenamefont {Tsuzuki}, \citenamefont
  {Miura}, \citenamefont {Isei}, \citenamefont {Kimura}, \citenamefont
  {Nakayama}, \citenamefont {Abe}, \citenamefont {Darrow},\ and\ \citenamefont
  {the JFT-2M~Group}}]{Kawashima2001}%
  \BibitemOpen
  \bibfield  {author} {\bibinfo {author} {\bibfnamefont {H.}~\bibnamefont
  {Kawashima}}, \bibinfo {author} {\bibfnamefont {M.}~\bibnamefont {Sato}},
  \bibinfo {author} {\bibfnamefont {K.}~\bibnamefont {Tsuzuki}}, \bibinfo
  {author} {\bibfnamefont {Y.}~\bibnamefont {Miura}}, \bibinfo {author}
  {\bibfnamefont {N.}~\bibnamefont {Isei}}, \bibinfo {author} {\bibfnamefont
  {H.}~\bibnamefont {Kimura}}, \bibinfo {author} {\bibfnamefont
  {T.}~\bibnamefont {Nakayama}}, \bibinfo {author} {\bibfnamefont
  {M.}~\bibnamefont {Abe}}, \bibinfo {author} {\bibfnamefont {D.}~\bibnamefont
  {Darrow}}, \ and\ \bibinfo {author} {\bibnamefont {the JFT-2M~Group}},\
  }\href {https://doi.org/10.1088/0029-5515/41/3/302} {\bibfield  {journal}
  {\bibinfo  {journal} {Nuclear Fusion}\ }\textbf {\bibinfo {volume} {41}},\
  \bibinfo {pages} {257} (\bibinfo {year} {2001})}\BibitemShut {NoStop}%
\bibitem [{\citenamefont {Stringer}(1972)}]{Stringer1972}%
  \BibitemOpen
  \bibfield  {author} {\bibinfo {author} {\bibfnamefont {T.~E.}\ \bibnamefont
  {Stringer}},\ }\href {\doibase 10.1088/0029-5515/12/6/010} {\bibfield
  {journal} {\bibinfo  {journal} {Nuclear Fusion}\ }\textbf {\bibinfo {volume}
  {12}},\ \bibinfo {pages} {689} (\bibinfo {year} {1972})}\BibitemShut
  {NoStop}%
\bibitem [{\citenamefont {Shaing}(2003)}]{Shaing2003}%
  \BibitemOpen
  \bibfield  {author} {\bibinfo {author} {\bibfnamefont {K.~C.}\ \bibnamefont
  {Shaing}},\ }\href {\doibase 10.1063/1.1567285} {\bibfield  {journal}
  {\bibinfo  {journal} {Physics of Plasmas}\ }\textbf {\bibinfo {volume}
  {10}},\ \bibinfo {pages} {1443} (\bibinfo {year} {2003})}\BibitemShut
  {NoStop}%
\bibitem [{\citenamefont {Shaing}\ \emph
  {et~al.}(2009{\natexlab{a}})\citenamefont {Shaing}, \citenamefont {Sabbagh},\
  and\ \citenamefont {Chu}}]{Shaing2009}%
  \BibitemOpen
  \bibfield  {author} {\bibinfo {author} {\bibfnamefont {K.~C.}\ \bibnamefont
  {Shaing}}, \bibinfo {author} {\bibfnamefont {S.~A.}\ \bibnamefont {Sabbagh}},
  \ and\ \bibinfo {author} {\bibfnamefont {M.~S.}\ \bibnamefont {Chu}},\ }\href
  {\doibase 10.1088/0741-3335/51/5/055003} {\bibfield  {journal} {\bibinfo
  {journal} {Plasma Physics and Controlled Fusion}\ }\textbf {\bibinfo {volume}
  {51}},\ \bibinfo {pages} {035004} (\bibinfo {year}
  {2009}{\natexlab{a}})}\BibitemShut {NoStop}%
\bibitem [{\citenamefont {Shaing}\ and\ \citenamefont
  {Callen}(1982{\natexlab{a}})}]{Shaing1982a}%
  \BibitemOpen
  \bibfield  {author} {\bibinfo {author} {\bibfnamefont {K.~C.}\ \bibnamefont
  {Shaing}}\ and\ \bibinfo {author} {\bibfnamefont {J.~D.}\ \bibnamefont
  {Callen}},\ }\href
  {http://iopscience.iop.org/article/10.1088/0029-5515/22/8/005/meta}
  {\bibfield  {journal} {\bibinfo  {journal} {Nuclear Fusion}\ }\textbf
  {\bibinfo {volume} {22}},\ \bibinfo {pages} {1061} (\bibinfo {year}
  {1982}{\natexlab{a}})}\BibitemShut {NoStop}%
\bibitem [{\citenamefont {Shaing}\ and\ \citenamefont
  {Callen}(1982{\natexlab{b}})}]{Shaing1982b}%
  \BibitemOpen
  \bibfield  {author} {\bibinfo {author} {\bibfnamefont {K.~C.}\ \bibnamefont
  {Shaing}}\ and\ \bibinfo {author} {\bibfnamefont {J.~D.}\ \bibnamefont
  {Callen}},\ }\href {\doibase 10.1063/1.863857} {\bibfield  {journal}
  {\bibinfo  {journal} {Physics of Fluids}\ }\textbf {\bibinfo {volume} {25}},\
  \bibinfo {pages} {1012} (\bibinfo {year} {1982}{\natexlab{b}})}\BibitemShut
  {NoStop}%
\bibitem [{\citenamefont {Shaing}\ \emph
  {et~al.}(2009{\natexlab{b}})\citenamefont {Shaing}, \citenamefont {Sabbagh},\
  and\ \citenamefont {Chu}}]{Shaing2009_sbp}%
  \BibitemOpen
  \bibfield  {author} {\bibinfo {author} {\bibfnamefont {K.~C.}\ \bibnamefont
  {Shaing}}, \bibinfo {author} {\bibfnamefont {S.~A.}\ \bibnamefont {Sabbagh}},
  \ and\ \bibinfo {author} {\bibfnamefont {M.~S.}\ \bibnamefont {Chu}},\ }\href
  {\doibase 10.1088/0741-3335/51/3/035009} {\bibfield  {journal} {\bibinfo
  {journal} {Plasma Physics and Controlled Fusion}\ }\textbf {\bibinfo {volume}
  {51}},\ \bibinfo {pages} {035009} (\bibinfo {year}
  {2009}{\natexlab{b}})}\BibitemShut {NoStop}%
\bibitem [{\citenamefont {Linsker}\ and\ \citenamefont
  {Boozer}(1982)}]{Linsker1982}%
  \BibitemOpen
  \bibfield  {author} {\bibinfo {author} {\bibfnamefont {R.}~\bibnamefont
  {Linsker}}\ and\ \bibinfo {author} {\bibfnamefont {A.~H.}\ \bibnamefont
  {Boozer}},\ }\href {\doibase 10.1063/1.863635} {\bibfield  {journal}
  {\bibinfo  {journal} {Phys. Fluids}\ }\textbf {\bibinfo {volume} {25}},\
  \bibinfo {pages} {143} (\bibinfo {year} {1982})}\BibitemShut {NoStop}%
\bibitem [{\citenamefont {Park}\ \emph {et~al.}(2009)\citenamefont {Park},
  \citenamefont {Boozer}, \citenamefont {Menard}, \citenamefont {Garofalo},
  \citenamefont {Schaffer}, \citenamefont {Hawryluk}, \citenamefont {Kaye},
  \citenamefont {Gerhardt},\ and\ \citenamefont {Sabbagh}}]{Park2009}%
  \BibitemOpen
  \bibfield  {author} {\bibinfo {author} {\bibfnamefont {J.-K.}\ \bibnamefont
  {Park}}, \bibinfo {author} {\bibfnamefont {A.~H.}\ \bibnamefont {Boozer}},
  \bibinfo {author} {\bibfnamefont {J.~E.}\ \bibnamefont {Menard}}, \bibinfo
  {author} {\bibfnamefont {A.~M.}\ \bibnamefont {Garofalo}}, \bibinfo {author}
  {\bibfnamefont {M.~J.}\ \bibnamefont {Schaffer}}, \bibinfo {author}
  {\bibfnamefont {R.~J.}\ \bibnamefont {Hawryluk}}, \bibinfo {author}
  {\bibfnamefont {S.~M.}\ \bibnamefont {Kaye}}, \bibinfo {author}
  {\bibfnamefont {S.~P.}\ \bibnamefont {Gerhardt}}, \ and\ \bibinfo {author}
  {\bibfnamefont {S.~A.}\ \bibnamefont {Sabbagh}},\ }\href
  {http://aip.scitation.org/doi/abs/10.1063/1.3122862} {\bibfield  {journal}
  {\bibinfo  {journal} {Physics of Plasmas}\ }\textbf {\bibinfo {volume}
  {16}},\ \bibinfo {pages} {056115} (\bibinfo {year} {2009})}\BibitemShut
  {NoStop}%
\bibitem [{\citenamefont {Galeev}\ and\ \citenamefont
  {Sagdeev}(1969)}]{Galeev1969}%
  \BibitemOpen
  \bibfield  {author} {\bibinfo {author} {\bibfnamefont {A.}~\bibnamefont
  {Galeev}}\ and\ \bibinfo {author} {\bibfnamefont {R.}~\bibnamefont
  {Sagdeev}},\ }\href {https://doi.org/10.1103/PhysRevLett.22.511} {\bibfield
  {journal} {\bibinfo  {journal} {Physical Review Letters}\ }\textbf {\bibinfo
  {volume} {22}},\ \bibinfo {pages} {511} (\bibinfo {year} {1969})}\BibitemShut
  {NoStop}%
\bibitem [{\citenamefont {Ho}\ and\ \citenamefont {Kulsrud}(1987)}]{Ho1987}%
  \BibitemOpen
  \bibfield  {author} {\bibinfo {author} {\bibfnamefont {D.~D.}\ \bibnamefont
  {Ho}}\ and\ \bibinfo {author} {\bibfnamefont {R.~M.}\ \bibnamefont
  {Kulsrud}},\ }\href {\doibase 10.1063/1.866395} {\bibfield  {journal}
  {\bibinfo  {journal} {Physics of Fluids}\ }\textbf {\bibinfo {volume} {30}},\
  \bibinfo {pages} {442} (\bibinfo {year} {1987})}\BibitemShut {NoStop}%
\bibitem [{\citenamefont {Frieman}(1970)}]{Frieman1970}%
  \BibitemOpen
  \bibfield  {author} {\bibinfo {author} {\bibfnamefont {E.~A.}\ \bibnamefont
  {Frieman}},\ }\href {\doibase 10.1063/1.1692944} {\bibfield  {journal}
  {\bibinfo  {journal} {Physics of Fluids}\ }\textbf {\bibinfo {volume} {13}},\
  \bibinfo {pages} {490} (\bibinfo {year} {1970})}\BibitemShut {NoStop}%
\bibitem [{\citenamefont {Connor}\ and\ \citenamefont
  {Hastie}(1973)}]{Connor1973}%
  \BibitemOpen
  \bibfield  {author} {\bibinfo {author} {\bibfnamefont {J.}~\bibnamefont
  {Connor}}\ and\ \bibinfo {author} {\bibfnamefont {R.}~\bibnamefont
  {Hastie}},\ }\href
  {http://iopscience.iop.org/article/10.1088/0029-5515/13/2/011/meta}
  {\bibfield  {journal} {\bibinfo  {journal} {Nuclear Fusion}\ }\textbf
  {\bibinfo {volume} {13}},\ \bibinfo {pages} {221} (\bibinfo {year}
  {1973})}\BibitemShut {NoStop}%
\bibitem [{\citenamefont {Yushmanov}(1982)}]{Yushmanov1982}%
  \BibitemOpen
  \bibfield  {author} {\bibinfo {author} {\bibfnamefont {P.}~\bibnamefont
  {Yushmanov}},\ }\href {\doibase 10.1088/0029-5515/22/3/001} {\bibfield
  {journal} {\bibinfo  {journal} {Nuclear Fusion}\ }\textbf {\bibinfo {volume}
  {22}},\ \bibinfo {pages} {315} (\bibinfo {year} {1982})}\BibitemShut
  {NoStop}%
\bibitem [{\citenamefont {Kadomtsev}\ and\ \citenamefont
  {Pogutse}(1971)}]{Kadomtsev1971}%
  \BibitemOpen
  \bibfield  {author} {\bibinfo {author} {\bibfnamefont {B.}~\bibnamefont
  {Kadomtsev}}\ and\ \bibinfo {author} {\bibfnamefont {O.}~\bibnamefont
  {Pogutse}},\ }\href {\doibase 10.1088/0029-5515/11/1/010} {\bibfield
  {journal} {\bibinfo  {journal} {Nuclear Fusion}\ }\textbf {\bibinfo {volume}
  {11}},\ \bibinfo {pages} {67} (\bibinfo {year} {1971})}\BibitemShut {NoStop}%
\bibitem [{\citenamefont {Davidson}(1976)}]{Davidson1976}%
  \BibitemOpen
  \bibfield  {author} {\bibinfo {author} {\bibfnamefont {J.~N.}\ \bibnamefont
  {Davidson}},\ }\href {\doibase 10.1088/0029-5515/16/5/001} {\bibfield
  {journal} {\bibinfo  {journal} {Nuclear Fusion}\ }\textbf {\bibinfo {volume}
  {16}},\ \bibinfo {pages} {731} (\bibinfo {year} {1976})}\BibitemShut
  {NoStop}%
\bibitem [{\citenamefont {Tsang}(1977)}]{Tsang1977}%
  \BibitemOpen
  \bibfield  {author} {\bibinfo {author} {\bibfnamefont {K.~T.}\ \bibnamefont
  {Tsang}},\ }\href {\doibase 10.1088/0029-5515/17/3/014} {\bibfield  {journal}
  {\bibinfo  {journal} {Nuclear Fusion}\ }\textbf {\bibinfo {volume} {17}},\
  \bibinfo {pages} {557} (\bibinfo {year} {1977})}\BibitemShut {NoStop}%
\bibitem [{\citenamefont {Shaing}\ and\ \citenamefont
  {Callen}(1983)}]{Shaing1983}%
  \BibitemOpen
  \bibfield  {author} {\bibinfo {author} {\bibfnamefont {K.~C.}\ \bibnamefont
  {Shaing}}\ and\ \bibinfo {author} {\bibfnamefont {J.~D.}\ \bibnamefont
  {Callen}},\ }\href {\doibase 10.1063/1.864108} {\bibfield  {journal}
  {\bibinfo  {journal} {Physics of Fluids}\ }\textbf {\bibinfo {volume} {26}},\
  \bibinfo {pages} {3315} (\bibinfo {year} {1983})}\BibitemShut {NoStop}%
\bibitem [{\citenamefont {Shaing}\ \emph {et~al.}(2008)\citenamefont {Shaing},
  \citenamefont {Cahyna}, \citenamefont {Becoulet}, \citenamefont {Park},
  \citenamefont {Sabbagh},\ and\ \citenamefont {Chu}}]{Shaing2008}%
  \BibitemOpen
  \bibfield  {author} {\bibinfo {author} {\bibfnamefont {K.~C.}\ \bibnamefont
  {Shaing}}, \bibinfo {author} {\bibfnamefont {P.}~\bibnamefont {Cahyna}},
  \bibinfo {author} {\bibfnamefont {M.}~\bibnamefont {Becoulet}}, \bibinfo
  {author} {\bibfnamefont {J.-K.}\ \bibnamefont {Park}}, \bibinfo {author}
  {\bibfnamefont {S.~A.}\ \bibnamefont {Sabbagh}}, \ and\ \bibinfo {author}
  {\bibfnamefont {M.~S.}\ \bibnamefont {Chu}},\ }\href
  {http://aip.scitation.org/doi/abs/10.1063/1.2969434} {\bibfield  {journal}
  {\bibinfo  {journal} {Physics of Plasmas}\ }\textbf {\bibinfo {volume}
  {15}},\ \bibinfo {pages} {082506} (\bibinfo {year} {2008})}\BibitemShut
  {NoStop}%
\bibitem [{\citenamefont {Shaing}\ \emph {et~al.}(2010)\citenamefont {Shaing},
  \citenamefont {Sabbagh},\ and\ \citenamefont {Chu}}]{Shaing2010}%
  \BibitemOpen
  \bibfield  {author} {\bibinfo {author} {\bibfnamefont {K.}~\bibnamefont
  {Shaing}}, \bibinfo {author} {\bibfnamefont {S.}~\bibnamefont {Sabbagh}}, \
  and\ \bibinfo {author} {\bibfnamefont {M.}~\bibnamefont {Chu}},\ }\href
  {\doibase 10.1088/0029-5515/50/2/025022} {\bibfield  {journal} {\bibinfo
  {journal} {Nuclear Fusion}\ }\textbf {\bibinfo {volume} {50}},\ \bibinfo
  {pages} {025022} (\bibinfo {year} {2010})}\BibitemShut {NoStop}%
\bibitem [{\citenamefont {Jardin}\ \emph {et~al.}(2008)\citenamefont {Jardin},
  \citenamefont {Ferraro}, \citenamefont {Luo}, \citenamefont {Chen},
  \citenamefont {Breslau}, \citenamefont {Jansen},\ and\ \citenamefont
  {Shephard}}]{Jardin2008}%
  \BibitemOpen
  \bibfield  {author} {\bibinfo {author} {\bibfnamefont {S.~C.}\ \bibnamefont
  {Jardin}}, \bibinfo {author} {\bibfnamefont {N.}~\bibnamefont {Ferraro}},
  \bibinfo {author} {\bibfnamefont {X.}~\bibnamefont {Luo}}, \bibinfo {author}
  {\bibfnamefont {J.}~\bibnamefont {Chen}}, \bibinfo {author} {\bibfnamefont
  {J.}~\bibnamefont {Breslau}}, \bibinfo {author} {\bibfnamefont {K.~E.}\
  \bibnamefont {Jansen}}, \ and\ \bibinfo {author} {\bibfnamefont {M.~S.}\
  \bibnamefont {Shephard}},\ }\href {\doibase 10.1088/1742-6596/125/1/012044}
  {\bibfield  {journal} {\bibinfo  {journal} {Journal of Physics: Conference
  Series}\ }\textbf {\bibinfo {volume} {125}},\ \bibinfo {pages} {012044}
  (\bibinfo {year} {2008})}\BibitemShut {NoStop}%
\bibitem [{\citenamefont {Hirshman}\ \emph
  {et~al.}(1986{\natexlab{a}})\citenamefont {Hirshman}, \citenamefont {Shaing},
  \citenamefont {van Rij}, \citenamefont {Beasley},\ and\ \citenamefont
  {Crume}}]{Hirshman1986a}%
  \BibitemOpen
  \bibfield  {author} {\bibinfo {author} {\bibfnamefont {S.~P.}\ \bibnamefont
  {Hirshman}}, \bibinfo {author} {\bibfnamefont {K.~C.}\ \bibnamefont
  {Shaing}}, \bibinfo {author} {\bibfnamefont {W.~I.}\ \bibnamefont {van Rij}},
  \bibinfo {author} {\bibfnamefont {C.~O.}\ \bibnamefont {Beasley}}, \ and\
  \bibinfo {author} {\bibfnamefont {E.~C.}\ \bibnamefont {Crume}},\ }\href
  {\doibase 10.1063/1.865495} {\bibfield  {journal} {\bibinfo  {journal}
  {Physics of Fluids}\ }\textbf {\bibinfo {volume} {29}},\ \bibinfo {pages}
  {2951} (\bibinfo {year} {1986}{\natexlab{a}})}\BibitemShut {NoStop}%
\bibitem [{\citenamefont {Ferraro}\ \emph {et~al.}(2010)\citenamefont
  {Ferraro}, \citenamefont {Jardin},\ and\ \citenamefont
  {Snyder}}]{Ferraro2010}%
  \BibitemOpen
  \bibfield  {author} {\bibinfo {author} {\bibfnamefont {N.~M.}\ \bibnamefont
  {Ferraro}}, \bibinfo {author} {\bibfnamefont {S.~C.}\ \bibnamefont {Jardin}},
  \ and\ \bibinfo {author} {\bibfnamefont {P.~B.}\ \bibnamefont {Snyder}},\
  }\href {http://aip.scitation.org/doi/abs/10.1063/1.3492727} {\bibfield
  {journal} {\bibinfo  {journal} {Physics of Plasmas}\ }\textbf {\bibinfo
  {volume} {17}} (\bibinfo {year} {2010})}\BibitemShut {NoStop}%
\bibitem [{\citenamefont {Cole}\ \emph {et~al.}(2011)\citenamefont {Cole},
  \citenamefont {Callen}, \citenamefont {Solomon}, \citenamefont {Garofalo},
  \citenamefont {Hegna}, \citenamefont {Lanctot},\ and\ \citenamefont
  {Reimerdes}}]{Cole2011}%
  \BibitemOpen
  \bibfield  {author} {\bibinfo {author} {\bibfnamefont {A.~J.}\ \bibnamefont
  {Cole}}, \bibinfo {author} {\bibfnamefont {J.~D.}\ \bibnamefont {Callen}},
  \bibinfo {author} {\bibfnamefont {W.~M.}\ \bibnamefont {Solomon}}, \bibinfo
  {author} {\bibfnamefont {A.~M.}\ \bibnamefont {Garofalo}}, \bibinfo {author}
  {\bibfnamefont {C.~C.}\ \bibnamefont {Hegna}}, \bibinfo {author}
  {\bibfnamefont {M.~J.}\ \bibnamefont {Lanctot}}, \ and\ \bibinfo {author}
  {\bibfnamefont {H.}~\bibnamefont {Reimerdes}},\ }\href
  {http://dx.doi.org/10.1063/1.3590933} {\bibfield  {journal} {\bibinfo
  {journal} {Physics of Plasmas}\ }\textbf {\bibinfo {volume} {18}} (\bibinfo
  {year} {2011})}\BibitemShut {NoStop}%
\bibitem [{\citenamefont {Sun}\ \emph {et~al.}(2010)\citenamefont {Sun},
  \citenamefont {Liang}, \citenamefont {Shaing}, \citenamefont {Koslowski},
  \citenamefont {Wiegmann},\ and\ \citenamefont {Zhang}}]{Sun2010}%
  \BibitemOpen
  \bibfield  {author} {\bibinfo {author} {\bibfnamefont {Y.}~\bibnamefont
  {Sun}}, \bibinfo {author} {\bibfnamefont {Y.}~\bibnamefont {Liang}}, \bibinfo
  {author} {\bibfnamefont {K.~C.}\ \bibnamefont {Shaing}}, \bibinfo {author}
  {\bibfnamefont {H.~R.}\ \bibnamefont {Koslowski}}, \bibinfo {author}
  {\bibfnamefont {C.}~\bibnamefont {Wiegmann}}, \ and\ \bibinfo {author}
  {\bibfnamefont {T.}~\bibnamefont {Zhang}},\ }\href
  {https://doi.org/10.1103/PhysRevLett.105.145002} {\bibfield  {journal}
  {\bibinfo  {journal} {Physical Review Letters}\ }\textbf {\bibinfo {volume}
  {105}},\ \bibinfo {pages} {145002} (\bibinfo {year} {2010})}\BibitemShut
  {NoStop}%
\bibitem [{\citenamefont {Satake}\ \emph
  {et~al.}(2011{\natexlab{a}})\citenamefont {Satake}, \citenamefont {Sugama},
  \citenamefont {Kanno},\ and\ \citenamefont {Park}}]{Satake2011a}%
  \BibitemOpen
  \bibfield  {author} {\bibinfo {author} {\bibfnamefont {S.}~\bibnamefont
  {Satake}}, \bibinfo {author} {\bibfnamefont {H.}~\bibnamefont {Sugama}},
  \bibinfo {author} {\bibfnamefont {R.}~\bibnamefont {Kanno}}, \ and\ \bibinfo
  {author} {\bibfnamefont {J.-K.}\ \bibnamefont {Park}},\ }\href {\doibase
  10.1088/0741-3335/53/5/054018} {\bibfield  {journal} {\bibinfo  {journal}
  {Plasma Physics and Controlled Fusion}\ }\textbf {\bibinfo {volume} {53}},\
  \bibinfo {pages} {054018} (\bibinfo {year} {2011}{\natexlab{a}})}\BibitemShut
  {NoStop}%
\bibitem [{\citenamefont {Satake}\ \emph
  {et~al.}(2011{\natexlab{b}})\citenamefont {Satake}, \citenamefont {Park},
  \citenamefont {Sugama},\ and\ \citenamefont {Kanno}}]{Satake2011b}%
  \BibitemOpen
  \bibfield  {author} {\bibinfo {author} {\bibfnamefont {S.}~\bibnamefont
  {Satake}}, \bibinfo {author} {\bibfnamefont {J.-K.}\ \bibnamefont {Park}},
  \bibinfo {author} {\bibfnamefont {H.}~\bibnamefont {Sugama}}, \ and\ \bibinfo
  {author} {\bibfnamefont {R.}~\bibnamefont {Kanno}},\ }\href {\doibase
  10.1103/PhysRevLett.107.055001} {\bibfield  {journal} {\bibinfo  {journal}
  {Physical Review Letters}\ }\textbf {\bibinfo {volume} {107}},\ \bibinfo
  {pages} {1} (\bibinfo {year} {2011}{\natexlab{b}})}\BibitemShut {NoStop}%
\bibitem [{\citenamefont {Martitsch}\ \emph {et~al.}(2016)\citenamefont
  {Martitsch}, \citenamefont {Kasilov}, \citenamefont {Kernbichler},
  \citenamefont {Kapper}, \citenamefont {Albert}, \citenamefont {Heyn},
  \citenamefont {Smith}, \citenamefont {Strumberger}, \citenamefont {Fietz},
  \citenamefont {Suttrop},\ and\ \citenamefont {Landreman}}]{Martitsch2016}%
  \BibitemOpen
  \bibfield  {author} {\bibinfo {author} {\bibfnamefont {A.~F.}\ \bibnamefont
  {Martitsch}}, \bibinfo {author} {\bibfnamefont {S.~V.}\ \bibnamefont
  {Kasilov}}, \bibinfo {author} {\bibfnamefont {W.}~\bibnamefont
  {Kernbichler}}, \bibinfo {author} {\bibfnamefont {G.}~\bibnamefont {Kapper}},
  \bibinfo {author} {\bibfnamefont {C.~G.}\ \bibnamefont {Albert}}, \bibinfo
  {author} {\bibfnamefont {M.~F.}\ \bibnamefont {Heyn}}, \bibinfo {author}
  {\bibfnamefont {H.~M.}\ \bibnamefont {Smith}}, \bibinfo {author}
  {\bibfnamefont {E.}~\bibnamefont {Strumberger}}, \bibinfo {author}
  {\bibfnamefont {S.}~\bibnamefont {Fietz}}, \bibinfo {author} {\bibfnamefont
  {W.}~\bibnamefont {Suttrop}}, \ and\ \bibinfo {author} {\bibfnamefont
  {M.}~\bibnamefont {Landreman}},\ }\href {\doibase
  10.1088/0741-3335/58/7/074007} {\bibfield  {journal} {\bibinfo  {journal}
  {Plasma Physics and Controlled Fusion}\ }\textbf {\bibinfo {volume} {58}},\
  \bibinfo {pages} {074007} (\bibinfo {year} {2016})}\BibitemShut {NoStop}%
\bibitem [{\citenamefont {Landreman}\ \emph {et~al.}(2014)\citenamefont
  {Landreman}, \citenamefont {Smith}, \citenamefont {Moll{\'{e}}n},\ and\
  \citenamefont {Helander}}]{Landreman2014}%
  \BibitemOpen
  \bibfield  {author} {\bibinfo {author} {\bibfnamefont {M.}~\bibnamefont
  {Landreman}}, \bibinfo {author} {\bibfnamefont {H.~M.}\ \bibnamefont
  {Smith}}, \bibinfo {author} {\bibfnamefont {A.}~\bibnamefont {Moll{\'{e}}n}},
  \ and\ \bibinfo {author} {\bibfnamefont {P.}~\bibnamefont {Helander}},\
  }\href {http://aip.scitation.org/doi/abs/10.1063/1.4870077} {\bibfield
  {journal} {\bibinfo  {journal} {Physics of Plasmas}\ }\textbf {\bibinfo
  {volume} {21}} (\bibinfo {year} {2014})}\BibitemShut {NoStop}%
\bibitem [{\citenamefont {Hirshman}\ \emph
  {et~al.}(1986{\natexlab{b}})\citenamefont {Hirshman}, \citenamefont {van
  Rij},\ and\ \citenamefont {Merkel}}]{Hirshman1986b}%
  \BibitemOpen
  \bibfield  {author} {\bibinfo {author} {\bibfnamefont {S.}~\bibnamefont
  {Hirshman}}, \bibinfo {author} {\bibfnamefont {W.}~\bibnamefont {van Rij}}, \
  and\ \bibinfo {author} {\bibfnamefont {P.}~\bibnamefont {Merkel}},\ }\href
  {\doibase 10.1016/0010-4655(86)90058-5} {\bibfield  {journal} {\bibinfo
  {journal} {Computer Physics Communications}\ }\textbf {\bibinfo {volume}
  {43}},\ \bibinfo {pages} {143} (\bibinfo {year}
  {1986}{\natexlab{b}})}\BibitemShut {NoStop}%
\bibitem [{\citenamefont {van Rij}\ and\ \citenamefont
  {Hirshman}(1989)}]{Rij1989}%
  \BibitemOpen
  \bibfield  {author} {\bibinfo {author} {\bibfnamefont {W.~I.}\ \bibnamefont
  {van Rij}}\ and\ \bibinfo {author} {\bibfnamefont {S.~P.}\ \bibnamefont
  {Hirshman}},\ }\href {\doibase 10.1063/1.859116} {\bibfield  {journal}
  {\bibinfo  {journal} {Physics of Fluids B: Plasma Physics}\ }\textbf
  {\bibinfo {volume} {1}},\ \bibinfo {pages} {563} (\bibinfo {year}
  {1989})}\BibitemShut {NoStop}%
\bibitem [{\citenamefont {{\'{C}}iri{\'{c}}}\ \emph {et~al.}(2011)\citenamefont
  {{\'{C}}iri{\'{c}}}, \citenamefont {Ash}, \citenamefont {Crowley},
  \citenamefont {Day}, \citenamefont {Gee}, \citenamefont {Hackett},
  \citenamefont {Homfray}, \citenamefont {Jenkins}, \citenamefont {Jones},
  \citenamefont {Keeling}, \citenamefont {King}, \citenamefont {King},
  \citenamefont {Kovari}, \citenamefont {McAdams}, \citenamefont {Surrey},
  \citenamefont {Young},\ and\ \citenamefont {Zacks}}]{Ciric2011}%
  \BibitemOpen
  \bibfield  {author} {\bibinfo {author} {\bibfnamefont {D.}~\bibnamefont
  {{\'{C}}iri{\'{c}}}}, \bibinfo {author} {\bibfnamefont {A.~D.}\ \bibnamefont
  {Ash}}, \bibinfo {author} {\bibfnamefont {B.}~\bibnamefont {Crowley}},
  \bibinfo {author} {\bibfnamefont {I.~E.}\ \bibnamefont {Day}}, \bibinfo
  {author} {\bibfnamefont {S.~J.}\ \bibnamefont {Gee}}, \bibinfo {author}
  {\bibfnamefont {L.~J.}\ \bibnamefont {Hackett}}, \bibinfo {author}
  {\bibfnamefont {D.~A.}\ \bibnamefont {Homfray}}, \bibinfo {author}
  {\bibfnamefont {I.}~\bibnamefont {Jenkins}}, \bibinfo {author} {\bibfnamefont
  {T.~T.~C.}\ \bibnamefont {Jones}}, \bibinfo {author} {\bibfnamefont
  {D.}~\bibnamefont {Keeling}}, \bibinfo {author} {\bibfnamefont {D.~B.}\
  \bibnamefont {King}}, \bibinfo {author} {\bibfnamefont {R.~F.}\ \bibnamefont
  {King}}, \bibinfo {author} {\bibfnamefont {M.}~\bibnamefont {Kovari}},
  \bibinfo {author} {\bibfnamefont {R.}~\bibnamefont {McAdams}}, \bibinfo
  {author} {\bibfnamefont {E.}~\bibnamefont {Surrey}}, \bibinfo {author}
  {\bibfnamefont {D.}~\bibnamefont {Young}}, \ and\ \bibinfo {author}
  {\bibfnamefont {J.}~\bibnamefont {Zacks}},\ }\href {\doibase
  10.1016/j.fusengdes.2010.11.035} {\bibfield  {journal} {\bibinfo  {journal}
  {Fusion Engineering and Design}\ }\textbf {\bibinfo {volume} {86}},\ \bibinfo
  {pages} {509} (\bibinfo {year} {2011})}\BibitemShut {NoStop}%
\bibitem [{\citenamefont {Parra}\ \emph {et~al.}(2012)\citenamefont {Parra},
  \citenamefont {Nave}, \citenamefont {Schekochihin}, \citenamefont {Giroud},
  \citenamefont {{De Grassie}}, \citenamefont {Severo}, \citenamefont {{De
  Vries}},\ and\ \citenamefont {Zastrow}}]{Parra2012}%
  \BibitemOpen
  \bibfield  {author} {\bibinfo {author} {\bibfnamefont {F.~I.}\ \bibnamefont
  {Parra}}, \bibinfo {author} {\bibfnamefont {M.~F.~F.}\ \bibnamefont {Nave}},
  \bibinfo {author} {\bibfnamefont {A.~A.}\ \bibnamefont {Schekochihin}},
  \bibinfo {author} {\bibfnamefont {C.}~\bibnamefont {Giroud}}, \bibinfo
  {author} {\bibfnamefont {J.~S.}\ \bibnamefont {{De Grassie}}}, \bibinfo
  {author} {\bibfnamefont {J.~H.~F.}\ \bibnamefont {Severo}}, \bibinfo {author}
  {\bibfnamefont {P.}~\bibnamefont {{De Vries}}}, \ and\ \bibinfo {author}
  {\bibfnamefont {K.~D.}\ \bibnamefont {Zastrow}},\ }\href {\doibase
  10.1103/PhysRevLett.108.095001} {\bibfield  {journal} {\bibinfo  {journal}
  {Physical Review Letters}\ }\textbf {\bibinfo {volume} {108}},\ \bibinfo
  {pages} {1} (\bibinfo {year} {2012})}\BibitemShut {NoStop}%
\bibitem [{\citenamefont {Rice}\ \emph {et~al.}(2007)\citenamefont {Rice},
  \citenamefont {Ince-Cushman}, \citenamefont {DeGrassie}, \citenamefont
  {Eriksson}, \citenamefont {Sakamoto}, \citenamefont {Scarabosio},
  \citenamefont {Bortolon}, \citenamefont {Burrell}, \citenamefont {Duval},
  \citenamefont {Fenzi-Bonizec}, \citenamefont {Greenwald}, \citenamefont
  {Groebner}, \citenamefont {Hoang}, \citenamefont {Koide}, \citenamefont
  {Marmar}, \citenamefont {Pochelon},\ and\ \citenamefont
  {Podpaly}}]{Rice2007}%
  \BibitemOpen
  \bibfield  {author} {\bibinfo {author} {\bibfnamefont {J.}~\bibnamefont
  {Rice}}, \bibinfo {author} {\bibfnamefont {A.}~\bibnamefont {Ince-Cushman}},
  \bibinfo {author} {\bibfnamefont {J.}~\bibnamefont {DeGrassie}}, \bibinfo
  {author} {\bibfnamefont {L.}~\bibnamefont {Eriksson}}, \bibinfo {author}
  {\bibfnamefont {Y.}~\bibnamefont {Sakamoto}}, \bibinfo {author}
  {\bibfnamefont {A.}~\bibnamefont {Scarabosio}}, \bibinfo {author}
  {\bibfnamefont {A.}~\bibnamefont {Bortolon}}, \bibinfo {author}
  {\bibfnamefont {K.}~\bibnamefont {Burrell}}, \bibinfo {author} {\bibfnamefont
  {B.}~\bibnamefont {Duval}}, \bibinfo {author} {\bibfnamefont
  {C.}~\bibnamefont {Fenzi-Bonizec}}, \bibinfo {author} {\bibfnamefont
  {M.}~\bibnamefont {Greenwald}}, \bibinfo {author} {\bibfnamefont
  {R.}~\bibnamefont {Groebner}}, \bibinfo {author} {\bibfnamefont
  {G.}~\bibnamefont {Hoang}}, \bibinfo {author} {\bibfnamefont
  {Y.}~\bibnamefont {Koide}}, \bibinfo {author} {\bibfnamefont
  {E.}~\bibnamefont {Marmar}}, \bibinfo {author} {\bibfnamefont
  {A.}~\bibnamefont {Pochelon}}, \ and\ \bibinfo {author} {\bibfnamefont
  {Y.}~\bibnamefont {Podpaly}},\ }\href {\doibase 10.1088/0029-5515/47/11/025}
  {\bibfield  {journal} {\bibinfo  {journal} {Nuclear Fusion}\ }\textbf
  {\bibinfo {volume} {47}},\ \bibinfo {pages} {1618} (\bibinfo {year}
  {2007})}\BibitemShut {NoStop}%
\bibitem [{\citenamefont {DeGrassie}\ \emph {et~al.}(2007)\citenamefont
  {DeGrassie}, \citenamefont {Rice}, \citenamefont {Burrell}, \citenamefont
  {Groebner},\ and\ \citenamefont {Solomon}}]{DeGrassie2007}%
  \BibitemOpen
  \bibfield  {author} {\bibinfo {author} {\bibfnamefont {J.~S.}\ \bibnamefont
  {DeGrassie}}, \bibinfo {author} {\bibfnamefont {J.~E.}\ \bibnamefont {Rice}},
  \bibinfo {author} {\bibfnamefont {K.~H.}\ \bibnamefont {Burrell}}, \bibinfo
  {author} {\bibfnamefont {R.~J.}\ \bibnamefont {Groebner}}, \ and\ \bibinfo
  {author} {\bibfnamefont {W.~M.}\ \bibnamefont {Solomon}},\ }\href
  {http://dx.doi.org/10.1063/1.2539055} {\bibfield  {journal} {\bibinfo
  {journal} {Physics of Plasmas}\ }\textbf {\bibinfo {volume} {14}},\ \bibinfo
  {pages} {056115} (\bibinfo {year} {2007})}\BibitemShut {NoStop}%
\bibitem [{\citenamefont {Noterdaeme}\ \emph {et~al.}(2003)\citenamefont
  {Noterdaeme}, \citenamefont {Righi}, \citenamefont {Chan}, \citenamefont
  {DeGrassie}, \citenamefont {Kirov}, \citenamefont {Mantsinen}, \citenamefont
  {Nave}, \citenamefont {Testa}, \citenamefont {Zastrow}, \citenamefont
  {Budny}, \citenamefont {Cesario}, \citenamefont {Gondhalekar}, \citenamefont
  {Hawkes}, \citenamefont {Hellsten}, \citenamefont {Lamalle}, \citenamefont
  {Meo}, \citenamefont {Nguyen},\ and\ \citenamefont {the
  EFDA-JET-EFDA~Contributors}}]{Noterdaeme2003}%
  \BibitemOpen
  \bibfield  {author} {\bibinfo {author} {\bibfnamefont {J.~M.}\ \bibnamefont
  {Noterdaeme}}, \bibinfo {author} {\bibfnamefont {E.}~\bibnamefont {Righi}},
  \bibinfo {author} {\bibfnamefont {V.}~\bibnamefont {Chan}}, \bibinfo {author}
  {\bibfnamefont {J.}~\bibnamefont {DeGrassie}}, \bibinfo {author}
  {\bibfnamefont {K.}~\bibnamefont {Kirov}}, \bibinfo {author} {\bibfnamefont
  {M.}~\bibnamefont {Mantsinen}}, \bibinfo {author} {\bibfnamefont {M.~F.~F.}\
  \bibnamefont {Nave}}, \bibinfo {author} {\bibfnamefont {D.}~\bibnamefont
  {Testa}}, \bibinfo {author} {\bibfnamefont {K.~D.}\ \bibnamefont {Zastrow}},
  \bibinfo {author} {\bibfnamefont {R.}~\bibnamefont {Budny}}, \bibinfo
  {author} {\bibfnamefont {R.}~\bibnamefont {Cesario}}, \bibinfo {author}
  {\bibfnamefont {A.}~\bibnamefont {Gondhalekar}}, \bibinfo {author}
  {\bibfnamefont {N.}~\bibnamefont {Hawkes}}, \bibinfo {author} {\bibfnamefont
  {T.}~\bibnamefont {Hellsten}}, \bibinfo {author} {\bibfnamefont
  {P.}~\bibnamefont {Lamalle}}, \bibinfo {author} {\bibfnamefont
  {F.}~\bibnamefont {Meo}}, \bibinfo {author} {\bibfnamefont {F.}~\bibnamefont
  {Nguyen}}, \ and\ \bibinfo {author} {\bibnamefont {the
  EFDA-JET-EFDA~Contributors}},\ }\href {\doibase 10.1088/0029-5515/43/4/309}
  {\bibfield  {journal} {\bibinfo  {journal} {Nuclear Fusion}\ }\textbf
  {\bibinfo {volume} {43}},\ \bibinfo {pages} {274} (\bibinfo {year}
  {2003})}\BibitemShut {NoStop}%
\bibitem [{\citenamefont {Lee}\ \emph {et~al.}(2014)\citenamefont {Lee},
  \citenamefont {Parra},\ and\ \citenamefont {Barnes}}]{Lee2014}%
  \BibitemOpen
  \bibfield  {author} {\bibinfo {author} {\bibfnamefont {J.}~\bibnamefont
  {Lee}}, \bibinfo {author} {\bibfnamefont {F.~I.}\ \bibnamefont {Parra}}, \
  and\ \bibinfo {author} {\bibfnamefont {M.}~\bibnamefont {Barnes}},\ }\href
  {\doibase 10.1088/0029-5515/54/2/022002} {\bibfield  {journal} {\bibinfo
  {journal} {Nuclear Fusion}\ }\textbf {\bibinfo {volume} {54}},\ \bibinfo
  {pages} {022002} (\bibinfo {year} {2014})}\BibitemShut {NoStop}%
\bibitem [{\citenamefont {Barnes}\ \emph {et~al.}(2013)\citenamefont {Barnes},
  \citenamefont {Parra}, \citenamefont {Lee}, \citenamefont {Belli},
  \citenamefont {Nave},\ and\ \citenamefont {White}}]{Barnes2013}%
  \BibitemOpen
  \bibfield  {author} {\bibinfo {author} {\bibfnamefont {M.}~\bibnamefont
  {Barnes}}, \bibinfo {author} {\bibfnamefont {F.~I.}\ \bibnamefont {Parra}},
  \bibinfo {author} {\bibfnamefont {J.~P.}\ \bibnamefont {Lee}}, \bibinfo
  {author} {\bibfnamefont {E.~A.}\ \bibnamefont {Belli}}, \bibinfo {author}
  {\bibfnamefont {M.~F.~F.}\ \bibnamefont {Nave}}, \ and\ \bibinfo {author}
  {\bibfnamefont {A.~E.}\ \bibnamefont {White}},\ }\href {\doibase
  10.1103/PhysRevLett.111.055005} {\bibfield  {journal} {\bibinfo  {journal}
  {Physical Review Letters}\ }\textbf {\bibinfo {volume} {111}},\ \bibinfo
  {pages} {1} (\bibinfo {year} {2013})}\BibitemShut {NoStop}%
\bibitem [{\citenamefont {Poli}\ \emph {et~al.}(2014)\citenamefont {Poli},
  \citenamefont {Kessel}, \citenamefont {Bonoli}, \citenamefont {Batchelor},
  \citenamefont {Harvey},\ and\ \citenamefont {Snyder}}]{Poli2014}%
  \BibitemOpen
  \bibfield  {author} {\bibinfo {author} {\bibfnamefont {F.}~\bibnamefont
  {Poli}}, \bibinfo {author} {\bibfnamefont {C.}~\bibnamefont {Kessel}},
  \bibinfo {author} {\bibfnamefont {P.}~\bibnamefont {Bonoli}}, \bibinfo
  {author} {\bibfnamefont {D.}~\bibnamefont {Batchelor}}, \bibinfo {author}
  {\bibfnamefont {R.}~\bibnamefont {Harvey}}, \ and\ \bibinfo {author}
  {\bibfnamefont {P.}~\bibnamefont {Snyder}},\ }\href {\doibase
  10.1088/0029-5515/54/7/073007} {\bibfield  {journal} {\bibinfo  {journal}
  {Nuclear Fusion}\ }\textbf {\bibinfo {volume} {54}},\ \bibinfo {pages}
  {073007} (\bibinfo {year} {2014})}\BibitemShut {NoStop}%
\bibitem [{\citenamefont {Elwasif}\ \emph {et~al.}(2010)\citenamefont
  {Elwasif}, \citenamefont {Bernholdt}, \citenamefont {Shet}, \citenamefont
  {Foley}, \citenamefont {Bramley}, \citenamefont {Batchelor},\ and\
  \citenamefont {Berry}}]{Elwasif2010}%
  \BibitemOpen
  \bibfield  {author} {\bibinfo {author} {\bibfnamefont {W.~R.}\ \bibnamefont
  {Elwasif}}, \bibinfo {author} {\bibfnamefont {D.~E.}\ \bibnamefont
  {Bernholdt}}, \bibinfo {author} {\bibfnamefont {A.~G.}\ \bibnamefont {Shet}},
  \bibinfo {author} {\bibfnamefont {S.~S.}\ \bibnamefont {Foley}}, \bibinfo
  {author} {\bibfnamefont {R.}~\bibnamefont {Bramley}}, \bibinfo {author}
  {\bibfnamefont {D.~B.}\ \bibnamefont {Batchelor}}, \ and\ \bibinfo {author}
  {\bibfnamefont {L.~A.}\ \bibnamefont {Berry}},\ }in\ \href {\doibase
  10.1109/PDP.2010.63} {\emph {\bibinfo {booktitle} {Proceedings of the 18th
  Euromicro Conference on Parallel, Distributed and \\ Network-Based
  Processing}}}\ (\bibinfo {year} {2010})\ pp.\ \bibinfo {pages}
  {419--427}\BibitemShut {NoStop}%
\bibitem [{\citenamefont {Jardin}\ \emph {et~al.}(1986)\citenamefont {Jardin},
  \citenamefont {Pomphrey},\ and\ \citenamefont {Delucia}}]{Jardin1986}%
  \BibitemOpen
  \bibfield  {author} {\bibinfo {author} {\bibfnamefont {S.~C.}\ \bibnamefont
  {Jardin}}, \bibinfo {author} {\bibfnamefont {N.}~\bibnamefont {Pomphrey}}, \
  and\ \bibinfo {author} {\bibfnamefont {J.}~\bibnamefont {Delucia}},\ }\href
  {\doibase 10.1016/0021-9991(86)90077-X} {\bibfield  {journal} {\bibinfo
  {journal} {Journal of Computational Physics}\ }\textbf {\bibinfo {volume}
  {66}},\ \bibinfo {pages} {481} (\bibinfo {year} {1986})}\BibitemShut
  {NoStop}%
\bibitem [{\citenamefont {Hawryluk}(1979)}]{Hawryluk1980}%
  \BibitemOpen
  \bibfield  {author} {\bibinfo {author} {\bibfnamefont {R.}~\bibnamefont
  {Hawryluk}},\ }in\ \href {http://w3.pppl.gov/transp/papers/Hawryluk.pdf}
  {\emph {\bibinfo {booktitle} {Physics of Plasmas Close to Thermonuclear
  Conditions}}}\ (\bibinfo {year} {1979})\ pp.\ \bibinfo {pages}
  {19--43}\BibitemShut {NoStop}%
\bibitem [{\citenamefont {Fukuyama}\ \emph {et~al.}(1995)\citenamefont
  {Fukuyama}, \citenamefont {Itoh}, \citenamefont {Itoh}, \citenamefont
  {Yagi},\ and\ \citenamefont {Azumi}}]{Fukuyama1995}%
  \BibitemOpen
  \bibfield  {author} {\bibinfo {author} {\bibfnamefont {A.}~\bibnamefont
  {Fukuyama}}, \bibinfo {author} {\bibfnamefont {K.}~\bibnamefont {Itoh}},
  \bibinfo {author} {\bibfnamefont {S.~I.}\ \bibnamefont {Itoh}}, \bibinfo
  {author} {\bibfnamefont {M.}~\bibnamefont {Yagi}}, \ and\ \bibinfo {author}
  {\bibfnamefont {M.}~\bibnamefont {Azumi}},\ }\href
  {http://iopscience.iop.org/article/10.1088/0741-3335/37/6/002/meta}
  {\bibfield  {journal} {\bibinfo  {journal} {Plasma Physics and Controlled
  Fusion}\ }\textbf {\bibinfo {volume} {37}},\ \bibinfo {pages} {611} (\bibinfo
  {year} {1995})}\BibitemShut {NoStop}%
\bibitem [{\citenamefont {Fukuyama}\ \emph {et~al.}(1998)\citenamefont
  {Fukuyama}, \citenamefont {Takatsuka}, \citenamefont {Itoh}, \citenamefont
  {Yagi},\ and\ \citenamefont {Itoh}}]{Fukuyama1998}%
  \BibitemOpen
  \bibfield  {author} {\bibinfo {author} {\bibfnamefont {A.}~\bibnamefont
  {Fukuyama}}, \bibinfo {author} {\bibfnamefont {S.}~\bibnamefont {Takatsuka}},
  \bibinfo {author} {\bibfnamefont {S.-I.}\ \bibnamefont {Itoh}}, \bibinfo
  {author} {\bibfnamefont {M.}~\bibnamefont {Yagi}}, \ and\ \bibinfo {author}
  {\bibfnamefont {K.}~\bibnamefont {Itoh}},\ }\href {\doibase
  10.1088/0741-3335/40/5/016} {\bibfield  {journal} {\bibinfo  {journal}
  {Plasma Physics and Controlled Fusion}\ }\textbf {\bibinfo {volume} {40}},\
  \bibinfo {pages} {653} (\bibinfo {year} {1998})}\BibitemShut {NoStop}%
\bibitem [{\citenamefont {Snyder}\ \emph {et~al.}(2011)\citenamefont {Snyder},
  \citenamefont {Groebner}, \citenamefont {Hughes}, \citenamefont {Osborne},
  \citenamefont {Beurskens}, \citenamefont {Leonard}, \citenamefont {Wilson},\
  and\ \citenamefont {Xu}}]{Snyder2011}%
  \BibitemOpen
  \bibfield  {author} {\bibinfo {author} {\bibfnamefont {P.}~\bibnamefont
  {Snyder}}, \bibinfo {author} {\bibfnamefont {R.}~\bibnamefont {Groebner}},
  \bibinfo {author} {\bibfnamefont {J.}~\bibnamefont {Hughes}}, \bibinfo
  {author} {\bibfnamefont {T.}~\bibnamefont {Osborne}}, \bibinfo {author}
  {\bibfnamefont {M.}~\bibnamefont {Beurskens}}, \bibinfo {author}
  {\bibfnamefont {A.}~\bibnamefont {Leonard}}, \bibinfo {author} {\bibfnamefont
  {H.}~\bibnamefont {Wilson}}, \ and\ \bibinfo {author} {\bibfnamefont
  {X.}~\bibnamefont {Xu}},\ }\href {\doibase 10.1088/0029-5515/51/10/103016}
  {\bibfield  {journal} {\bibinfo  {journal} {Nuclear Fusion}\ }\textbf
  {\bibinfo {volume} {51}},\ \bibinfo {pages} {103016} (\bibinfo {year}
  {2011})}\BibitemShut {NoStop}%
\bibitem [{\citenamefont {Goldston}\ \emph {et~al.}(1981)\citenamefont
  {Goldston}, \citenamefont {McCune}, \citenamefont {Towner}, \citenamefont
  {Davis}, \citenamefont {Hawryluk},\ and\ \citenamefont
  {Schmidt}}]{Goldston1981}%
  \BibitemOpen
  \bibfield  {author} {\bibinfo {author} {\bibfnamefont {R.~J.}\ \bibnamefont
  {Goldston}}, \bibinfo {author} {\bibfnamefont {D.~C.}\ \bibnamefont
  {McCune}}, \bibinfo {author} {\bibfnamefont {H.~H.}\ \bibnamefont {Towner}},
  \bibinfo {author} {\bibfnamefont {S.~L.}\ \bibnamefont {Davis}}, \bibinfo
  {author} {\bibfnamefont {R.~J.}\ \bibnamefont {Hawryluk}}, \ and\ \bibinfo
  {author} {\bibfnamefont {G.~L.}\ \bibnamefont {Schmidt}},\ }\href {\doibase
  10.1016/0021-9991(81)90111-X} {\bibfield  {journal} {\bibinfo  {journal}
  {Journal of Computational Physics}\ }\textbf {\bibinfo {volume} {43}},\
  \bibinfo {pages} {61} (\bibinfo {year} {1981})}\BibitemShut {NoStop}%
\bibitem [{\citenamefont {Pankin}\ \emph {et~al.}(2004)\citenamefont {Pankin},
  \citenamefont {McCune}, \citenamefont {Andre}, \citenamefont {Bateman},\ and\
  \citenamefont {Kritz}}]{Pankin2004}%
  \BibitemOpen
  \bibfield  {author} {\bibinfo {author} {\bibfnamefont {A.}~\bibnamefont
  {Pankin}}, \bibinfo {author} {\bibfnamefont {D.}~\bibnamefont {McCune}},
  \bibinfo {author} {\bibfnamefont {R.}~\bibnamefont {Andre}}, \bibinfo
  {author} {\bibfnamefont {G.}~\bibnamefont {Bateman}}, \ and\ \bibinfo
  {author} {\bibfnamefont {A.}~\bibnamefont {Kritz}},\ }\href {\doibase
  10.1016/j.cpc.2003.11.002} {\bibfield  {journal} {\bibinfo  {journal}
  {Computer Physics Communications}\ }\textbf {\bibinfo {volume} {159}},\
  \bibinfo {pages} {157} (\bibinfo {year} {2004})}\BibitemShut {NoStop}%
\bibitem [{\citenamefont {Shinohara}\ \emph {et~al.}(2009)\citenamefont
  {Shinohara}, \citenamefont {Oikawa}, \citenamefont {Urano}, \citenamefont
  {Oyama}, \citenamefont {Lonnroth}, \citenamefont {Saibene}, \citenamefont
  {Parail},\ and\ \citenamefont {Kamada}}]{Shinohara2009}%
  \BibitemOpen
  \bibfield  {author} {\bibinfo {author} {\bibfnamefont {K.}~\bibnamefont
  {Shinohara}}, \bibinfo {author} {\bibfnamefont {T.}~\bibnamefont {Oikawa}},
  \bibinfo {author} {\bibfnamefont {H.}~\bibnamefont {Urano}}, \bibinfo
  {author} {\bibfnamefont {N.}~\bibnamefont {Oyama}}, \bibinfo {author}
  {\bibfnamefont {J.}~\bibnamefont {Lonnroth}}, \bibinfo {author}
  {\bibfnamefont {G.}~\bibnamefont {Saibene}}, \bibinfo {author} {\bibfnamefont
  {V.}~\bibnamefont {Parail}}, \ and\ \bibinfo {author} {\bibfnamefont
  {Y.}~\bibnamefont {Kamada}},\ }\href {\doibase
  10.1016/j.fusengdes.2008.08.040} {\bibfield  {journal} {\bibinfo  {journal}
  {Fusion Engineering and Design}\ }\textbf {\bibinfo {volume} {84}},\ \bibinfo
  {pages} {24} (\bibinfo {year} {2009})}\BibitemShut {NoStop}%
\bibitem [{\citenamefont {Rosenbluth}\ and\ \citenamefont
  {Hinton}(1996)}]{Rosenbluth1996}%
  \BibitemOpen
  \bibfield  {author} {\bibinfo {author} {\bibfnamefont {M.~N.}\ \bibnamefont
  {Rosenbluth}}\ and\ \bibinfo {author} {\bibfnamefont {F.~L.}\ \bibnamefont
  {Hinton}},\ }\href {\doibase 10.1088/0029-5515/36/1/I04} {\bibfield
  {journal} {\bibinfo  {journal} {Nuclear Fusion}\ }\textbf {\bibinfo {volume}
  {36}},\ \bibinfo {pages} {55} (\bibinfo {year} {1996})}\BibitemShut {NoStop}%
\bibitem [{\citenamefont {Lee}(2013)}]{LeeThesis}%
  \BibitemOpen
  \bibfield  {author} {\bibinfo {author} {\bibfnamefont {J.}~\bibnamefont
  {Lee}},\ }\emph {\bibinfo {title} {Theoretical Study of Ion Toroidal Rotation
  in the Presence of Lower Hybrid Current Drive in a Tokamak}},\ \href@noop {}
  {Ph.D. thesis},\ \bibinfo  {school} {Massachusetts Institute of Technology}
  (\bibinfo {year} {2013})\BibitemShut {NoStop}%
\bibitem [{\citenamefont {Hillesheim}\ \emph {et~al.}(2015)\citenamefont
  {Hillesheim}, \citenamefont {Parra}, \citenamefont {Barnes}, \citenamefont
  {Crocker}, \citenamefont {Meyer}, \citenamefont {Peebles}, \citenamefont
  {Scannell},\ and\ \citenamefont {Thornton}}]{Hillesheim2015}%
  \BibitemOpen
  \bibfield  {author} {\bibinfo {author} {\bibfnamefont {J.}~\bibnamefont
  {Hillesheim}}, \bibinfo {author} {\bibfnamefont {F.}~\bibnamefont {Parra}},
  \bibinfo {author} {\bibfnamefont {M.}~\bibnamefont {Barnes}}, \bibinfo
  {author} {\bibfnamefont {N.}~\bibnamefont {Crocker}}, \bibinfo {author}
  {\bibfnamefont {H.}~\bibnamefont {Meyer}}, \bibinfo {author} {\bibfnamefont
  {W.}~\bibnamefont {Peebles}}, \bibinfo {author} {\bibfnamefont
  {R.}~\bibnamefont {Scannell}}, \ and\ \bibinfo {author} {\bibfnamefont
  {A.}~\bibnamefont {Thornton}},\ }\href {\doibase
  10.1088/0029-5515/55/3/032003} {\bibfield  {journal} {\bibinfo  {journal}
  {Nuclear Fusion}\ }\textbf {\bibinfo {volume} {55}},\ \bibinfo {pages}
  {032003} (\bibinfo {year} {2015})}\BibitemShut {NoStop}%
\bibitem [{\citenamefont {Chrystal}\ \emph {et~al.}(2017)\citenamefont
  {Chrystal}, \citenamefont {Grierson}, \citenamefont {Solomon}, \citenamefont
  {Tala}, \citenamefont {DeGrassie}, \citenamefont {Petty}, \citenamefont
  {Salmi},\ and\ \citenamefont {Burrell}}]{Chrystal2017}%
  \BibitemOpen
  \bibfield  {author} {\bibinfo {author} {\bibfnamefont {C.}~\bibnamefont
  {Chrystal}}, \bibinfo {author} {\bibfnamefont {B.~A.}\ \bibnamefont
  {Grierson}}, \bibinfo {author} {\bibfnamefont {W.~M.}\ \bibnamefont
  {Solomon}}, \bibinfo {author} {\bibfnamefont {T.}~\bibnamefont {Tala}},
  \bibinfo {author} {\bibfnamefont {J.~S.}\ \bibnamefont {DeGrassie}}, \bibinfo
  {author} {\bibfnamefont {C.~C.}\ \bibnamefont {Petty}}, \bibinfo {author}
  {\bibfnamefont {A.}~\bibnamefont {Salmi}}, \ and\ \bibinfo {author}
  {\bibfnamefont {K.~H.}\ \bibnamefont {Burrell}},\ }\href {\doibase
  10.1063/1.4978563} {\bibfield  {journal} {\bibinfo  {journal} {Physics of
  Plasmas}\ }\textbf {\bibinfo {volume} {24}},\ \bibinfo {pages} {042501}
  (\bibinfo {year} {2017})}\BibitemShut {NoStop}%
\bibitem [{\citenamefont {Berkery}\ \emph {et~al.}(2010)\citenamefont
  {Berkery}, \citenamefont {Sabbagh}, \citenamefont {Betti}, \citenamefont
  {Hu}, \citenamefont {Bell}, \citenamefont {Gerhardt}, \citenamefont
  {Manickam},\ and\ \citenamefont {Tritz}}]{Berkery2010}%
  \BibitemOpen
  \bibfield  {author} {\bibinfo {author} {\bibfnamefont {J.~W.}\ \bibnamefont
  {Berkery}}, \bibinfo {author} {\bibfnamefont {S.~A.}\ \bibnamefont
  {Sabbagh}}, \bibinfo {author} {\bibfnamefont {R.}~\bibnamefont {Betti}},
  \bibinfo {author} {\bibfnamefont {B.}~\bibnamefont {Hu}}, \bibinfo {author}
  {\bibfnamefont {R.~E.}\ \bibnamefont {Bell}}, \bibinfo {author}
  {\bibfnamefont {S.~P.}\ \bibnamefont {Gerhardt}}, \bibinfo {author}
  {\bibfnamefont {J.}~\bibnamefont {Manickam}}, \ and\ \bibinfo {author}
  {\bibfnamefont {K.}~\bibnamefont {Tritz}},\ }\href {\doibase
  10.1103/PhysRevLett.104.035003} {\bibfield  {journal} {\bibinfo  {journal}
  {Physical Review Letters}\ }\textbf {\bibinfo {volume} {104}},\ \bibinfo
  {pages} {1} (\bibinfo {year} {2010})}\BibitemShut {NoStop}%
\bibitem [{\citenamefont {Liu}\ \emph {et~al.}(2009)\citenamefont {Liu},
  \citenamefont {Chu}, \citenamefont {Chapman},\ and\ \citenamefont
  {Hender}}]{Liu2009}%
  \BibitemOpen
  \bibfield  {author} {\bibinfo {author} {\bibfnamefont {Y.}~\bibnamefont
  {Liu}}, \bibinfo {author} {\bibfnamefont {M.}~\bibnamefont {Chu}}, \bibinfo
  {author} {\bibfnamefont {I.~T.}\ \bibnamefont {Chapman}}, \ and\ \bibinfo
  {author} {\bibfnamefont {T.}~\bibnamefont {Hender}},\ }\href {\doibase
  10.1088/0029-5515/49/3/035004} {\bibfield  {journal} {\bibinfo  {journal}
  {Nuclear Fusion}\ }\textbf {\bibinfo {volume} {49}},\ \bibinfo {pages}
  {035004} (\bibinfo {year} {2009})}\BibitemShut {NoStop}%
\bibitem [{\citenamefont {Hinton}\ and\ \citenamefont
  {Hazeltine}(1976)}]{Hinton1976}%
  \BibitemOpen
  \bibfield  {author} {\bibinfo {author} {\bibfnamefont {L.}~\bibnamefont
  {Hinton}}\ and\ \bibinfo {author} {\bibfnamefont {R.~D.}\ \bibnamefont
  {Hazeltine}},\ }\href {https://doi.org/10.1103/RevModPhys.48.239} {\bibfield
  {journal} {\bibinfo  {journal} {Reviews of Modern Physics}\ }\textbf
  {\bibinfo {volume} {48}},\ \bibinfo {pages} {239} (\bibinfo {year}
  {1976})}\BibitemShut {NoStop}%
\bibitem [{\citenamefont {Hirshman}\ and\ \citenamefont
  {Sigmar}(1981)}]{Hirshman1981}%
  \BibitemOpen
  \bibfield  {author} {\bibinfo {author} {\bibfnamefont {S.}~\bibnamefont
  {Hirshman}}\ and\ \bibinfo {author} {\bibfnamefont {D.}~\bibnamefont
  {Sigmar}},\ }\href {http://iopscience.iop.org/0029-5515/21/9/003} {\bibfield
  {journal} {\bibinfo  {journal} {Nuclear Fusion}\ }\textbf {\bibinfo {volume}
  {21}},\ \bibinfo {pages} {1079} (\bibinfo {year} {1981})}\BibitemShut
  {NoStop}%
\bibitem [{\citenamefont {Callen}(2011)}]{Callen2011}%
  \BibitemOpen
  \bibfield  {author} {\bibinfo {author} {\bibfnamefont {J.~D.}\ \bibnamefont
  {Callen}},\ }\href {\doibase 10.1088/0029-5515/51/9/094026} {\bibfield
  {journal} {\bibinfo  {journal} {Nuclear Fusion}\ }\textbf {\bibinfo {volume}
  {51}},\ \bibinfo {pages} {094026} (\bibinfo {year} {2011})}\BibitemShut
  {NoStop}%
\bibitem [{\citenamefont {Sun}\ \emph {et~al.}(2011)\citenamefont {Sun},
  \citenamefont {Liang}, \citenamefont {Shaing}, \citenamefont {Koslowski},
  \citenamefont {Wiegmann},\ and\ \citenamefont {Zhang}}]{Sun2011}%
  \BibitemOpen
  \bibfield  {author} {\bibinfo {author} {\bibfnamefont {Y.}~\bibnamefont
  {Sun}}, \bibinfo {author} {\bibfnamefont {Y.}~\bibnamefont {Liang}}, \bibinfo
  {author} {\bibfnamefont {K.}~\bibnamefont {Shaing}}, \bibinfo {author}
  {\bibfnamefont {H.}~\bibnamefont {Koslowski}}, \bibinfo {author}
  {\bibfnamefont {C.}~\bibnamefont {Wiegmann}}, \ and\ \bibinfo {author}
  {\bibfnamefont {T.}~\bibnamefont {Zhang}},\ }\href {\doibase
  10.1088/0029-5515/51/5/053015} {\bibfield  {journal} {\bibinfo  {journal}
  {Nuclear Fusion}\ }\textbf {\bibinfo {volume} {51}},\ \bibinfo {pages}
  {053015} (\bibinfo {year} {2011})}\BibitemShut {NoStop}%
\bibitem [{\citenamefont {Sauter}\ \emph {et~al.}(1999)\citenamefont {Sauter},
  \citenamefont {Angioni},\ and\ \citenamefont {Lin-Liu}}]{Sauter1999}%
  \BibitemOpen
  \bibfield  {author} {\bibinfo {author} {\bibfnamefont {O.}~\bibnamefont
  {Sauter}}, \bibinfo {author} {\bibfnamefont {C.}~\bibnamefont {Angioni}}, \
  and\ \bibinfo {author} {\bibfnamefont {Y.~R.}\ \bibnamefont {Lin-Liu}},\
  }\href {http://dx.doi.org/10.1063/1.873240} {\bibfield  {journal} {\bibinfo
  {journal} {Physics of Plasmas}\ }\textbf {\bibinfo {volume} {6}},\ \bibinfo
  {pages} {2834} (\bibinfo {year} {1999})}\BibitemShut {NoStop}%
\bibitem [{\citenamefont {Albert}\ \emph {et~al.}(2016)\citenamefont {Albert},
  \citenamefont {Heyn}, \citenamefont {Kapper}, \citenamefont {Kasilov},
  \citenamefont {Kernbichler},\ and\ \citenamefont {Martitsch}}]{Albert2016}%
  \BibitemOpen
  \bibfield  {author} {\bibinfo {author} {\bibfnamefont {C.~G.}\ \bibnamefont
  {Albert}}, \bibinfo {author} {\bibfnamefont {M.~F.}\ \bibnamefont {Heyn}},
  \bibinfo {author} {\bibfnamefont {G.}~\bibnamefont {Kapper}}, \bibinfo
  {author} {\bibfnamefont {S.~V.}\ \bibnamefont {Kasilov}}, \bibinfo {author}
  {\bibfnamefont {W.}~\bibnamefont {Kernbichler}}, \ and\ \bibinfo {author}
  {\bibfnamefont {A.~F.}\ \bibnamefont {Martitsch}},\ }\href
  {http://dx.doi.org/10.1063/1.4961084} {\bibfield  {journal} {\bibinfo
  {journal} {Physics of Plasmas}\ }\textbf {\bibinfo {volume} {23}} (\bibinfo
  {year} {2016})}\BibitemShut {NoStop}%
\bibitem [{\citenamefont {Shaing}(1986)}]{Shaing1986}%
  \BibitemOpen
  \bibfield  {author} {\bibinfo {author} {\bibfnamefont {K.~C.}\ \bibnamefont
  {Shaing}},\ }\href {http://dx.doi.org/10.1063/1.865561} {\bibfield  {journal}
  {\bibinfo  {journal} {Physics of Fluids}\ }\textbf {\bibinfo {volume} {29}},\
  \bibinfo {pages} {2231} (\bibinfo {year} {1986})}\BibitemShut {NoStop}%
\bibitem [{\citenamefont {Shaing}(2006)}]{Shaing2006}%
  \BibitemOpen
  \bibfield  {author} {\bibinfo {author} {\bibfnamefont {K.~C.}\ \bibnamefont
  {Shaing}},\ }\href {\doibase 10.1063/1.2198214} {\bibfield  {journal}
  {\bibinfo  {journal} {Physics of Plasmas}\ }\textbf {\bibinfo {volume} {3}},\
  \bibinfo {pages} {4276} (\bibinfo {year} {2006})}\BibitemShut {NoStop}%
\bibitem [{\citenamefont {Connor}\ and\ \citenamefont
  {Hastie}(1974)}]{Connor1974}%
  \BibitemOpen
  \bibfield  {author} {\bibinfo {author} {\bibfnamefont {J.~W.}\ \bibnamefont
  {Connor}}\ and\ \bibinfo {author} {\bibfnamefont {R.~J.}\ \bibnamefont
  {Hastie}},\ }\href {\doibase 10.1063/1.1694573} {\bibfield  {journal}
  {\bibinfo  {journal} {Physics of Fluids}\ }\textbf {\bibinfo {volume} {17}},\
  \bibinfo {pages} {114} (\bibinfo {year} {1974})}\BibitemShut {NoStop}%
\bibitem [{\citenamefont {Shaing}\ \emph
  {et~al.}(2009{\natexlab{c}})\citenamefont {Shaing}, \citenamefont {Sabbagh},\
  and\ \citenamefont {Chu}}]{Shaing2009_sb}%
  \BibitemOpen
  \bibfield  {author} {\bibinfo {author} {\bibfnamefont {K.~C.}\ \bibnamefont
  {Shaing}}, \bibinfo {author} {\bibfnamefont {S.~A.}\ \bibnamefont {Sabbagh}},
  \ and\ \bibinfo {author} {\bibfnamefont {M.~S.}\ \bibnamefont {Chu}},\ }\href
  {\doibase 10.1088/0741-3335/51/5/055003} {\bibfield  {journal} {\bibinfo
  {journal} {Plasma Physics and Controlled Fusion}\ }\textbf {\bibinfo {volume}
  {51}},\ \bibinfo {pages} {055003} (\bibinfo {year}
  {2009}{\natexlab{c}})}\BibitemShut {NoStop}%
\bibitem [{\citenamefont {Calvo}\ \emph {et~al.}(2014)\citenamefont {Calvo},
  \citenamefont {Parra}, \citenamefont {Alonso},\ and\ \citenamefont
  {Velasco}}]{Calvo2014}%
  \BibitemOpen
  \bibfield  {author} {\bibinfo {author} {\bibfnamefont {I.}~\bibnamefont
  {Calvo}}, \bibinfo {author} {\bibfnamefont {F.~I.}\ \bibnamefont {Parra}},
  \bibinfo {author} {\bibfnamefont {J.~A.}\ \bibnamefont {Alonso}}, \ and\
  \bibinfo {author} {\bibfnamefont {J.~L.}\ \bibnamefont {Velasco}},\ }\href
  {\doibase 10.1088/0741-3335/56/9/094003} {\bibfield  {journal} {\bibinfo
  {journal} {Plasma Physics and Controlled Fusion}\ }\textbf {\bibinfo {volume}
  {56}},\ \bibinfo {pages} {094003} (\bibinfo {year} {2014})}\BibitemShut
  {NoStop}%
\bibitem [{\citenamefont {Calvo}\ \emph {et~al.}(2017)\citenamefont {Calvo},
  \citenamefont {Parra}, \citenamefont {Velasco},\ and\ \citenamefont
  {Alonso}}]{Calvo2016}%
  \BibitemOpen
  \bibfield  {author} {\bibinfo {author} {\bibfnamefont {I.}~\bibnamefont
  {Calvo}}, \bibinfo {author} {\bibfnamefont {F.~I.}\ \bibnamefont {Parra}},
  \bibinfo {author} {\bibfnamefont {J.~L.}\ \bibnamefont {Velasco}}, \ and\
  \bibinfo {author} {\bibfnamefont {J.~A.}\ \bibnamefont {Alonso}},\ }\href
  {http://stacks.iop.org/0741-3335/59/i=5/a=055014} {\bibfield  {journal}
  {\bibinfo  {journal} {Plasma Physics and Controlled Fusion}\ }\textbf
  {\bibinfo {volume} {59}},\ \bibinfo {pages} {055014} (\bibinfo {year}
  {2017})}\BibitemShut {NoStop}%
\bibitem [{\citenamefont {Matsuoka}\ \emph {et~al.}(2015)\citenamefont
  {Matsuoka}, \citenamefont {Satake}, \citenamefont {Kanno},\ and\
  \citenamefont {Sugama}}]{Matsuoka2015}%
  \BibitemOpen
  \bibfield  {author} {\bibinfo {author} {\bibfnamefont {S.}~\bibnamefont
  {Matsuoka}}, \bibinfo {author} {\bibfnamefont {S.}~\bibnamefont {Satake}},
  \bibinfo {author} {\bibfnamefont {R.}~\bibnamefont {Kanno}}, \ and\ \bibinfo
  {author} {\bibfnamefont {H.}~\bibnamefont {Sugama}},\ }\href
  {http://dx.doi.org/10.1063/1.4923434} {\bibfield  {journal} {\bibinfo
  {journal} {Physics of Plasmas}\ }\textbf {\bibinfo {volume} {22}} (\bibinfo
  {year} {2015})}\BibitemShut {NoStop}%
\bibitem [{\citenamefont {Sugama}\ \emph {et~al.}(2016)\citenamefont {Sugama},
  \citenamefont {Matsuoka}, \citenamefont {Satake},\ and\ \citenamefont
  {Kanno}}]{Sugama2016}%
  \BibitemOpen
  \bibfield  {author} {\bibinfo {author} {\bibfnamefont {H.}~\bibnamefont
  {Sugama}}, \bibinfo {author} {\bibfnamefont {S.}~\bibnamefont {Matsuoka}},
  \bibinfo {author} {\bibfnamefont {S.}~\bibnamefont {Satake}}, \ and\ \bibinfo
  {author} {\bibfnamefont {R.}~\bibnamefont {Kanno}},\ }\href {\doibase
  10.1063/1.4945618} {\bibfield  {journal} {\bibinfo  {journal} {Physics of
  Plasmas}\ }\textbf {\bibinfo {volume} {23}},\ \bibinfo {pages} {042502}
  (\bibinfo {year} {2016})}\BibitemShut {NoStop}%
\bibitem [{\citenamefont {Shaing}(2015)}]{Shaing2015}%
  \BibitemOpen
  \bibfield  {author} {\bibinfo {author} {\bibfnamefont {K.~C.}\ \bibnamefont
  {Shaing}},\ }\href {\doibase 10.1017/S0022377814001068} {\bibfield  {journal}
  {\bibinfo  {journal} {Journal of Plasma Physics}\ }\textbf {\bibinfo {volume}
  {81}},\ \bibinfo {pages} {905810203} (\bibinfo {year} {2015})}\BibitemShut
  {NoStop}%
\bibitem [{\citenamefont {Connor}\ \emph {et~al.}(2004)\citenamefont {Connor},
  \citenamefont {Fukuda}, \citenamefont {Garbet}, \citenamefont {Gormezano},
  \citenamefont {Mukhovatov}, \citenamefont {Wakatani}, \citenamefont {{{the
  ITB Database Group}}},\ and\ \citenamefont {{{the ITPA Topical Group on
  Transport and Internal Barrier Physics}}}}]{Connor2004}%
  \BibitemOpen
  \bibfield  {author} {\bibinfo {author} {\bibfnamefont {J.}~\bibnamefont
  {Connor}}, \bibinfo {author} {\bibfnamefont {T.}~\bibnamefont {Fukuda}},
  \bibinfo {author} {\bibfnamefont {X.}~\bibnamefont {Garbet}}, \bibinfo
  {author} {\bibfnamefont {C.}~\bibnamefont {Gormezano}}, \bibinfo {author}
  {\bibfnamefont {V.}~\bibnamefont {Mukhovatov}}, \bibinfo {author}
  {\bibfnamefont {M.}~\bibnamefont {Wakatani}}, \bibinfo {author} {\bibnamefont
  {{{the ITB Database Group}}}}, \ and\ \bibinfo {author} {\bibnamefont {{{the
  ITPA Topical Group on Transport and Internal Barrier Physics}}}},\ }\href
  {\doibase 10.1088/0029-5515/44/4/R01} {\bibfield  {journal} {\bibinfo
  {journal} {Nuclear Fusion}\ }\textbf {\bibinfo {volume} {44}},\ \bibinfo
  {pages} {R1} (\bibinfo {year} {2004})}\BibitemShut {NoStop}%
\bibitem [{\citenamefont {Waltz}\ \emph {et~al.}(1994)\citenamefont {Waltz},
  \citenamefont {Kerbel},\ and\ \citenamefont {Milovich}}]{Waltz1994}%
  \BibitemOpen
  \bibfield  {author} {\bibinfo {author} {\bibfnamefont {R.~E.}\ \bibnamefont
  {Waltz}}, \bibinfo {author} {\bibfnamefont {G.~D.}\ \bibnamefont {Kerbel}}, \
  and\ \bibinfo {author} {\bibfnamefont {J.}~\bibnamefont {Milovich}},\ }\href
  {\doibase 10.1063/1.870934} {\bibfield  {journal} {\bibinfo  {journal}
  {Physics of Plasmas}\ }\textbf {\bibinfo {volume} {1}},\ \bibinfo {pages}
  {2229} (\bibinfo {year} {1994})}\BibitemShut {NoStop}%
\end{thebibliography}%

\end{document}